# Optimization and Evaluation of Nested Queries and Procedures

Submitted in partial fulfillment of the requirements
for the degree of

Doctor of Philosophy

by

**Ravindra Guravannavar**
**Roll No. 03405702**

Advisor
**Prof. S. Sudarshan**

DEPARTMENT OF COMPUTER SCIENCE & ENGINEERING
INDIAN INSTITUTE OF TECHNOLOGY–BOMBAY

2009

To all my teachers and my grandfather.

# Abstract

Many database applications perform complex data retrieval and update tasks. Nested queries, and queries that invoke user-defined functions, which are written using a mix of procedural and SQL constructs, are often used in such applications. A straight-forward evaluation of such queries involves repeated execution of parameterized sub-queries or blocks containing queries and procedural code. Repeated execution of queries and updates also happens when external batch programs call database stored procedures repeatedly with different parameter bindings.

Iterative execution of queries and updates is often inefficient due to lack of opportunities for sharing of work, random IO, and network round-trip delays. Query decorrelation is an important technique which addresses the problem of iterative evaluation of nested queries, by rewriting them using set operations such as joins and outer-joins. Thereby, decorrelation enables the use of set-oriented plans with reduced random IO, which are often more efficient than the alternative iterative plans. However, decorrelation is not applicable to complex nested blocks such as user-defined functions and stored procedures.

The focus of this thesis is to develop query evaluation, optimization and program transformation techniques to improve the performance of repeatedly invoked tasks such as parameterized database queries, updates, stored-procedures and user-defined functions.

To do so, we first propose enhancements to iterative query execution plans which improve their efficiency by exploiting sorted parameter bindings and state retention. For several queries, even when decorrelation is applicable, an iterative plan can be the most efficient alternative. Hence, speeding up the execution of iterative plans and their inclusion in the optimizer's search space of alternative plans is important. We show how to extend a cost-based query optimizer so that the effects of sorted parameter bindings and state retention of plans are taken into account.

An important problem that arises while optimizing nested queries as well as queries with joins, aggregates and set operations is the problem of finding an optimal sort order from a factorial number of possible sort orders. Our second contribution is to show that even a special case of this problem is NP-Hard, and present practical heuristics that are effective and easy to incorporate in existing query optimizers.

We then consider iterative execution of queries and updates inside complex procedural blocks such as user-defined functions and stored procedures. Parameter batching is an important means of improving performance as it enables set-orientated processing. The key challenge to parameter batching lies in rewriting a given procedure/function to process a batch of parameter values. Our third contribution is a solution, based on program analysis and rewrite rules, to automate the generation of batched forms of procedures and replace iterative database calls within imperative loops with a single call to the batched form.

We present experimental results for all the proposed techniques, and the results show significant gains in performance.

# Contents

# List of Figures

# Chapter 1

# Introduction

Database applications with complex queries have become commonplace. For example, nested queries, queries containing complex joins and grouping, and queries that make use of procedural extensions to SQL are frequently encountered in database applications. Applications also make use of stored procedures, which use a mix of procedural language constructs and SQL.

In such applications, queries and updates to a database are often executed repeatedly, with different values for their parameters. Repeated invocations of queries and updates can occur due to several reasons. For example, consider a nested query in which a sub-query or inner query would be nested below an outer query block. Example 1.1 shows a nested query on the TPC-H schema [55]. The query retrieves *orders* none of whose *lineitems* were shipped on the day the order was placed.

***Example 1.1*** *A Nested Query*

```
SELECT  o_orderkey, o_orderdate FROM ORDERS
WHERE   o_orderdate NOT IN (SELECT l_shipdate FROM LINEITEM
                                    WHERE l_orderkey = o_orderkey);
```

Sub-queries can use parameters whose values are bound by the outer query block as illustrated in Example 1.1. The default way of executing nested queries is to iteratively invoke the sub-query for each parameter binding produced by the outer query block. Similarly, when a query makes use of a user-defined function (UDF) in its WHERE/SELECT clause, the UDF will be invoked multiple times with different values for its parameters. As a result, queries used inside the UDF get executed repeatedly. External programs that perform batch processing are another important reason for repeated invocation of queries/updates. Such programs call database stored-procedures repeatedly by iterating over a set of parameters, and as a result, cause repeated invocations of queries and updates contained in the body of the stored-procedure. Application programs, stored-procedures and user-defined functions can also make explicit use of looping constructs, such as FOR/WHILE/CURSOR loops, and invoke queries/updates inside the loop.

## 1.1 Problem Overview and Motivation

Iterative execution plans for queries and updates are often very inefficient and result in poor application performance. Lack of opportunity for sharing of work (*e.g.*, disk IO) between multiple invocations of the query/update is a key reason for the inefficiency of iterative plans. Random disk IO and network round-trip delays also degrade the performance of iterative plans. Consider the query shown in Example 1.1. A naïve nested iteration plan for the query would invoke the sub-query on the LINEITEM table for every tuple of the ORDERS relation. If a useful index is not present to answer the sub-query, the LINEITEM table would be scanned once for each invocation of the sub-query. In practice, an index on the *o_orderkey* column is expected to exist and the sub-query can be evaluated with an index lookup. However, repeated index lookups result in a large number of random IOs and can lead to poor performance.

An important technique used to address poor performance of nested queries is query decorrelation, also known as query unnesting[31, 30, 19, 9, 38, 49, 17, 12]. Query decorrelation aims at rewriting a given nested query as an equivalent non-nested query by making use of set operations such as joins and outer-joins. Example 1.2[1] shows a decorrelated form of the query in Example 1.1.

---

***Example 1.2*** *A Decorrelated Form of the Nested Query in Example 1.1*

---

```
SELECT  o_orderkey, o_orderdate
FROM    ORDERS ANTI SEMI JOIN LINEITEM ON
        o_orderdate=l_shipdate AND o_orderkey=l_orderkey;
```

---

[1]Standard SQL does not provide a construct for anti semi-join. The syntax used here is merely to illustrate the internal representation after decorrelation

By rewriting nested queries using set operations, decorrelation expands the space of alternative execution strategies, which otherwise would be restricted only to iterative execution plans. The query optimizer, can now consider set-oriented strategies, such as hash-join and merge-join to answer a nested query. The query optimizer estimates the cost of alternative plans, including the iterative execution plans, and chooses the one with the least expected cost.

---

***Example 1.3*** *A Query with UDF Invocation*

---

```
SELECT orderid FROM sellorders
WHERE mkt='NSE' AND count_offers(itemcode, amount, curcode) > 0;

INT count_offers(INT itemcode, FLOAT amount, VARCHAR curcode)
DECLARE
      FLOAT amount_usd;
BEGIN
      IF (curcode == "USD")
            amount_usd := amount;
      ELSE
            amount_usd := amount * (SELECT exchrate FROM curexch
                                    WHERE ccode = curcode);          // (q1)
      END IF

      RETURN (SELECT count(*) FROM buyoffers
                  WHERE itemid = itemcode AND price >= amount_usd);     // (q2)
END;
```

---

Although decorrelation techniques are applicable for a wide variety of nested queries, iterative execution of queries and updates can still occur due to the following reasons:

1. For several queries, even when decorrelation is applicable, an iterative execution plan might be the most efficient of the available alternatives [22, 44].

2. Known decorrelation techniques are not applicable for complex nested blocks such as those containing procedural code. For example, consider the query shown in Example 1.3. The user-defined function *count-offers* used in the *where* clause of the query forms a nested block with procedural code and sub-queries inside it. Decorrelation techniques proposed till date are not applicable to such queries, except in special cases when the body of the UDF is very simple.

3. Decorrelation is also not applicable when queries and updates are invoked from

imperative loops (*e.g.,* a *while* loop) of external programs and user-defined functions or stored procedures.

The focus of this thesis is to develop query evaluation, optimization and program transformation techniques to improve the performance of repeatedly invoked parameterized sub-queries, updates, stored-procedures and user-defined functions. To this end we make the following contributions.

## 1.2 Summary of Contributions

1. We propose new query evaluation techniques, which make use of *sort order* of parameter bindings and *state retention* across calls, to speed up the evaluation of nested queries. These alternative techniques augment the optimizer's plan space with improved nested iteration plans. We show how to extend a cost-based query optimizer so that the effects of sorted parameter bindings and state retention are taken into account.

2. We address the problem of choosing optimal sort orders during query plan generation. The problem of choosing optimal sort orders is important not only for nested queries, but also for queries containing joins, aggregates and other set operations.

3. We address the problem of efficient evaluation of repeatedly called user-defined functions and stored-procedures, which contain SQL queries and updates embedded within procedural code. We present an approach, based on program analysis and transformation, to automatically generate batched forms of procedures. Batched forms of procedures work with a set of parameter bindings and thereby enable set-oriented processing of queries/updates inside the procedure body.

We provide details of each of these contributions, below.

### 1.2.1 Improved Iterative Execution with Parameter Sorting

Parameterized sub-queries being pure (side-effect free) functions, it is possible to reorder their invocations without affecting the results. Ordered (sorted) parameter bindings provide several opportunities for speeding up the sub-query evaluation. For example, it is known [22] that ordered parameter bindings reduce random disk access and yield better cache hit ratio. We propose additional query evaluation techniques, which exploit sorted parameter bindings by retaining state across calls.

*Restartable Scan:* Nested sub-queries or queries inside user-defined functions often select tuples from a relation, one of whose columns matches the parameter value. The selection predicate involving the parameter is called *correlation predicate*. In the query of Example 1.1, the selection predicate *l_orderkey=o_orderkey* in the sub-query is a correlation predicate. A relation referenced by the sub-query is called an inner relation. If the inner relation is stored sorted (clustered) on the column appearing in the correlation predicate, making the sub-query invocations in the sorted order of parameter values allows the following: at the end of each invocation, the scan for the inner relation can remember its position in the relation and restart from that point in the next invocation. Thus, the restartable scan enables us to achieve the efficiency of set-oriented algorithms like merge-join to situations where a merge-join is not directly applicable.

*Incremental Computation of Aggregates:* Nested queries where the sub-query computes an aggregate value are often encountered in practice and are known as nested aggregate queries. For nested aggregate queries having non-equality correlation predicates ($<, \leq, >$ or $\geq$), known query processing strategies are very inefficient. We describe a strategy, which by employing a combination of restartable scan and a state retaining aggregate operator, computes the result of the nested aggregate query very efficiently.

*Plan Generation:* While considering alternative plans for a given query, it is imperative that the query optimizer must take into account query execution plans that exploit ordered parameter bindings and estimate their cost appropriately. We present a cost-model of operators, which takes into account the effect of sorted parameter bindings. We then address the problem of extending a cost-based query optimizer to take into account state retention of operators and the effect of sorted parameter bindings. We have implemented the proposed ideas and present experimental results, which show significant benefits for several classes of queries.

A key challenge in optimizing queries by taking parameter sort orders into account, is to decide the optimal sort order of parameters. This problem is important not only for nested queries, but also for queries containing joins, aggregates and other set operations. We describe it next.

### 1.2.2 Choosing Sort Orders in Query Optimization

For a given set of sub-query parameters, several sort orders are possible. Different sort orders can result in different plan costs and the query optimizer must make a judicious choice of sort orders to use. Sort orders, in general, play an important role in query processing. Algorithms that rely on sorted inputs are widely used to implement joins, grouping, duplicate elimination and other set operations. The notion of *interesting orders* [48] has allowed query optimizers to consider plans that could be locally sub-optimal, but produce ordered output beneficial for other operators, and thus be part of a globally optimal plan. However, the number of interesting orders for most operators is factorial in the number of attributes involved. For example, all possible sort orders on the set of join attributes are of interest to a merge-join. Considering the exhaustive set of sort orders is prohibitively expensive as the input sub-expressions must be optimized for each of these sort orders. The following factors make the problem of choosing sort orders non-trivial.

- Clustering and secondary indices that cover the query (an index is said to cover a query if it includes all the attributes required to answer the query) make it possible to produce some sort orders at much lesser cost than others.

- Partially available sort orders can greatly reduce the cost of intermediate sorting step. For example, if the input is sorted on attribute $a$, it is more efficient to obtain the sort order $(a, b)$ as compared to the sort order $(b, a)$.

- Attributes common to multiple joins, group-by and set operations must be taken into account for choosing globally optimal sort orders. For example, consider the following query.

  *SELECT R.a, S.b, T.c FROM R, S, T*
  *WHERE R.a = S.a AND R.b = S.b AND S.b = T.b AND S.c = T.c*
  *GROUP BY R.a, S.b, T.c;*

  A good query execution plan makes a coordinated choice of sort orders for the two joins and the group-by, so that the cost of intermediate sorting steps is minimized.

- Binary operators like merge-join, can accept any sort order on the attributes involved, but require a matching sort order from their two inputs.

Previous work on sort orders has mainly focused on inferring sort orders from functional dependencies and predicates in the input sub-expression. Simmen et.al. [51] highlight the importance of choosing good sort orders for operators like group-by, which have flexible order requirements. But their approach cannot be used for binary operators such as merge-join, which require a matching sort order from their two inputs.

We show that even a simplified version of the problem of choosing sort orders is *NP-Hard*. We then give an approach to address the general case of the problem of choosing sort orders.

1. We introduce the notion of *favorable* orders, which characterizes the set of sort orders easily producible on the the result of an expression. The notion of favorable orders allows us to consider promising sort orders during plan generation.

2. We show how a cost-based query optimizer can be extended to identify favorable orders and make use them to choose good sort orders during plan generation. Our optimizer extensions also take into account plans that may not completely produce a required sort order but produce only a part (prefix) of the required sort order. When a sort order is partly available, a partial sort operator is used to obtain the complete (desired) sort order. The partial sort operation uses a modified version of the standard external-sorting algorithm.

3. We introduce a plan refinement phase in which the sort orders chosen during plan generation are further refined to take into account attributes common to different operators, and thus reduce the cost of intermediate sorts further, when possible.

We have implemented the proposed techniques in a cost-based query optimizer, and we carry out a performance comparison of the plans generated by our optimizer with those of two widely used database systems. The results show performance benefits up to 50%.

### 1.2.3 Rewriting Procedures for Batched Bindings

Several data retrieval and update task need more expressive power than what standard SQL offers. Therefore, many applications perform database queries and updates from within procedural application code. Database systems also support stored procedures and user-defined functions (UDFs), which can use a mix of procedural constructs and SQL. Repeated execution of UDFs and stored procedures can occur if queries make use of UDFs in their WHERE or SELECT clause, or if a stored procedure is invoked from external batch processing applications. In Example 1.3, the function *counter_offers* gets invoked for every tuple in the *sellorders* table. This in turn results in multiple invocations of queries inside the function body. Known decorrelation techniques do not apply to such cases and hence most database systems are forced to choose iterative execution plans.

An important technique to speed up repeated invocation of such procedures/functions is parameter batching. Parameter batching allows the choice of efficient set-oriented plans for queries and updates. For example, the the *batched form* of the UDF *count_offers* would take a set of triplets *(itemcode, amount, curcode)* and return a set comprising of the function's results for each triplet. By working with a set of parameters, the batched form can avoid repeated calls to the queries contained inside the function body. For instance, the batched form of the UDF *count_offers*, can issue a single join query to obtain the exchange rates for all the required currencies. Assuming the set of parameters to the UDF is available in a temporary relation *pbatch*, the query issued by the batched form of *count_offers* can be as follows:

*SELECT pb.curcode, cx.exchrate FROM curexch cx, pbatch pb*
*WHERE cx.ccode = pb.curcode;*

The two key challenges in exploiting batching lie in developing techniques to

1. automatically rewrite a given procedure to work with a batch of parameters bindings. In other words, to generate the *batched form* of a given procedure. It is possible for an application developer to manually create the batched form of a procedure, but the process is time consuming and error prone.

2. automatically rewrite a program, which iteratively invokes a database procedure from within an imperative loop, such as a *while* loop, so that the rewritten program makes a single invocation of the batched form of the procedure.

We present an approach, based on program analysis, to automate the generation of batched forms of procedures and replace iterative calls by code to create parameter batch and then call the batched form. The approach comprises of a set of program transformation rules, which make use of conditions on the data dependence graph of the given program. The data dependence graph gives information about various types dependencies between program statements and is obtained through program analysis.

To the best of our knowledge no previously published work addresses the problem of rewriting procedures to accept batched bindings. Lieuwen and DeWitt[34] consider the problem of optimizing set iteration loops in database programming languages. The goal of their work is to convert nested set iteration loops arising in object-oriented database languages into joins, and they propose a program transformation based approach to achieve this. The rewrite rules proposed in our thesis share some similarities with their work, but are designed to address the problem of batching database calls made from programs written in general procedural languages. Our rewriting technique can be used with complex programs, written using a rich set of language constructs such as set-iteration (or cursor) loops, *while* loops, control flow statements (*if-then-else*) and assignment statements. Our rewrite rules make use of some of the techniques, such as *loop splitting*, known in the context of parallelizing compilers [29].

In many situations, the inter-statement data dependencies within a program may not permit set-oriented evaluation of a specific query inside the program body. We show that by appropriately reordering the program statements and by introducing temporary variables, we can enable set-oriented evaluation in large number of cases.

Our rewrite rules can conceptually be used with any language. We have prototyped the rewrite techniques for a subset of Java. We present our performance study on examples taken from three real-world applications. The results are very promising and show performance benefits up to 75% due to the proposed rewriting techniques.

## 1.3 Organization of the Thesis

This thesis is organized as follows. Chapter 2 describes our work on optimizing nested queries by exploiting parameter sort orders. This is followed by Chapter 3, which addresses the problem of choosing interesting sort orders. Chapter 4 addresses the problem of parameter batching for procedure calls. Chapter 5 concludes the thesis with some comments on possible directions for future work.

# Chapter 2

# Iterative Plans with Parameter Sorting

In many cases of iterative execution, such as those involving sub-queries or functions, the order of calls to the nested sub-query or the function does not affect the end result. Ordering the query/function invocations on the parameter values gives several opportunities for improving the efficiency. For example, System R[48] caches and reuses inner sub-query results for duplicate parameter bindings, and it uses sorted parameter bindings so that only a single result of the inner sub-query is required to be held in memory at any given time. Graefe [22] emphasizes the importance of nested iteration plans, and highlights sorting of outer tuples as an important technique to improve buffer effects.

This chapter describes additional query evaluation techniques that exploit sort order of parameter bindings. The task of a query optimizer is to consider alternative plans for a given query and choose the best, *i.e.*, the least cost plan. The optimizer must therefore consider iterative plans that exploit sort order of parameters, and estimate their cost appropriately. The second part of this chapter address the problem of extending a cost-based query optimizer to consider iterative plans exploiting parameter sort orders.

The rest of this chapter is structured as follows. After stating some basic definitions in Section 2.1, in Section 2.2 we illustrate how to make use of ordered parameter bindings to improve the efficiency of iterative query execution plans. Section 2.3 illustrates how a Volcano style cost-based optimizer [23] can be extended to consider the proposed techniques, and Section 2.4 presents experimental results and analysis. We discuss related work in Section 2.5, and summarize our work in Section 2.6.

## 2.1 Definitions

**Definition 2.1 Sort Order**

*Let s be a relational schema having n attributes,* $a_1, \ldots, a_n$*. A sort order* o *on schema s is an ordered sequence of attributes from any subset of* $\{a_1, \ldots, a_n\}$*.*

We denote sort orders by enclosing the sequence of attributes in parentheses as in $(a_1, a_4, a_5)$. A sort order $s = (a_{s1}, a_{s2}, \ldots, a_{sk})$ is said to hold on a sequence of tuples conforming to schema $s$, if the sequence has the tuples sorted on attribute $a_{s1}$ and for a given value of $a_{s1}$ the tuples are sorted on $a_{s2}$, and so on up to $a_{sk}$. Note that we ignore the sort direction (ascending/descending) in our description. Sort direction can be represented by using the complement notation, *e.g.*, $(a_1, \bar{a_2})$ can be used to represent a sequence sorted in non-decreasing order on attribute $a_1$ and then in non-increasing order on attribute $a_2$. The techniques we propose in this chapter are applicable independent of the sort direction, and hence we omit the sort direction in all our description.

**Definition 2.2 Subsumption of Sort Orders**

*Sort order* $o_1$ *is said to subsume sort order* $o_2$ *iff* $o_2$ *is a prefix of* $o_1$*.*

For example, sort order $(a, b, c)$ subsumes sort order $(a, b)$. Note that the subsumption relation forms a partial order on the set of sort orders.

## 2.2 Query Evaluation with Ordered Parameters

Query evaluation algorithms for standard relational operators, such as selection, join and grouping are well studied [20]. For example, a relational selection can be implemented as a table scan or an index lookup, and a relational join can use hashing, sorting or repeated index lookups (index nested loops join). A query execution plan comprises of operators, each of which implements such an algorithm. In the case of iterative execution of query plans, the operators are initialized with a different parameter in each iteration and then executed, which gives no opportunities for reordering or sharing of disk IO. In this section, we illustrate how the standard query evaluation techniques can be extended for improved performance by making use of the sort order of query parameters.

### 2.2.1 Restartable Table Scan

Nested sub-queries or queries inside user-defined functions often select tuples from a relation, one of whose columns matches the parameter value. The selection predicate involving the parameter is called the *correlation predicate.* In the query of Example 1.1, the selection predicate *l_orderkey=o_orderkey* in the sub-query is a correlation predicate. A relation referenced by the sub-query is called an inner relation. In Example 1.1, LINEITEM is an inner relation.

Consider an inner relation, which is stored sorted on the column that appears in an equality correlation predicate. For instance, if the LINEITEM table in Example 1.1 has a clustering index on the *l_orderkey* column, it would be stored sorted on the *l_orderkey* column. Now, if the sub-query invocations are made in the order of the parameter values, we can employ a *restartable table scan* for the inner relation. The *restartable table scan* works as described next. In the following description we assume the parameter bindings to be duplicate free. For the first value of the parameter, the scan starts from the beginning of the relation (or the first matching tuple in the clustering index) and returns all the records that match the equality correlation predicate. The scan stops on encountering a record that does not satisfy the correlation predicate. The scan *remembers* this position. For subsequent bindings of the parameter, the scan continues from the remembered position, *i.e.,* the position at which it left off in the previous binding. This allows the complete query to be evaluated with at most one full scan of the inner relation. In the above explanation, we assumed the parameter bindings to be duplicate free. Duplicate parameter values can be handled in two ways: (a) Cache the sub-query result and reuse it for subsequent invocations with the same parameter value, thus avoiding re-execution of the subquery with the same parameter values. When the parameter values are sorted, at most one result needs to be cached at any given time [48], and (b) remember the most recent parameter value *(v)* and two positions *segstart* and *segend* for the inner relation scan - *segstart* positioned at the first tuple which matched the parameter value and *segend* positioned at the last tuple that matched the parameter value. If a new call has the same parameter value *(v)*, continue the scan from *segstart*, otherwise continue the scan from *segend.* In general, the approach of caching the sub-query result is more efficient unless the subquery result for each invocation is too large to fit in memory.

Apart from clustering index, *query covering* secondary indices also allow the use of *restartable table scan.* An index is said to *cover* a query, if the index leaf pages contain all the columns required to answer the query. By having additional columns (columns in addition to the index key) in the index leaf pages, the index supports reading tuples in the index key order without incurring random IO to fetch data pages. Thus query covering indices make it possible to efficiently obtain alternative sort orders for the same relation, and are being used increasingly in read intensive applications.

### Cost Model

Given an iterative plan, traditional query optimizers estimate the plan cost as follows. The plan cost for the inner and outer sub-plans are computed independently by adding the individual operator costs. The cost of the inner sub-plan is then multiplied by the estimated number of times it is invoked, which is the number of distinct parameter values bound from the outer query block. With operators such as restartable scan, which retain state across calls, such a model of computing the plan cost cannot be used. The solution is to cost a plan for $n$ invocations at once. Accordingly, each operator in the plan must have a cost function, which takes into account the number of invocations.

Thus, the cost of restartable scan for $n$ invocations with parameterized selection predicate $p$ on relation $r$ having $B_r$ disk blocks is computed as follows.

$$\text{restart-scan::cost}(r, p, n) = \begin{cases} B_r & \text{if } r \text{ is sorted to support } p \\ INF & \text{otherwise. (the operator cannot be used)} \end{cases}$$

On the other hand, if a plain relation scan were to be employed, the relation would be scanned $n$ times over all the iterations, amounting to a cost of $n \times B_r$.

A plan that employs ordered parameter bindings and restartable scan has the following advantages over a plan that employs naïve iterative index lookups.

1. Performs sequential reads and hence incurs reduced seek time and permits prefetching of disk blocks.

2. If more than one record from the same data page are needed, parameter sorting guarantees that the page is accessed exactly once irrespective of the buffer replacement policy.

However, if a query requires very small number of tuples from the inner relation, an index lookup plan is generally more efficient than the restartable scan. In such cases the index lookup plan avoids reading most of the relation while the restartable scan has to perform a complete scan of the relation. The iterative index lookup plan can however benefit from sorted parameter bindings as we shall see in Section 2.2.2.

The effect produced by a restartable scan is similar to that of a merge join. In essence, the restartable scan extends the benefits of merge-join to iterative plans. This is important since merge-join is not directly applicable to complex nested blocks such as user-defined functions with embedded queries.

### 2.2.2 Clustered Index Scan with Parameter Sorting

If a clustering index exists for the inner relation on the column that is involved in the correlation predicate and if the query requires a small number of tuples from the inner relation, it is often more efficient to employ iterative index lookups as against a restartable scan. However, a naïve iterative index lookup plan leads to random IO. Performance of clustered index lookups in the evaluation of correlated nested queries can be greatly improved by producing the parameter bindings in sorted order [22]. Sorting ensures sequential I/O and therefore permits prefetching. Further, sorting of parameters ensures each data page is fetched at most once irrespective of the buffer replacement policy. In this section, we derive a cost model for iterative clustered index lookups with sorted keys. An accurate cost-model, which takes into account the benefits of parameter sorting, is essential for the optimizer to pick the overall best plan.

#### Cost of Clustered Index Lookups with Sorted Parameters

For ease of illustration, we assume the outer query block references a single relation $R$ and the inner block references a single relation $S$. We assume the following statistics are available for the optimizer.

- Number of blocks occupied by the outer relation = $B_r$
- Number of tuples in the outer relation = $N_r$
- Number of blocks occupied by the inner relation = $B_s$
- Number of tuples in the inner relation = $N_s$
- Tuples per block for the inner relation = $F_s$

- Number of distinct values for the attribute of $S$ involved in the correlation predicate $= d$.
- Number of inner relation tuples that match each value of the correlation variable $= C_s$. Assuming uniform distribution, $C_s = N_s/d$.
- Combined selectivity of all the simple predicates in the outer block $= S_o$
- $t_t$: Transfer time for a block, default value 0.1 msec for a 4K block at 40 MB/s
- $t_s$: Seek time, default value 4 msec

**Cost Estimate Without Sorting**

When the correlation bindings that act as the lookup keys for the clustered index are not guaranteed to follow any order, each record fetch can potentially require a disk I/O. The number of records from the inner relation that match each correlation binding is $C_s$. Since the inner relation is clustered on the lookup column the records to be fetched would be stored contiguously, and hence occupy $\lceil C_s/F_s \rceil$ contiguous blocks. Let $k$ be the average number of cache misses on index nodes for each lookup. Then the estimated cost of each lookup and fetch is given by:

$$C_l = t_t(\lceil C_s/F_s \rceil \ + k) + t_s(k+1)$$

The total estimated cost (across all the iterations) would thus be $N_r \times S_o \times C_l$.

**Cost Estimate With Sorting**

We consider two cases:

1. When the outer predicate is such that the correlation bindings contain all the values in an interval of the index (*e.g.*, an outer predicate on the correlation attribute, which is also a foreign key of the inner relation), the inner relation records accessed over all the iterations lie on a set of contiguous blocks. Thus multiple lookups can be served from a single block fetch. The total number of records from the inner relation accessed over all the iterations will be $A_s = N_r \times S_o \times C_s$. As the records are stored contiguously, the total inner relation access cost will be: $t_t(\lceil A_s/F_s \rceil) + t_s$.

   Assuming $N_r = 100,000$, $S_o = 0.5$, $C_s = 10$ and $F_s = 100$, the expected number of inner relation's blocks accessed (across all iterations) with sorted correlation bindings will be 5000, and we pay only a transfer cost for each. On the other hand the expected number of blocks accessed without sorting will be 50,000 (ten times higher) even with no cache misses for the index pages (*i.e.,* $k = 0$). When the correlation bindings are not sorted multiple fetches can occur for the same block of the inner relation due to the interleaved access with other blocks. This is the reason for the higher estimated cost (which assumes a worst case cache behavior) when the correlation bindings are not sorted.

2. When the predicates of the outer query block are on attributes different from the ones used as correlation variables, the correlation bindings will be from disjoint intervals. Let $j$ be the expected number of times the inner query block is evaluated ($j$ is the number of distinct correlation bindings generated from the $N_r \times S_o$ tuples that qualify the outer predicates). Let $q = \lceil F_s/C_s \rceil$. $q$ denote the number of distinct values for the attribute of $S$ involved in the correlation predicate, which are stored on each block. Recall that $d$ is the total number of distinct values for the attribute of $S$ that is involved in the correlation predicate. We can estimate the number of inner relation's blocks accessed as follows: a block of the inner relation will be accessed if any of the $q$ distinct values in it is part of the $j$ distinct correlation bindings. Therefore, the probability that a block of $S$ gets accessed is given by $p = (1 - \frac{{}^{d-q}C_j}{{}^{d}C_j})$, where ${}^{m}C_n$ is the notation for choosing $n$ from $m$, i.e., $\frac{m!}{(m-n)!n!}$.

   Therefore, the expected number of blocks read, *num_blocks* will be: $p \times B_s$.

   The cost estimate will thus be $min(scantime, (t_s + t_t)num_blocks)$ where *scantime* is the time to scan the complete table ($t_s + B_s \times t_t$).

### 2.2.3 Incremental Computation of Aggregates

We now describe an efficient technique to evaluate nested aggregate queries having non-equality correlation predicates, using the restartable scan. Decorrelation is often very expensive for such queries. Consider the SQL query shown in Example 2.1. The query lists days on which the sales exceeded the sales seen on any day in the past.

***Example 2.1*** *A Nested Aggregate Query with Inequality Predicate*

```
SELECT  day, sales FROM DAILYSALES DS1
WHERE   sales > (SELECT MAX(sales) FROM DAILYSALES DS2
                 WHERE DS2.day < DS1.day);
```

A naïve nested iteration plan for the above query employs a sequential scan of the DAILYSALES table for both the outer and the inner block. Assuming the inner block scans an average of half of the table for each outer tuple, the cost of this plan would be $t_t(B_{ds} + N_{ds} \times B_{ds}/2) + t_s(1 + N_{ds})$, where $B_{ds}$ is the number of blocks occupied by DAILYSALES table, $N_{ds}$ is the number of tuples in the same table, and $t_t$ and $t_s$ are the block transfer time and seek time respectively.

Now, suppose the DAILYSALES relation (or materialized view) is stored, sorted on the *day* column. If the plan for the outer query block generates the bindings for the correlation variable (DAILYSALES.day) in non-decreasing order, we can see that the tuples that qualify for the aggregate (MAX) operator's input in the $i^{th}$ iteration will be a superset of the tuples that qualified in the $(i-1)^{th}$ iteration. The MAX operator, in its state, can retain the maximum value seen so far and use it for computing the maximum value for the next iteration by looking at only the additional tuples. So, the scan needs to return only those additional tuples that qualify the predicate since its previous evaluation. The technique proposed here is applicable for $<, \leq, >$ and $\geq$ predicates and the aggregate operators MIN, MAX, SUM, AVG and COUNT. The maximum cost of such a plan would be $2 \times B_{ds} \times t_t + 2 \times t_s$, which is significantly lesser than the cost of the naïve nested iteration plan.

When there are GROUP BY columns specified along with the aggregate, the aggregate operator has to maintain one result for each group. The aggregate operator can maintain its state in a hash table; the key for the hash table being the values for the GROUP BY columns and the value against each key being the aggregate computed so far for the corresponding group.

## 2.3 Parameter Sort Orders in Query Optimization

A query optimizer considers alternative execution plans for a given query, estimates the cost of each plan and chooses the plan with the least expected cost. To estimate the cost of an iterative plan, traditional optimizers first identify the best plan for the nested sub-query independently and multiply its cost by the expected number of iterations. Clearly, this approach does not take into account plans that exploit ordered parameter bindings. The optimizer must consider different sort orders on the sub-query parameters. For each sort order, there is an associated benefit for the sub-query plan that exploits the sort order and a cost for the outer query plan to generate the parameters in the required order. The optimizer must choose the optimal plan for the query by considering benefits and cost of various possible sort orders.

The Volcano query optimization framework [23] is a state of the art cost-based query optimizer. This section illustrates how a Volcano style query optimizer can be extended to take into account plans that make use of parameter sort orders. The rest of this section is organized as follows. Section 2.3.1 briefly describes the Volcano optimizer framework. Section 2.3.2 proposes extensions to the optimizer's high-level interface to support optimization of parameterized expressions. Section 2.3.3 describes the logical representation we adopt for nested queries. The changes to the to plan space and search algorithm are described in Section 2.3.4 and Section 2.3.5 respectively.

### 2.3.1 The Optimizer Framework

In this section we briefly describe the PYRO cost-based optimizer framework over which we propose our extensions. PYRO is an extension of the Volcano optimizer [23]. A detailed description of the PYRO optimizer can be found in in [46] and [47].
The optimizer performs three main tasks.

1. Logical Plan Space Generation

   In this first step the optimizer, by applying logical transformations such as join associativity and pushing down of selections through joins, generates all the semantically equivalent rewritings of the input query.

2. Physical Plan Space Generation

   This step generates several possible execution plans for each rewriting produced in the first step. An execution plan specifies the exact algorithm to be used for evaluating each logical operator in the query. Apart from selecting algorithms for each logical operation, this step also considers *enforcers* that help in producing required *physical properties* (such as sort order of tuples) on the output. Physical property requirements arise due to two reasons *(i)* the user/application may specify a physical property requirement as part of the query, *e.g.,* an order-by clause and *(ii)* algorithms that implement operations such as join and duplicate elimination may require their inputs to satisfy a sort order or grouping property. The *algorithms* for relational operators and the *enforcers* of physical properties are collectively referred to as *physical operators* as against the logical operators of the logical plan space.

3. Finding the Best Plan

   Given the cost estimates of different algorithms that implement the logical operations and the enforcers, the cost of each execution plan is estimated. The goal of this step is to find the plan with minimum cost.

The above three steps can either be executed in a depth-first order or in a breadth-first order [47]. For the purpose of our explanation we consider the breadth-first order, in which each step is performed completely before the next step is started. However, in our actual implementation, physical plan generation and search for the best plan are combined in a single phase.

An AND-OR graph representation called Logical Query DAG (LQDAG) is used to represent the logical plan space, *i.e.,* all the semantically equivalent rewritings of a given query. The LQDAG is a directed acyclic graph whose nodes can be divided into *equivalence nodes* and *operation nodes*; the equivalence nodes have only operation nodes as children and the operation nodes have only equivalence nodes as children. An operation node in the LQDAG corresponds to an algebraic operation, such as join ($\bowtie$) or select ($\sigma$). It represents the expression defined by the operation and its inputs. An equivalence node in the LQDAG represents the equivalence class of logical expressions (rewritings) that generate the same result set, each expression being defined by a child operation node of the equivalence node and its inputs. An example LQDAG is shown in Figure 2.1[1].

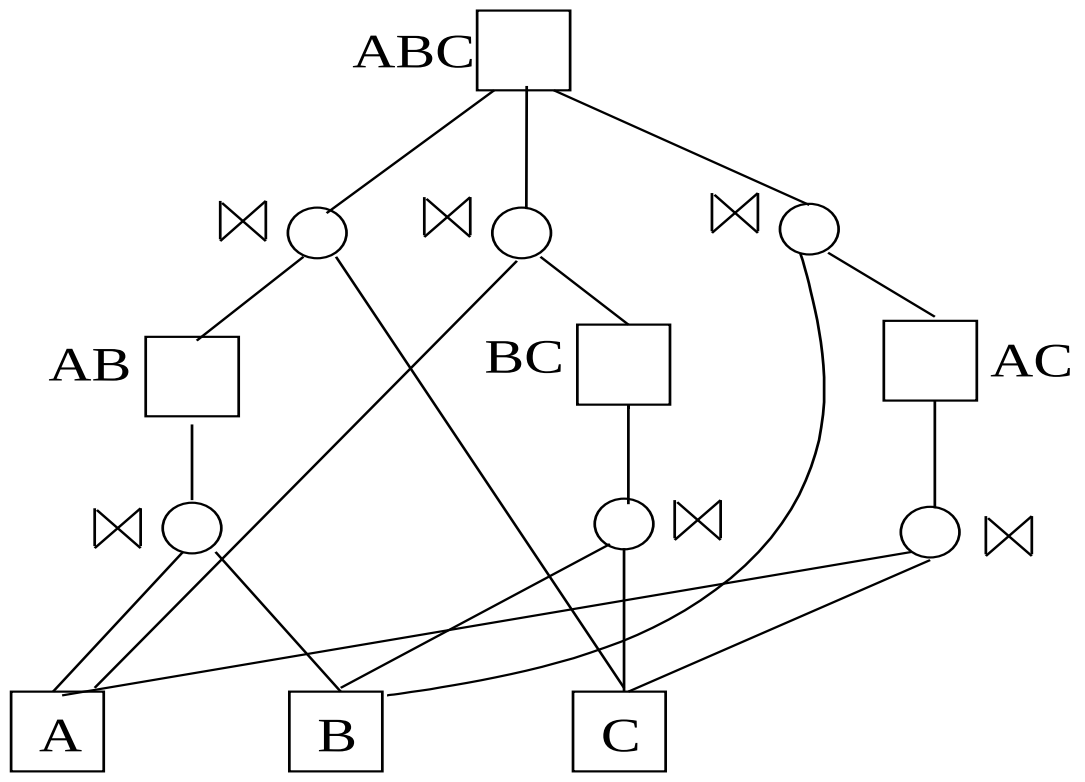

Figure 2.1: A Logical Query DAG for A $\bowtie$ B $\bowtie$ C

[1]This figure is taken from [47].

Once all the semantically equivalent rewritings of the query are generated, the Volcano optimizer generates the *physical plan space* by considering different algorithms for each logical operation and *enforcers* to generate required physical properties. In some optimizers, such as Cascades [21] and SQL Server [16], the logical and physical plan space generation stages are intermixed. The physical plan space is represented by an AND-OR graph called PQDAG which is a refinement of the LQDAG. Given an equivalence node $e$ in the LQDAG, and a physical property $p$ required on the result of $e$, there exists an equivalence node in the PQDAG representing the set of physical plans for computing the result of $e$ with the physical property $p$. A physical plan in this set is identified by a child operation node of the equivalence node and its input equivalence nodes. The equivalence nodes in a PQDAG are called *physical equivalence nodes* to distinguish them from the *logical equivalence nodes* of the LQDAG. Similarly, the operation nodes in a PQDAG are called *physical operation nodes* to distinguish them from the *logical operation nodes* of the LQDAG.

The optimizer framework we use models each of the logical operators, physical operators and transformation rules as separate classes, and this design permits the extensions we propose to be easily incorporated.

### 2.3.2 Extensions to the Optimizer Interface

Both Volcano [23] and PYRO [47] optimizers take the initial query (expression), a set of physical properties (such as sort order) required on the query result and a cost limit (the upper bound on plan cost) as inputs and return the execution plan with least expected cost. The following method-signature summarizes the Volcano optimizer's input and output.

Plan **FindBestPlan** (Expr $e$, PhyProp $p$, CostLimit c);

The optimizer makes an assumption that if the expression is evaluated multiple times the cost gets multiplied accordingly. This assumption ignores advantageous buffer effects due to sorted parameter bindings and the benefits due to state retention techniques proposed in the previous section. When the parameter bindings are sorted, the cost of evaluating an expression $n$ times can be significantly lesser than $n$ times the cost of evaluating the expression once. In order to consider these factors, we propose a new form of the *FindBestPlan* method. The following method-signature summarizes the new form of the *FindBestPlan* method.

Plan **FindBestPlan** (Expr *e*, PhysProp *p*, CostLimit *c*, SortOrder *pso*, int *callCount*);

The new *FindBestPlan* procedure takes two additional parameters. The first of these, is the sort order guaranteed on the parameters (outer variables) used by *e*. The second parameter, termed *callCount*, tells the number of times the expression is expected to be evaluated. The cost of the returned plan is the estimated cost for *callCount* invocations. Note that the original Volcano algorithm's interface can be thought of as a special case of this enhancement, where the expression *e* is assumed to have no unbound references (parameters) and the callCount is 1.

### 2.3.3 Logical Representation of Nested Queries

We now describe the way in which we represent nested queries in the LQDAG. A nested query, in the simplest case, contains two query blocks - an *outer* query block and an *inner* query block. The inner query block uses parameters whose values are bound from the tuples obtained by evaluating the outer query block. We adopt a variant of the *Apply* operator proposed in [17] for representing nested queries. In its simplest form, the *Apply* operator has two sub-expressions: the *bind* sub-expression corresponds to the outer query block and the *use* sub-expression corresponds to the parameterized inner query block. Conceptually, the *Apply* operator evaluates the use sub-expression for every tuple in the result of the bind sub-expression. After each evaluation of the use sub-expression, the *Apply* operator combines the tuple from the bind sub-expression and the result of the use sub-expression. Combining may involve evaluating a predicate such as IN or NOT IN that check for set membership, EXISTS or NOT EXISTS that check for set cardinality, a scalar comparison ($=, \neq, >, \geq, <, \leq$), or a comparison of a scalar with members of a set: *relop* ANY or *relop* ALL, where *relop* is one of the comparison operators. Figure 2.2 shows the logical representation of the query given in Example 1.1 of Chapter 1.

We refer to the *bind* sub-expression of an *Apply* operator as its *left* sub-expression and the *use* sub-expression as its *right* sub-expression. In general, an *Apply* operator can have multiple *use* expressions that represent multiple sub-queries nested at the same level. In a complex multi-level nested query a sub-expression *e* may use some variables and bind other variables. The variables that $e$ binds may be passed on to the *use* subexpressions of parent or ancestor *Apply* operators; $e$ must be in the left-most subtree of such *Apply* operators. The variables that $e$ uses must be defined at parent or ancestor *Apply* operators; $e$ must be in a use-subtree, *i.e.*, non-left-most subtree, of such *Apply* operators.

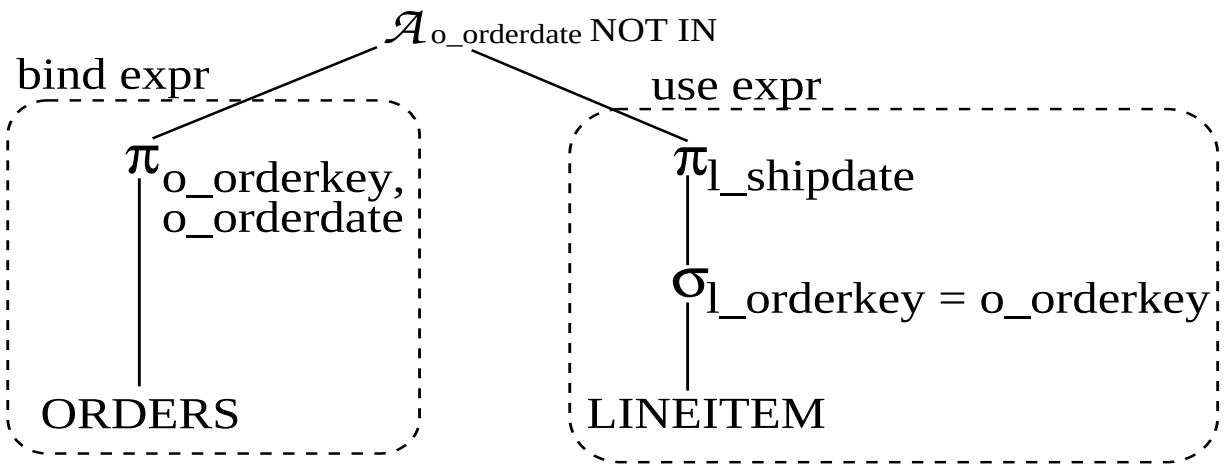

Figure 2.2: Example of Representing a Nested Query using *Apply* ($\mathcal{A}$)

### 2.3.4 Physical Plan Space Generation

The physical plan space generation involves generating alternative execution plans for a given logical expression and representing them in the PQDAG. In PYRO, two query execution plans $p_1$ and $p_2$ are considered equivalent (*i.e.,* they belong to the same physical equivalence class) *iff* the following conditions are met: *(i)* $p_1$ and $p_2$ evaluate the same logical expression $e$, and *(ii)* $p_1$ and $p_2$ produce the result of $e$ in the same sort order.

We redefine the notion of equivalence of execution plans in PYRO to include the parameter sort orders required by the plans. Two plans $p_1$ and $p_2$ belong to the same equivalence class *iff* $p_1$ and $p_2$ evaluate the same logical expression, guarantee the same physical properties on their output and require the same sort order on the parameter bindings, when invoked iteratively. Thus, for a given logical expression $e$ and physical property $p$, there exists a set of physical equivalence nodes in the PQDAG. Each equivalence node in this set corresponds to a distinct sort order requirement on the parameters used in $e$.

The physical plan space generation step therefore involves the following: given a logical equivalence node $e$, a physical property $p$ required on the result of $e$ and a sort order $s$ known to hold on the parameters in $e$, generate all the evaluation plans and representing them in the PQDAG. The search phase of optimization then takes the PQDAG and a *call count* as its inputs and finds the best plan.

### Generating Plans for Non-Apply Operators

The plan generation step in PYRO works as follows. The LQDAG is traversed, and at each logical operation node, all the applicable algorithms (physical operators) are considered. A physical operator is applicable if it implements the logical operation and ensures the required physical properties on the output.

Procedure **ProcLogicalOpNode**
Inputs: *o*, a logical operation node in the LQDAG
*p*, physical property required on the output
*s*, sort order guaranteed on the parameter bindings
*e*, the physical equivalence node for the new plans
Output: Expanded physical plan space. New plans are created under *e*.
BEGIN
For each algorithm *a* such that *a* implements *o*, ensures physical property *p* on the output and requires a sort order $s'$ on the parameters, where $s'$ is subsumed by *s*
Create an algorithm node $o_a$ under $e$
For each input *i* of $o_a$
Let $o_i$ be the $i^{th}$ input (logical equivalence node) of $o_a$
Let $p_i$ be the physical property required from input *i* by algorithm *a*
Set input *i* of $o_a$ = PhysDAGGen($o_i$, $p_i$, *s*) // Main plan generation procedure
// shown in Figure 2.6
END

Figure 2.3: Plan Generation at a Non-Apply Node

To take into account the sort order of parameters, we need a minor modification to the plan generation step. For a given logical operation, we consider only those physical operators (algorithms) as applicable, whose sort order requirement on the parameters is subsumed by the sort order known to hold (guaranteed) on the parameters. As an example, consider a selection logical operator $\sigma_{R1.a=p_1}(R1)$, where $p_1$ is a correlation (outer) variable. Suppose two algorithms, a plain table scan requiring no sort order and a state retaining scan requiring a sort order $(p_1)$ on the parameters, are available. Now, if the parameter sort order guaranteed by the parent block subsumes $(p_1)$, both the algorithms (physical operators) are applicable. However, if the parameter sort order guaranteed by the parent block does not subsume $(p_1)$, then only the plain table scan is applicable. Figure 2.3 shows the plan generation steps at a *Non-Apply* logical operator.

### Plan Generation for an Apply Operator

Earlier work on *Apply* operator [17] attempts to rewrite the *Apply* operator using joins, outer-joins or semi-joins before plan generation. Our goal here is to expand the plan space to include efficient iterative plans for the *Apply* operator. This involves two steps.

1. Identify a set of useful and valid sort orders on the sub-query parameters.
2. Generate iterative plans in which the plan for the sub-query makes use of the sort order of parameters produced by the plan for the outer-block.

**Identifying Valid Interesting Sort Orders**

If sub-query has $n$ parameters, there are potentially $n!$ sort orders on the parameters. Considering all possible sort orders of parameters used in an expression is prohibitively expensive. Only few sort orders on the parameters are expected to be useful. To consider selected sort orders in the optimization process, System R [48] introduced the notion of *interesting orders.* We extend this notion to sort orders of parameters and call them as *interesting parameter sort orders.* Our algorithm to generate the physical plan space creates physical equivalence nodes for only *interesting* parameter sort orders.

Intuitively, interesting parameter sort orders are the ones that are useful for efficient evaluation of one or more nested blocks (*use* expressions of an apply operator). Typically, the clustering order and query covering indices on base relations used inside the nested block(s) decide the interesting parameter sort orders. However, the optimizer must also consider plans that explicitly sort a relation in the inner block to match the sort order easily available from the outer block. The problem of identifying interesting orders is common to both nested queries and joins and deserves special attention. We address this problem in Chapter 3. In the rest of this chapter we assume that the set of interesting sort orders on unbound parameters of an expression, is available to us.

Every interesting sort order on the parameters may not be *valid* (feasible) under the given nesting structure of query blocks. For example, consider a query with two levels of nesting; $q_a(q_b(q_c))$, $q_a$ is the outer-most query block and $q_c$ is the inner most. Assume $q_c$ uses two parameters $a$ and $b$, where $a$ is bound by $q_a$ and $b$ is bound by $q_b$. Now, the sort order $(a, b)$ is a valid interesting order for $q_c$ but not the sort order $(b, a)$. As block $q_b$ that binds parameter $b$ is nested below the block $q_a$ that binds parameter $a$, the sort order $(b, a)$ cannot be generated and hence invalid.

**Definition 2.3 Valid Interesting Sort Orders**

*A sort order $s = (a_1, a_2, \ldots, a_n)$ on the parameters of a query block $b$ is* valid *(feasible)* iff *there is a nesting of query blocks such that the following two conditions are satisfied:*

1. *$level(a_i) \leq level(a_j)$ for all $i, j$ s.t. $i < j$, where $level(a_i)$ is the level of the query block in which $a_i$ is bound. The level of the outer most query block is considered as 0 and all the query blocks nested under a* level-i *query block have the level $i + 1$.*

2. *Let the BindAttrs of a block $b_k$ with respect to a sort order $s$ be defined as follows. BindAttrs($b_k$, $s$): Attributes in $s$ that are bound by either block $b_k$ or by an ancestor of $b_k$.*

   *Then, for each ancestor block $b_k$ of $b$ at level $k$ such that level($b$) $> k + 1$ (i.e., excluding the parent block of $b$), values of BindAttrs($b_k, s$) are distinct for any two invocations of $b_{k+1}$.*

The first condition in Definition 2.3 ensures that attributes bound by an outer query block always precede attributes bound by an inner query block in any valid sort order. The need for the second condition is best explained with an example. Consider a query with two levels of nesting; $q_a(q_b(q_c))$, $q_a$ being the outer-most query block and $q_c$ being the inner most. Suppose block $q_c$ makes use of two parameters: parameter $a$ bound at $q_a$ and parameter $b$ bound at $q_b$. If $q_a$ generates duplicate values for $a$, then $(a, b)$ is not a valid parameter sort order for $q_c$. This is because if $q_a$ generates duplicate bindings for $a$, then even if the plan for $q_b$ produces the bindings for $b$ in sorted order, $q_c$ cannot see a sorted sequence on $(a, b)$ across all its invocations; the bindings for attribute $b$ will cycle back for the same value of attribute $a$. Now, if $q_a$ is guaranteed to not generate duplicates for $a$, then $(a, b)$ is a valid parameter sort order for block $q_c$. However the sort order $(b)$ is not valid (unless $q_a$ invokes $q_b$ at most once) since even if the plan for block $q_b$ produces the bindings for $b$ in sorted order, block $q_c$ will see a sorted sequence of $b$ values for a single invocation from $q_a$, but may not see a sorted sequence on $b$ across multiple invocations from $q_a$.

**Generating Plans for the *Bind* and *Use* Expressions of an Apply Operator**

Consider a query block $q$ under which blocks $q_1, q_2, \ldots, q_n$ are nested. This is represented by an *Apply* expression with $q$ as the *bind* sub-expression and $q_1, q_2, \ldots, q_n$ as *use* sub-expressions. Generating plans at the *Apply* node involves the following steps:

1. For each use expression $q_i$, identify the set $s_i$ of interesting parameter sort orders. Identifying a good set of interesting sort orders is a topic of interest not only for nested queries but also for queries with joins, group-by and set operations. We address this problem in the next chapter.

2. For the outer query block $q$, identify the set $s_q$ of sort orders that are available readily or at a low cost.
3. Form the set *i_ords* as $s_1 \cup \ldots \cup s_n \cup s_q$.
4. From the set *i_ords* discard the sort orders that are not *valid*, under the given nesting structure of query blocks. The conditions for validity are specified in definition 2.3.
5. From the set *i_ords* we derive a set *l_ords* consisting of sort orders that are relevant to the *bind* expression $q$ of the *Apply* operator. Note that the sort orders in *i_ords* can contain some attributes that are bound higher up in the complete query structure. Deriving *l_ords* from *i_ords* involves extracting the suffix of each order $ord \in$ *i_ords* such that the suffix contains only those parameters that are bound in $q$.
6. For each sort order $o \in$ *l_ords* $\cup \{\epsilon\}$, where $\epsilon$ stands for the empty sort order, we generate plans for the *bind* expression by making $o$ as the required physical property on the result (output), and then generate plans for all the *use* expressions. We create a physical operation node $a$ for the *Apply* operation depending on the predicate associated with the *Apply* node. The plans generated for the *bind* and *use* expressions are added as the child plans for $a$.

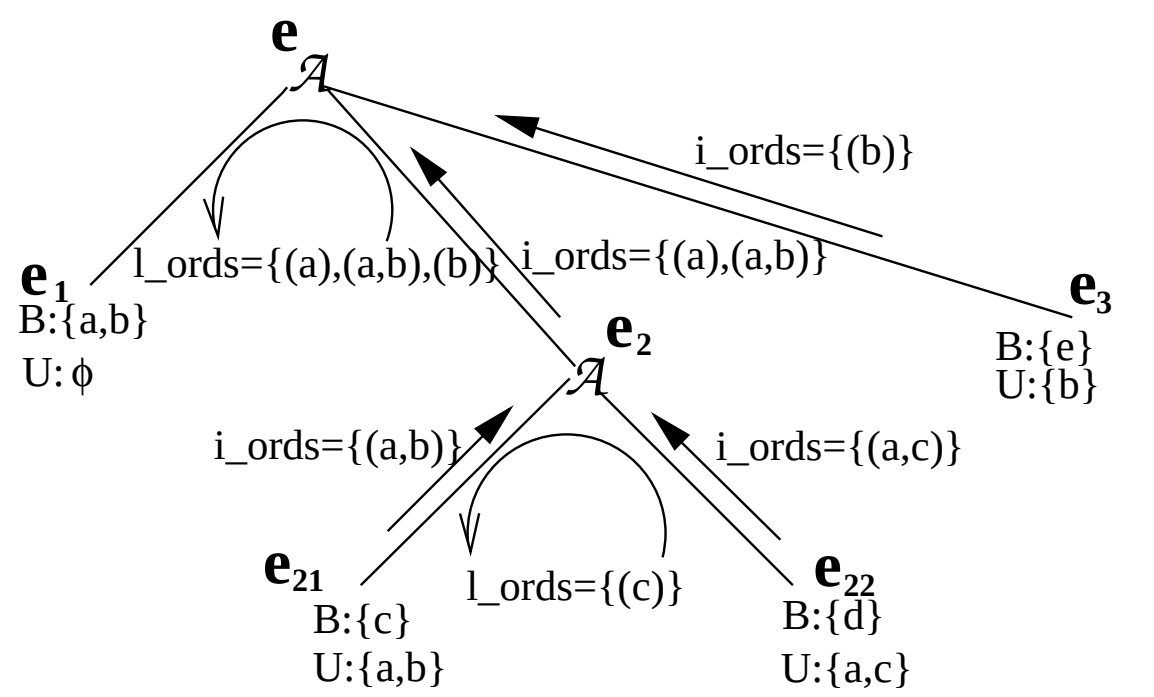

Figure 2.4: Sort Order Propagation for a Multi-Level Multi-Branch Expression

We now illustrate the working of the above steps using the example expression shown in Figure 2.4. Two sub-expressions $e_2 = (e_{21}$ *Apply* $e_{22})$ and $e_3$ are nested under the outer-most expression $e_1$. In the figure, we indicate the parameters bound and parameters used by each expression with the convention *B:* and *U:* respectively. Consider the outer-most apply operator present at the root of the expression tree. In step 1, for each of the *use* sub-expressions *i.e.*, for $e_2$ and $e_3$ we identify the set of interesting parameter sort orders. The interesting parameter sort orders of an expression depend upon the sort orders of base relations used in the expression and the correlation predicates. The details of deriving interesting orders is the topic of the next chapter. For now, suppose the sub-expression $e_2$ has two interesting sort orders $(a)$ and $(a, b)$ on the parameters it uses, and suppose the sub-expression $e_3$ has one interesting sort order $(b)$. In step 2, we identify the set of sort orders available at low cost on the output of expression $e_1$. Such sort orders are called *favorable* sort orders and the details of finding the favorable sort orders are given the next chapter. For this example, suppose there exists a single favorable sort order $(a)$ for expression $e_1$. In step 3, we compute the set *i_ords* as $\{(a), (a, b), (b)\}$. In step 4, we check for the validity of these sort orders as per Definition 2.3. All the three sort orders are valid in this case. We then derive the set *l_ords* in step 5, by extracting the sort order suffix relevant to the *bind* expression $e_1$. $e_1$ being the outer-most block, *l_ords* will be same as *i_ords*. In step 6, we generate plans for the *use* expression $e_1$ with each of the three sort orders $(a)$, $(a, b)$ and $(b)$ as the required output sort order, and also generate the plans for the *use* expressions with each of these sort orders as the guaranteed sort order on the parameter bindings. The corresponding plans of the *bind* and *use* expressions are then paired as child plans of a physical apply operator. Note how the set *l_ords* is computed for the apply operator at the root of sub-expression $e_2 = (e_{21}$ *Apply* $e_{22})$. The set *i_ords* of interesting orders for $e_{22}$ has a single element $(a, c)$, *i.e.*, *i_ords*=$\{(a, c)\}$. From this set we derive the set *l_ords* as $\{(c)\}$ since $c$ is the only parameter bound by the expression $e_{21}$, which is the *bind* expression for the *Apply* operator in consideration.

Procedure *ProcApplyNode* in Figure 2.5 shows the plan generation steps at an *Apply* operator node. The top level procedure for generating the physical plan space is given in Figure 2.6, and it makes use of the two procedures *ProcLogicalOpNode* and *ProcApplyNode*. For simplicity, we omit the cost based pruning from our description and return to it later. As a result the *callCount* parameter does not appear in the algorithm. Figures 2.7 and 2.8 show the logical query DAG and the resulting physical query DAG (assuming a very limited collection of algorithms) for the example of Figure 2.2.

To check if a sort order is valid, we need a mapping from each parameter to the level number of the block in which the parameter is bound. In the logical query DAG (LQDAG), due to the sharing of common sub-expressions, the mapping of parameters to the level of

Procedure **ProcApplyNode**
Inputs: *o*, logical operation node corresponding to the *Apply* operator in the LQDAG
*s*, sort order guaranteed on the parameter bindings
*e*, the physical equivalence node for the plans generated
Output: Expanded physical plan space. New plans are created under *e*.

```
BEGIN
    Form the set i_ords of valid interesting orders on parameters by considering all the input
    sub-expressions of o.
    From the set i_ords, create the set l_ords by extracting sort order prefixes of attributes bound
    by o.bindInput.
    For each order ord in l_ords and the empty sort order ϵ
        // Generate plans for the bind expression
        Let l_eq = PhysDAGGen(o.bindInput, ord, s)
        Let newParamOrd = concat(s, ord)
        Let iterOp = New iterator physical op for Apply
        iterOp.bindInput = l_eq
        For each use input u of o
            Let u_eq = PhysDAGGen(u, ϵ, newParamOrd)
            Add u_eq as a use input of iterOp
        Add iterOp as a child of e
END
```

Figure 2.5: Plan Generation at an *Apply* Node

Procedure **PhysDAGGen**
Inputs: *e*, logical equivalence node for the expression
*p*, physical property required on the output
*s*, sort order guaranteed on the parameter bindings
Output: Generates the physical plan space and returns the physical equivalence node

```
BEGIN
    If a physical equivalence node n_p exists for e, p, s in the memo structure
        return n_p
    Create an equivalence node n_p for e, p, s and add it to the memo structure
    For each child logical operation node o of e
        If(o is an instance of ApplyOp)
            ProcApplyNode(o, s, n_p)
        else
            ProcLogicalOpNode(o, p, s, n_p)
    For each enforcer f that generates property p
        Create an enforcer node o_f under n_p
        Set the input of o_f = PhysDAGGen(e, ϵ, s)
    return n_p
END
```

Figure 2.6: Main Algorithm for Physical Plan Space Generation

the query block that binds it cannot be fixed statically for each logical equivalence node. In fact, a single logical equivalence node can get different level numbers because of the

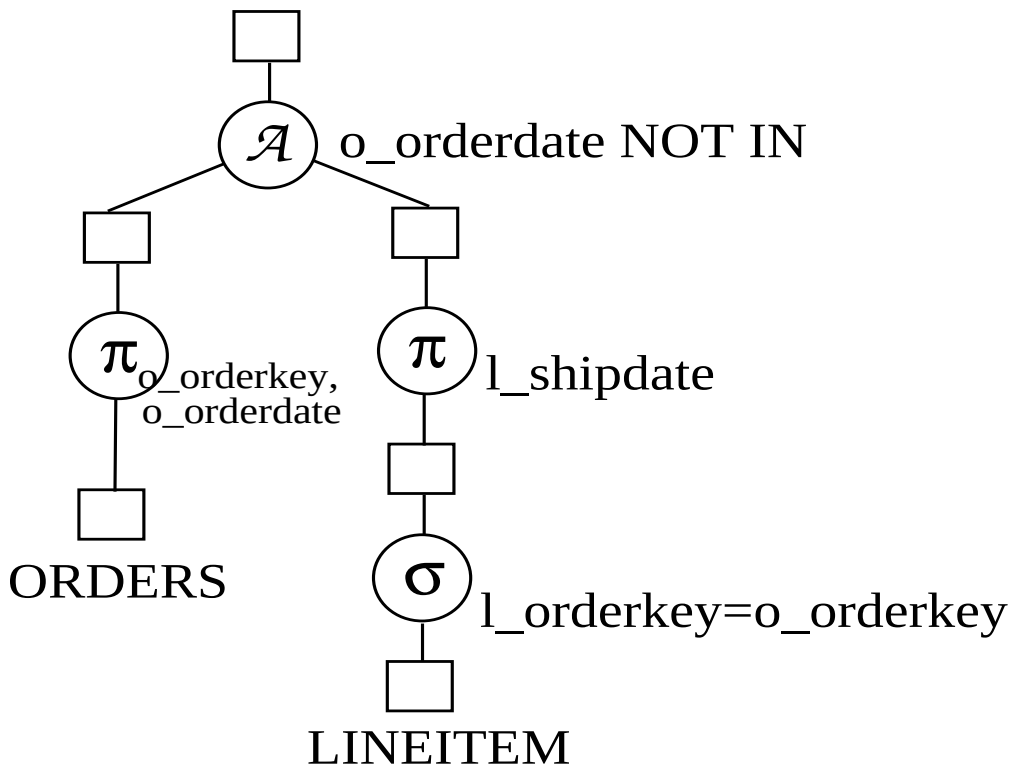

Figure 2.7: LQDAG for the Example of Figure 2.2

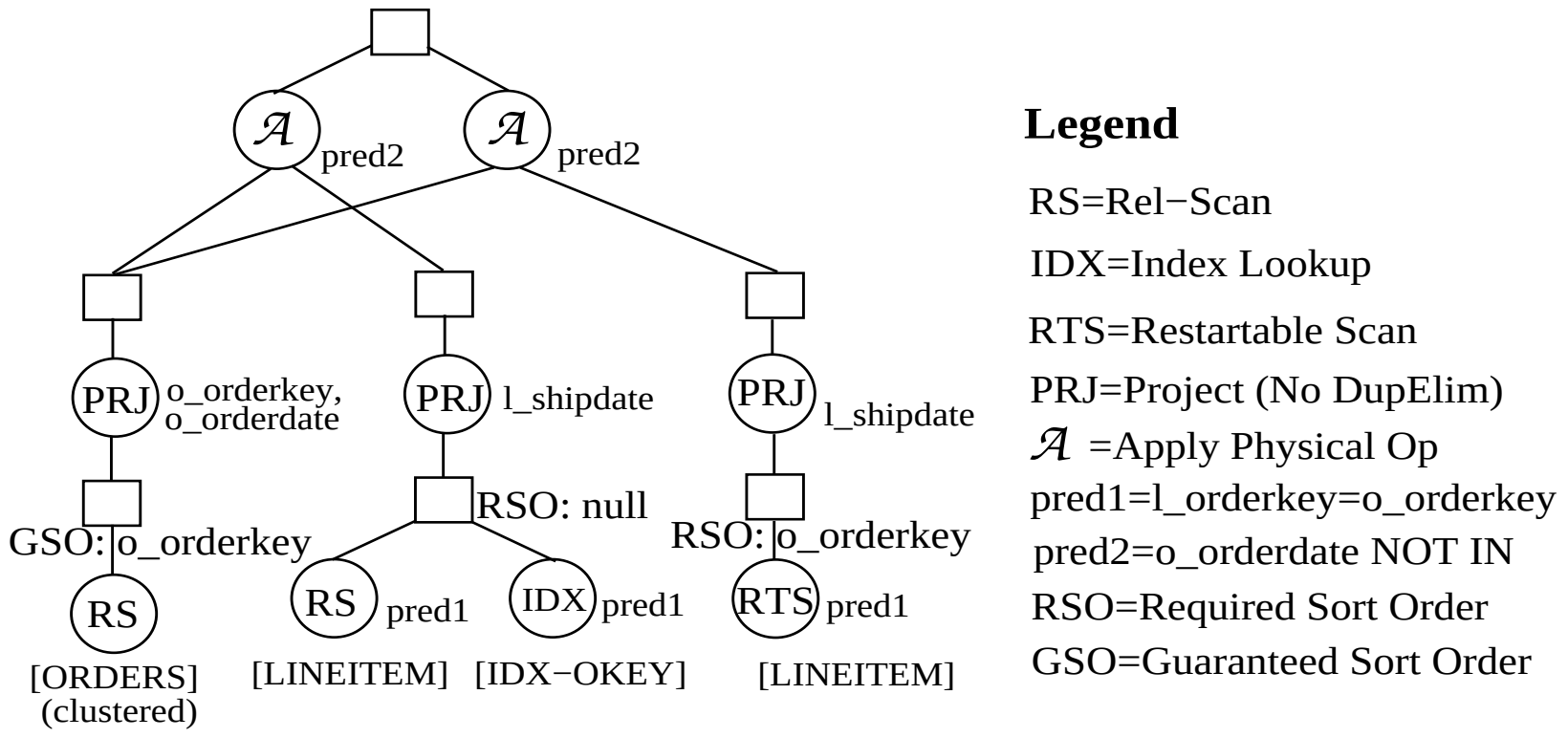

Figure 2.8: PQDAG for the Example of Figure 2.2

*level altering* transformations such as:

$$(R\ \mathcal{A}_{EXISTS}(\sigma_{S.c2=R.c1}(S)\mathcal{A}_{EXISTS}(\sigma_{T.c3=R.c1}T)))$$

$$\Longleftrightarrow ((R \ltimes_{R.c1=S.c2} S)\mathcal{A}_{EXISTS}(\sigma_{T.c3=R.c1}T))$$

In the LHS of the above equivalence rule relation $T$ gets a level number two levels higher than $R$, whereas in the RHS $T$ gets a level number one level higher than $R$. This happens because the RHS uses an outer join to remove the nesting of $S$ within $R$. Figure 2.9 gives a pictorial illustration.

In general with such transformations, a sub-expression can see a different mapping of parameters to levels depending on which expression is chosen above it in a logical plan. We can thus get multiple interesting parameter sort orders, corresponding to the different nesting structures. In our implementation, we address this issue by carrying along a map of parameters to levels when recursively traversing the LQDAG to find valid interesting sort orders. A node in the LQDAG could be traversed more than once, if there are alternative nesting structures above. Interesting sort orders computed at a node can be memoized against the nesting structure to avoid repeated computation.

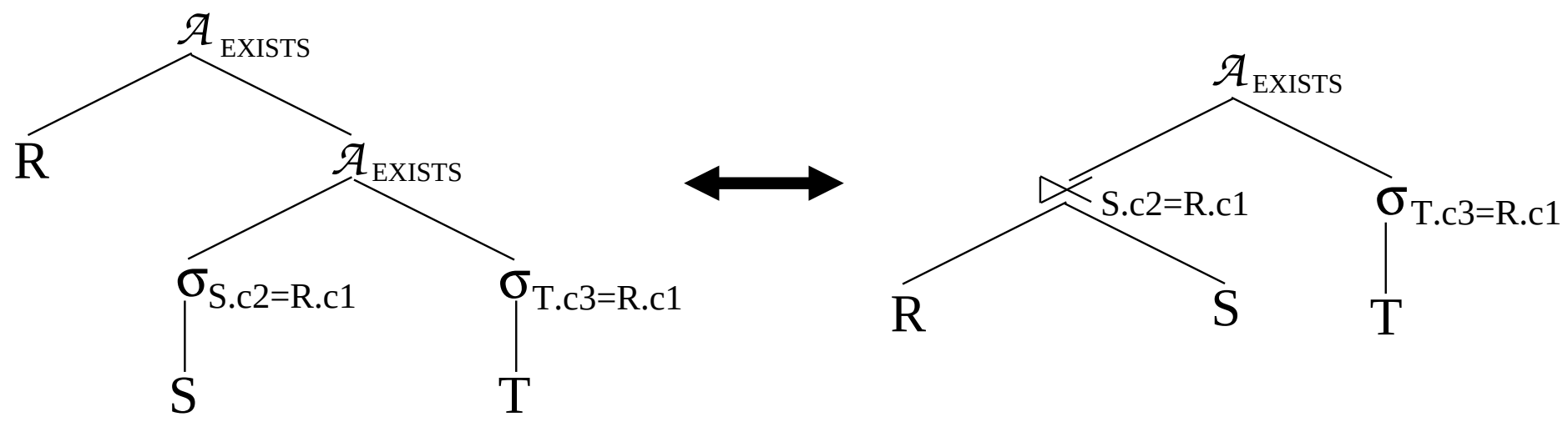

Figure 2.9: An Example of Level Altering Transformation

### 2.3.5 Search for Best Plan and Cost-Based Pruning

At the end of physical plan space generation we will have a physical query DAG with a root physical equivalence node. The best plan for the PQDAG is computed recursively by adding the cost of each physical operator to the cost of the best plans for its inputs and retaining the cheapest combination.

While computing the plan cost we take into account the fact that the *use* sub-expressions of an *Apply* operator are evaluated as many times as the cardinality of the *bind* sub-expression of the *Apply* operator. If caching of sub-query results is employed, then the number of distinct correlation bindings will be used in place of cardinality. Each physical operator's cost function is enhanced to take an integer $n$ as a parameter and return the cost for $n$ invocations of the operator. Memoization of the best plan is done for each 4-tuple (expression, output physical properties, input parameter sort order, call count). This is required since the best plan may be different for different call counts.

Optimization with different call counts can potentially increase the cost of optimization. However, if the plan is the same for two different call counts, we can assume that it would be the same for all intermediate call counts. The same plan can then be reused for all calls with an intermediate call count, with no further memoization required. Results from parametric query optimization [25] indicate that the number of different plans can be expected to be quite small. This helps in reducing both the number of plans memoized and the number of optimization calls. We apply all simple (non-nested) predicates before the nested predicate is applied. This further reduces the number of distinct call counts with which an expression is optimized.

**Cost-Based Pruning**

In our description, we ignored cost-based pruning for simplicity and separated the physical DAG generation and the search phases. In our actual implementation, the generation of the physical plan space and search for the best plan take place in a single phase. While generating the physical plan space, the cost of each plan is calculated and the best plan seen so far is memoized. We perform cost-based pruning as in the original Volcano algorithm [23].

## 2.4 Experimental Results

We implemented the state-retention techniques in PostgreSQL and carried out a performance study. The optimization techniques were implemented in our Volcano-style optimizer called PYRO, and these plans were forced on PostgreSQL bypassing it optimizer. We considered three algorithms: nested iteration(NI), magic decorrelation(MAG) [49] and nested iteration with state retention(NISR). In the case of nested iteration (NI) a suitable index was assumed to be present and used. Whenever a relation was assumed to be sorted, the NI plan used a clustered index. Magic decorrelation [49] involves partially evaluating the outer query block so as to identify the full set of parameters with which the subquery is to be executed. The partial result of the outer query block is called a *supplementary* table. The correlated subquery is then rewritten as a non-nested query by using an appropriate type of join with the supplementary table. The rewritten query produces the sub-query results for the set of parameters from the supplementary table. A join of the rewritten sub-query and the supplementary table to evaluate the remaining outer predicates gives the final result. For a more detailed description of the magic decorrelation technique we refer the reader to [49]. In our experiments, the plans employing magic decorrelation were composed with the supplementary table materialized.

PostgreSQL did not automatically decorrelate any of the queries we considered, and it always used a simple nested iteration plan. Hence, the results noted for the nested iteration (NI) algorithm also act as the baseline PostgreSQL measures. The plans employing state-retention techniques and magic decorrelation were forced through code changes, bypassing the PostgreSQL's optimizer.

For our experiments, we used the TPC-H [55] dataset on 1GB scale, and an additional relation, *dailysales* which had 2,500 records. The experiments are described below.

### Experiment 1

For this experiment, we used the query shown in Example 2.2, which is a minor variant of the query given in Example 1.1 of Chapter 1. The query in Example 1.1 uses a NOT IN predicate whose decorrelated form requires an implementation of anti-join, which is not available in PostgreSQL. Hence, we changed the predicate to an IN predicate.

***Example 2.2*** *Query Used in Experiment 1*

```
SELECT  o_orderkey, o_orderdate FROM ORDERS
WHERE   o_orderdate IN ( SELECT l_shipdate FROM LINEITEM
                         WHERE l_orderkey = o_orderkey);
```

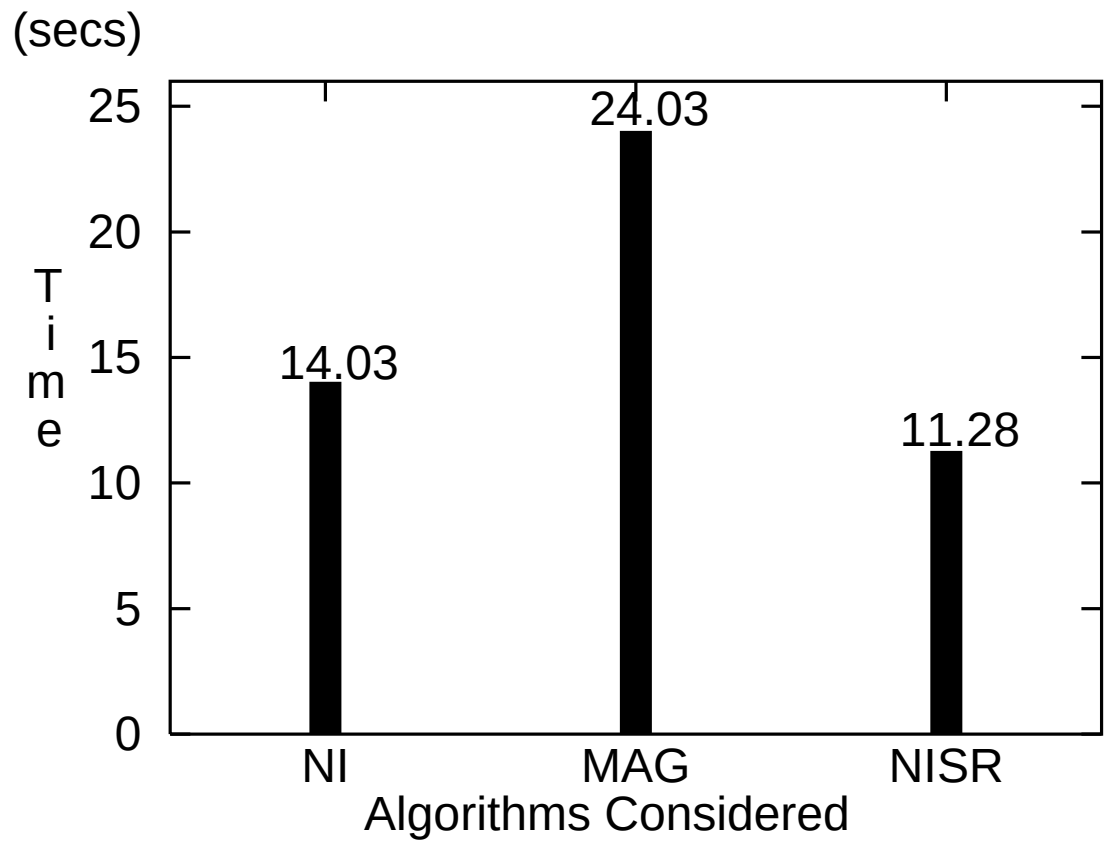

Figure 2.10: Performance Results for Experiment 1

Figure 2.10 shows the execution times for this query. Magic decorrelation performs poorly because there are no outer predicates and no duplicates. This leads to a large redundant join in the plan produced by magic decorrelation. Indexed nested loops join performs significantly better but is still less efficient than nested iteration with state retention. This is due to the overhead of index lookups. This overhead is significant even though most of the index pages above the leaf level are cached in memory.

### Experiment 2

For our second experiment we used the query shown in Example 2.1 of Section 2.2.3. The query lists the days on which the sales exceeded the maximum daily sales seen in the past. Figure 2.11 shows the execution times for the plans.

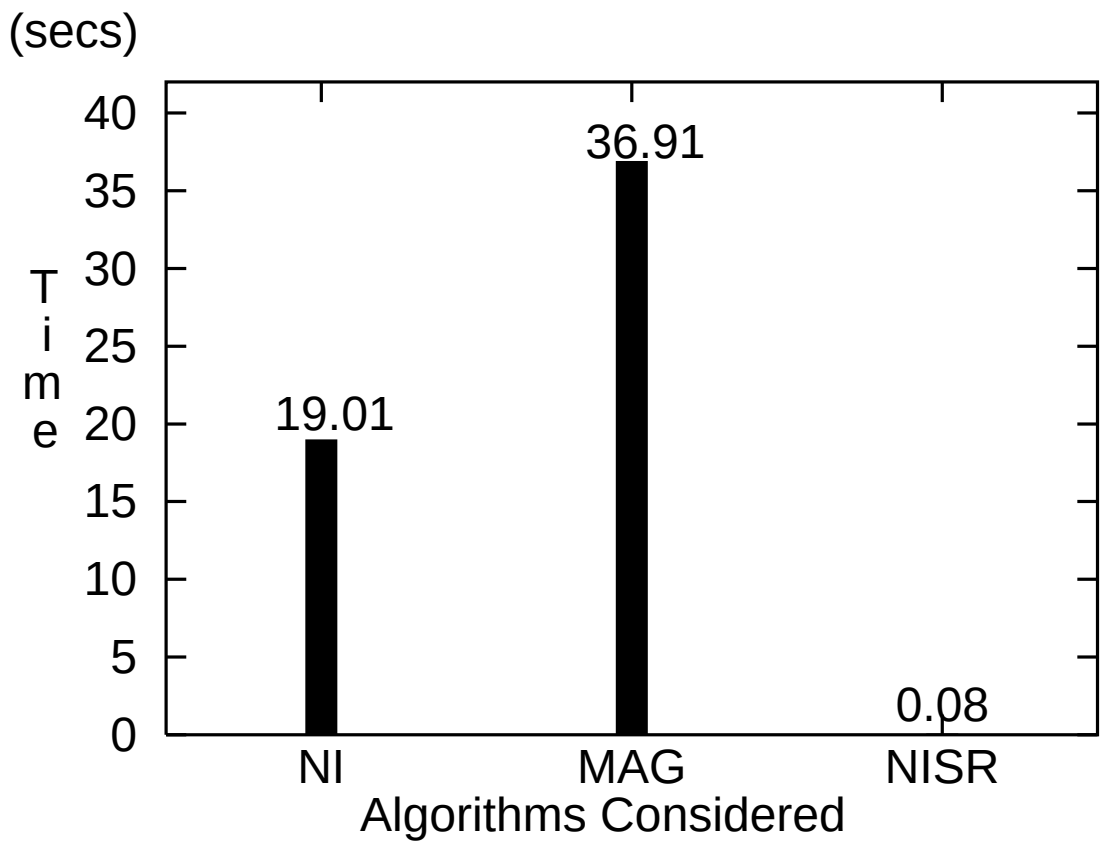

Figure 2.11: Performance Results for Experiment 2

For this query, nested iteration with state retention completely outperforms magic decorrelation and plain nested iteration. Due to the presence of a non-equality correlation predicate the cost of both magic decorrelation and plain nested iteration increase very rapidly with the increase in the number of outer block tuples. Nested iteration with state retention performs a single scan of the inner and the outer relations as described in Section 2.2.3.

#### Experiment 3

For our third experiment we used a query which is a modified version of the TPC-H min cost supplier query shown in Example 2.3 below.

***Example 2.3*** *Query Used in Experiment 3*

```
SELECT  s_sname, s_acctbal, s_address, s_phone
FROM    PARTS, SUPPLIER, PARTSUPP
WHERE   s_nation='FRANCE' AND p_size=15 AND p_type='BRASS' AND
        p_partkey=ps_partkey AND s_suppkey=ps_suppkey AND
        ps_supplycost =
                  ( SELECT min(PS1.ps_supplycost)
                    FROM PARTSUPP PS1, SUPPLIER S1
                    WHERE p_partkey=PS1.ps_partkey AND
                          S1.s_suppkey=PS1.ps_suppkey AND
                          S1.s_nation='FRANCE');
```

The results are shown in figure 2.12(a). Magic decorrelation performs the best because of the low selectivity of the outer predicates. There were only 108 distinct tuples satisfying the outer predicates. Restart scan performs poorly in this case as the entire re-lation is scanned where only small fraction of it was required. However, as the selectivity of the outer predicates increases, NISR becomes more attractive. This is evident from a second experiment that we carried out by dropping the predicate "p_size=15". As can be seen in Figure 2.12(b), NISR performs better in this case.

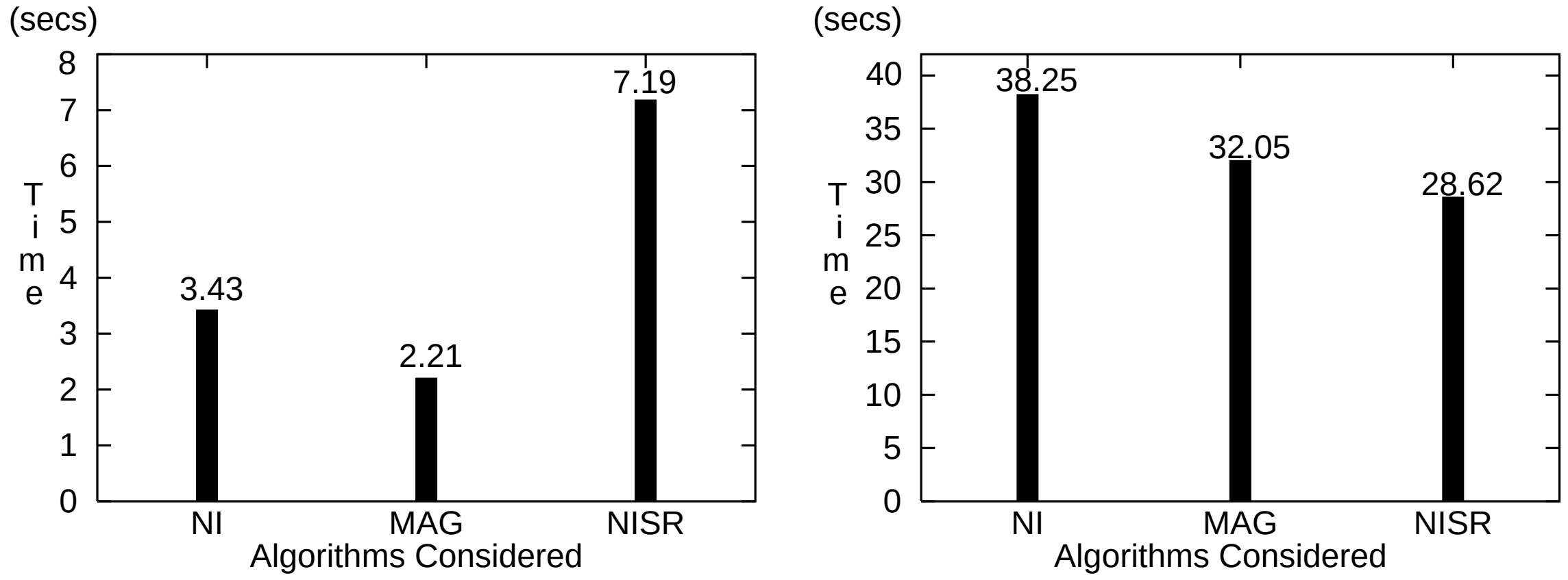

(a) With all outer predicates (b) Dropping 'p_size=15'

Figure 2.12: Performance Results for Experiment 3

---

***Example 2.4*** *Query Used in Experiment 4*

Find the turn around time for high priority orders. The turn around time of an order is calculated as the maximum of the differences between the ship date and placement date of all its line items, if the order price is < 2000 and it is calculated as the maximum of the differences between the commit date and placement date otherwise.

```
SELECT o_orderkey, turn_around_time(o_orderkey, o_totalprice, o_orderdate)
FROM   ORDERS WHERE o_orderpriority='HIGH';

DEFINE turn_around_time(@orderkey, @totalprice, @orderdate) {
    IF (@totalprice < 2000)
        SELECT max(l_shipdate − @orderdate) FROM LINEITEM
        WHERE l_orderkey=@orderkey;
    ELSE
        SELECT max(l_commitdate−@orderdate) FROM LINEITEM
        WHERE l_orderkey=@orderkey;
}
```

---

**Experiment 4**

The query used in our fourth experiment is shown in Example 2.4. For this query, we compare only NI with NISR since decorrelation techniques are not directly applicable. The nested iteration plan employed a clustered index lookup on the *lineitem* table, where as NISR employed two restartable scans. As can be seen in Figure 2.13, NISR performs significantly better than NI; this is because the inner relation (*lineitem*) is scanned at most twice with NISR, whereas NI performs an indexed lookup of the inner relation for each tuple in the outer relation.

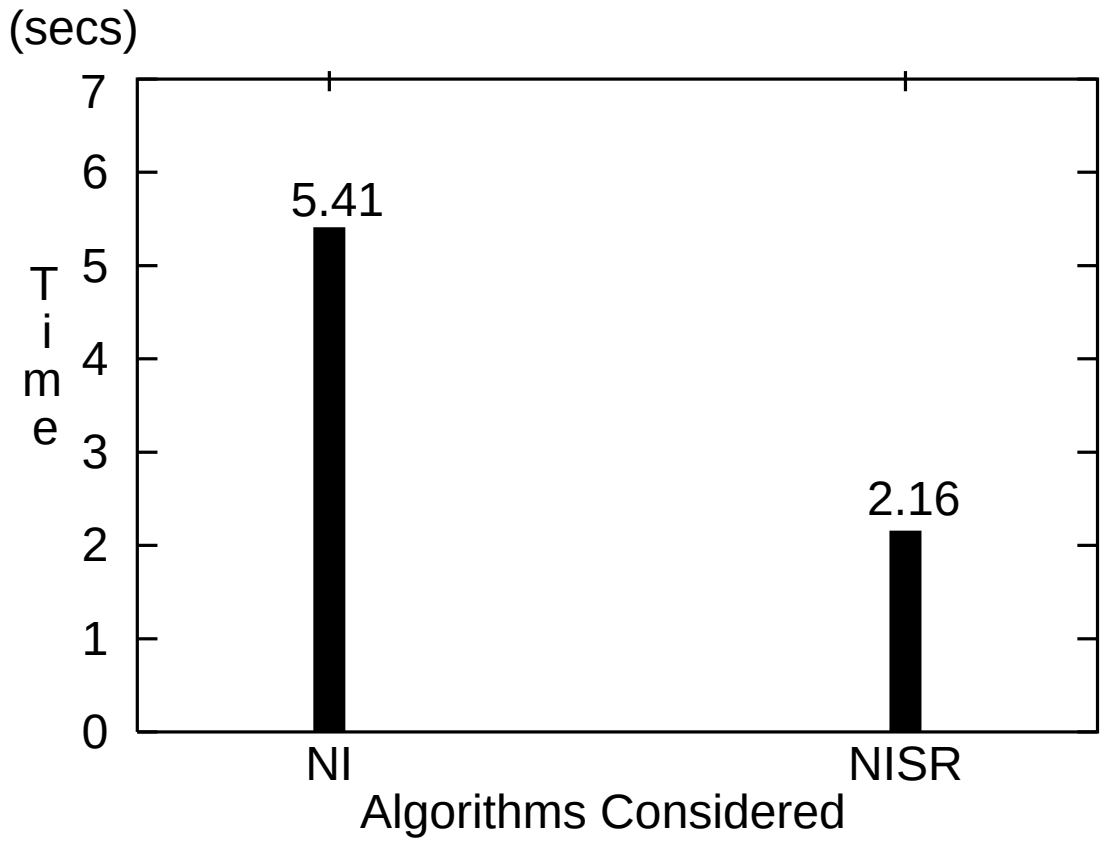

Figure 2.13: Performance Results for Experiment 4

### Optimization Overheads

The optimization overheads due to the proposed extensions in the Volcano optimizer were negligible. We measured the optimization time for complex nested queries, with up to 10 levels of nesting, and observed no measurable overheads.

## 2.5 Related Work

Since early '70s, there has been extensive work in query optimization. An excellent survey of query optimization techniques is [5]. Nested queries in particular, have received significant interest; however, most of the emphasis so far has been on decorrelation techniques [31, 30, 19, 9, 38, 49, 17, 12]. Decorrelation of queries on XML databases with XQuery as the query language is addressed in [36]. Decorrelation techniques try to rewrite a given nested query into a form that does not use the nested subquery construct. Decorrelation techniques allow an optimizer to consider alternative set oriented plans such as merge join or hash join for evaluating a nested query, and in most cases these methods perform better than the naïve nested iteration method. The techniques we proposed to speed up nested iteration are orthogonal to decorrelation, and a cost-based optimizer should consider both decorrelated evaluation and the improved nested iteration methods while choosing the best plan. In this chapter, we show several cases where the improved nested iteration methods can perform significantly better than the plans generated by decorrelation techniques.

Techniques for improving the performance of nested iteration have been proposed by Selinger et.al. [48] and Graefe [22]. System R [48] uses the idea of caching the inner sub-query result for distinct values of correlation variables and sorting the outer tuples which allows caching of only one result of the inner query at any point in time. Graefe [22] emphasizes the importance of nested iteration plans and discusses asynchronous IO, caching and sorting of outer tuples as techniques that can improve the performance of nested iteration. Asynchronous IO as a means to improve iterative query execution plans is also considered by Elhemali et.al. [12] and Iyengar et.al. [26]. Rao and Ross [44] propose an approach of improving the efficiency of iterative plans by identifying and reusing invariants (sub-expressions within the nested sub-query, which are not correlated to the outer query). The invariants can be computed just once, in the first invocation, and reused for subsequent invocations of the nested sub-query. Akinde and Böhlen [1] argue that decorrelation techniques may not *always* produce the most efficient plans, especially for complex OLAP queries, and propose generalized multi-dimensional join (GDMJ) operator as an alternative means for efficient evaluation of several types of OLAP queries.

The techniques we propose in this paper for improving the nested iteration method augment the techniques proposed in [48] and [22]. Sorting in System R is purely to ensure the cached result can be kept in memory (only one cached result needs be retained). Graefe [22] describes sorting of outer tuples to produce advantageous buffer effects in the inner query plan. Both [48] and [22] do not discuss the changes required in the optimizer to consider these options and generate an overall best plan. Database systems such as Microsoft SQL Server consider sorted correlation bindings and the expected number of times a query block is evaluated with the aim of efficiently caching the inner query results when duplicates are present, and to appropriately estimate the cost of nested query blocks [16]. To the best of our knowledge, the state-retention techniques and optimization of multi-branch, multi-level correlated queries considering parameter sort orders have not been proposed or implemented earlier.

## 2.6 Summary

In this chapter we revisited iterative execution plans for nested queries and showed how sort order of parameter bindings can be exploited through state retention to improve their execution time. For several queries, even when decorrelation is applicable, an iterative execution plan might be the most efficient of the available alternatives and hence the optimizer's search space should include these improved nested iteration plans and the optimizer must estimate their cost appropriately. We showed how a Volcano style cost-based optimizer can be extended to take into account state retention of operators and effects of parameter sort orders. We presented a performance study based on our implementation of the proposed techniques and the results show significant benefits for several types of queries, with no noticeable overheads in the optimization time.

# Chapter 3

# Sort Order Selection

The previous chapter showed the use of sort orders (of sub-query parameters) in improving the efficiency of iterative execution plans. Sort orders, in general, play an important role in query evaluation. Algorithms that rely on sorted inputs are widely used to implement joins, grouping, duplicate elimination and other set operations. The notion of *interesting orders* [48] has allowed query optimizers to consider plans that could be locally sub-optimal, but produce ordered output beneficial for other operators, and thus produce a globally optimal plan. However, the number of interesting orders for most operators is factorial in the number of attributes involved. For example, all possible sort orders on the set of join attributes are of interest to a merge join. Considering the exhaustive set of interesting orders is prohibitively expensive as the input sub-expressions must be optimized for each interesting sort order and the corresponding plan must be memoized.

Deciding a practical set of interesting sort orders is a crucial step in query optimization. The variation in plan cost due to the choice of different sort orders could be very high. A plan with carefully chosen sort orders, which exploit clustering/covering indices and commonalities between order requirements of multiple operators, can perform significantly better than a plan with naïvely chosen orders, due to the reduction in sorting cost. Sorting cost in a plan can also be reduced by exploiting partially available sort orders. For example, a primary index on a subset of attributes involved in the join predicate partially fulfills the sort order requirement of merge-join, and can greatly reduce the cost of the intermediate sorting stage. The optimizer must therefore consider partial sort orders while choosing interesting orders. Sorting based algorithms for binary operators such as join and union, although agnostic to the exact sort order of their inputs, require a matching order from the two inputs. These factors make the problem of choosing sort non-trivial.

In this chapter we address the problem of choosing interesting sort orders. We use join expressions for our description. However, the solution can be used for choosing input sort orders for any sorting based operator as well as for choosing sort orders on sub-query/UDF parameters.

The rest of this chapter is organized as follows. Section 3.1 describes how partial sort orders can be exploited during sorting and proposes extensions to a cost-based optimizer to account for their benefits during plan generation. In Section 3.2 we show that a special case of the problem of selecting globally optimal sort orders is NP-hard and give a 1/2-benefit approximation algorithm to handle the case. Although the problem is intractable, the knowledge of available indices and sort order propagation properties of physical operators allows us to provide a good heuristic approach, which we describe in Section 3.3. Section 3.4 shows how the solution can be used to obtain interesting parameter sort orders for nested sub-expressions. We present our experimental results in Section 3.5. We discuss related work in Section 3.6 and summarize the work in Section 3.7.

## 3.1 Exploiting Partial Sort Orders

Often, sort order requirements of operators are partially satisfied by indices or other operators in the input subexpressions. A prior knowledge of partial sort orders available from inputs allows us to efficiently produce the required (complete) sort order more efficiently. When operators have flexible order requirements, it is thus important to choose a sort order that makes maximum use of partial sort orders already available. We motivate the problem with an example. Consider the query shown in Example 3.1. Such queries frequently arise in consolidating data from multiple sources, *e.g.,* in extract-transform-load (ETL) tasks. The join predicate between the two *catalog* tables involves four attributes and two of these attributes are also involved in another join with the *rating* table. Further, the order-by clause asks for sorting on a large number of columns including the columns involved in the join predicate.

The two catalog tables contain 2 million records each, and have average tuple sizes of 100 and 80. We assume a disk block size of 4K bytes and 10000 blocks (40 MB) of main memory for sorting. The table *catalog1* is clustered on *year* and the table *catalog2*

**Example 3.1** *A Query with Many Join and Grouping Attributes*

*SELECT c1.make, c1.year, c1.city, c1.color, c1.sellreason, c2.breakdowns, r.rating*
*FROM catalog1 c1, catalog2 c2, rating r*
*WHERE c1.city=c2.city AND c1.make=c2.make AND c1.year=c2.year AND*
*c1.color=c2.color AND c1.make=r.make and c1.year=r.year*
*ORDER BY c1.make, c1.year, c1.color, c1.city, c1.sellreason, c2.breakdowns, r.rating;*

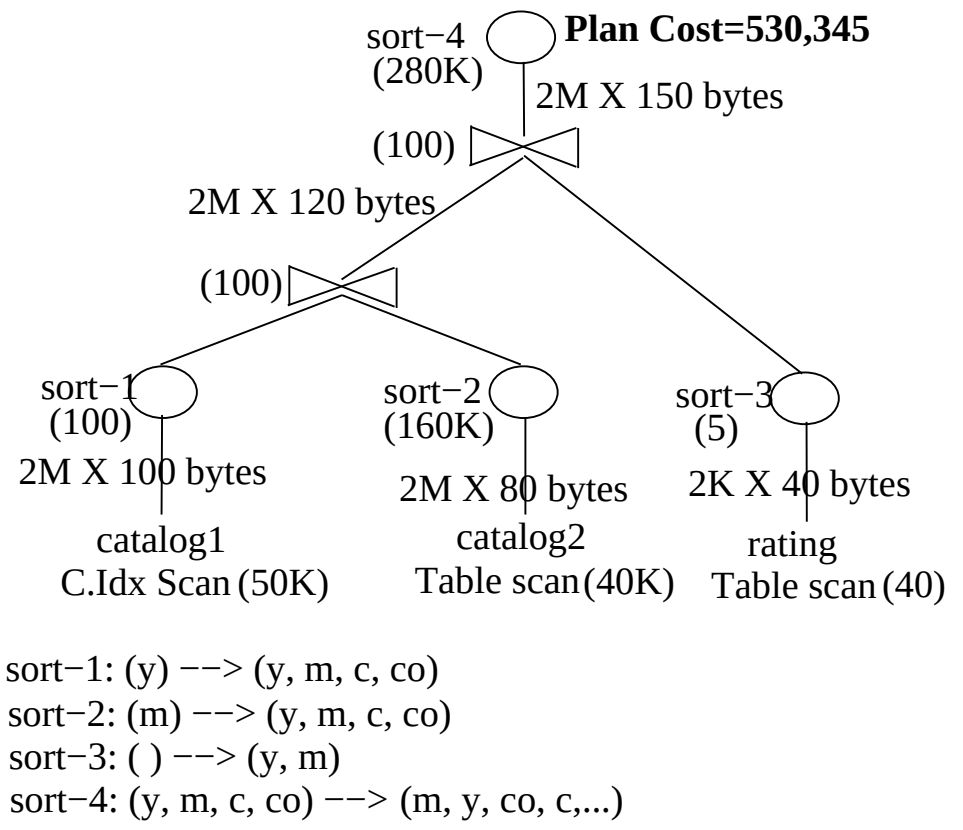

Figure 3.1: A Naïve Merge-Join Plan

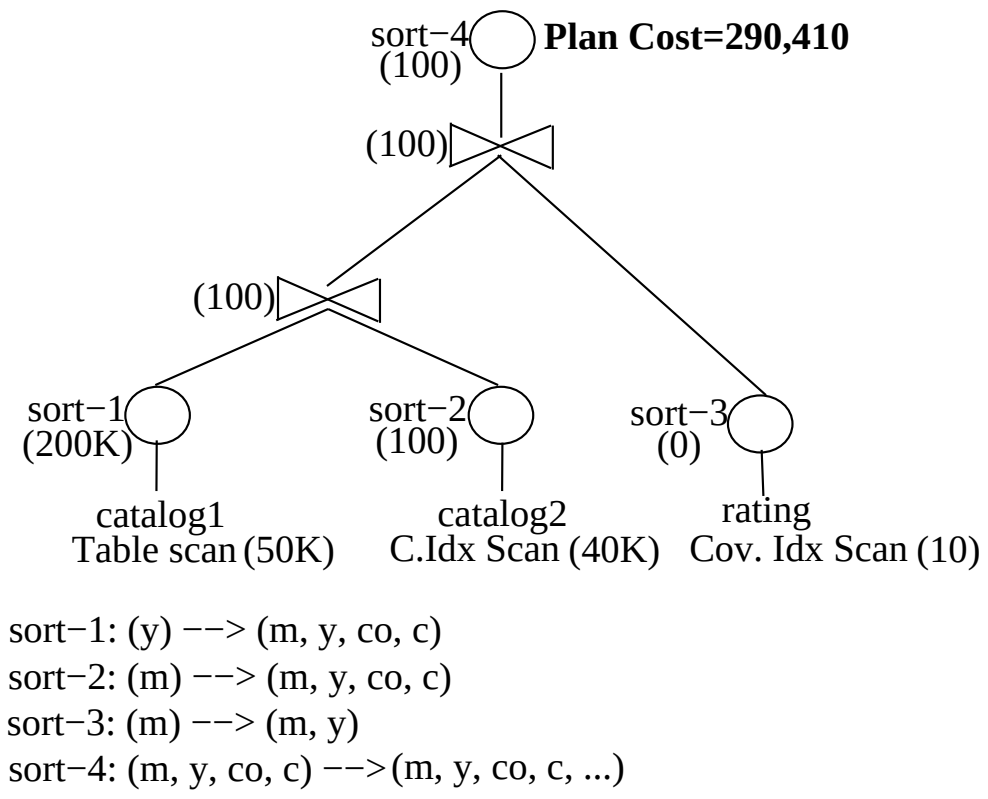

Figure 3.2: Optimal Merge-Join Plan

is clustered on *make*. The *rating* table has a secondary index on the *make* column, and the index includes the *year* and *rating* columns in its leaf pages (*i.e.,* the index covers the query). Figures 3.1 and 3.2 show two different plans for the example query. Numbers in the parentheses indicate estimated cost of the operators in number of I/Os (CPU cost is appropriately translated into I/O cost units). Edges are marked with the number of tuples expected to flow on that edge and their average size. For brevity, the input and output orders for the sort enforcers are shown using only the starting letters of the column names. Though both plans use the same join order and employ sort-merge joins, the second plan is expected to perform significantly better than the first.

### 3.1.1 Changes to External Sort

External sorting algorithms have been studied extensively but in isolation. The standard replacement selection [32] for run formation well adapts with the extent to which input is presorted. In the extreme case, when the input is fully sorted, it generates a single run on the disk and avoids merging altogether. Larson [33] revisits run formation in the context of query processing and extends the standard replacement selection to handle variable length keys and to improve locality of reference. Estivill-Castro and Wood [13] provide a survey of adaptive sorting algorithms. The technique we propose in this section to exploit partial sort orders is a specific optimization in the context of multi-key external sorting. We observe that, by exploiting prior knowledge of partial sort order of input, it is possible to eliminate disk I/O altogether and have a completely pipelined execution of the sort operator.

We use the following conventions: $o, o_1, o_2$ *etc.* refer to sort orders. A sort order of size $n$ is a sequence of attributes/columns $(a_1, a_2, \ldots, a_n)$. Sort direction (ascending/descending) is ignored; our techniques are applicable independent of the sort direction.

- $\epsilon$ : Empty (no) sort order
- $attrs(o)$ : The set of attributes in sort order $o$
- $|o|$ : Number of attributes in the sort order $o$
- $o_1 \leq o_2$ : Order $o_2$ subsumes order $o_1$ ($o_1$ is a prefix of $o_2$)
- $o_1 < o_2$ : Order $o_1$ is a strict prefix of $o_2$

Consider a case where the sort order to produce is $(col_1, col_2)$ and the input already has the order $(col_1)$. Standard replacement-selection writes a single run with all the tuples to the disk and reads it back again; this breaks the pipeline and incurs substantial I/O for large inputs. It is not difficult to see how the standard replacement selection can be modified to exploit the partial sort orders. Let $o = (a_1, a_2, \ldots, a_n)$ be the desired sort order and $o' = (a_1, a_2, \ldots, a_k)$, $k < n$ be the partial sort order known to hold on the input. At any point during sorting we need to retain only those tuples that have the same value for attributes $a_1, a_2, \ldots, a_k$. When a tuple with a new value for these set of attributes is read, all the tuples in the heap (or on disk if there are large number of tuples matching a given value of $a_1, a_2, \ldots, a_k$) can be sent to the next operator in sorted order. Thus in most cases, partial sort orders allow a completely pipelined execution of the sort. Exploiting partial sort orders in this way has several benefits:

1. Let $o = (a_1, a_2, \ldots, a_n)$ be the desired sort order and $o' = (a_1, a_2, \ldots, a_k)$, $k < n$ be the partial sort order known to already hold on the input. We call the set of tuples that have the same value for attributes $(a_1, a_2, \ldots, a_k)$ as a *partial sort segment.* If each *partial sort segment* fits in memory (which is quite often the case in practice), the entire sort operation can be completed without any disk I/O.

2. Exploiting partial sort orders allows us to output tuples early (as soon as a new segment starts). In a pipelined execution this can have large benefits. Moreover, producing tuples early has immense benefits for Top-K queries and situations where the user retrieves only some result tuples.

3. Since sorting of each *partial sort segment* is done independently, the number of comparisons are significantly reduced. Note that we empty the heap every time a new segment starts and hence insertions into heap will be faster. In general, independently sorting $k$ segments each of size $n/k$ elements, has the complexity $O(n\ log(n/k))$ as against $O(n\ log(n))$ for sorting all $n$ elements. Further, while sorting each *partial sort segment* comparisons need to be done on fewer attributes, $(a_{k+1}, \ldots, a_n)$ in the above case.

Our experimental study presented in Section 3.5 confirms that the benefits of exploiting partial sort orders can be substantial, and yet none of the systems we evaluated exploited the partial sort orders.

### 3.1.2 Optimizer Extensions for Partial Sort Orders

In this section we assume operators have fixed sort order requirements, and we focus only on incorporating partial sort orders. We deal with flexible sort order requirements of operators in subsequent sections.
We use the following notations:

- $o_1 \wedge o_2$ : Longest common prefix between sort orders $o_1$ and $o_2$
- $o \wedge s$ : Longest prefix of sort order $o$ such that each attribute in the prefix belongs to the attribute set $s$
- $o_1 + o_2$ : Sort order obtained by concatenating $o_1$ and $o_2$
- $o_1 - o_2$ : Sort order $o'$ such that $o_2 + o' = o_1$ (defined only when $o_2 \leq o_1$)
- $coe(e, o_1, o_2)$ : The cost of enforcing order $o_2$ on the result of expression $e$ which already has order $o_1$
- $N(e)$ : Expected size, in number of tuples, of the result of expression $e$

- $B(e)$ : Expected size, in number of blocks, of the result of expression $e$
- $D(e, s)$ : Distinct values for attribute(s) $s$ of expression $e$. $D(e, s) = N(\Pi_s(e))$
- $cpu_cost(e, o)$ : CPU cost of sorting the result of $e$ to get order $o$
- $M$ : Number of memory blocks available for sorting

The *Volcano* optimizer framework [23] assumes that an algorithm (physical operator) either guarantees a required sort order fully or it does not. Further, a physical property enforcer (such as sort) only knows the property to be enforced and has no information about the properties that hold on its input. The optimizer's cost estimate for the enforcer thus depends only on the required output property (sort order). In order to remedy these deficiencies we extended the optimizer in the following way: consider an optimization goal $(e, o)$, where $e$ is the expression and $o$ the required output sort order. If the physical operator being considered for the logical operator at the root of $e$ guarantees a sort order $o' < o$, then the optimizer adds a partial sort enforcer *enf* to enforce $o$ from $o'$. We use the following cost model to account for the benefits of partial sorting.

$$coe(e, \epsilon, o) = \begin{cases} cpu\text{-}cost(e, o) & \text{if } B(e) \leq M \\ B(e)(2\lceil log_{M-1}(B(e)/M)\rceil + 1) & \text{otherwise} \end{cases}$$

If $e$ is known to have the order $o_1$, we estimate the cost of obtaining an order $o_2$ as follows: $coe(e, o_1, o_2) = D(e, attrs(o_s)) * coe(e', \epsilon, o_r)$, where $o_s = o_2 \wedge o_1$, $o_r = o_2 - o_s$ and $e' = \sigma_p(e)$, where $p$ equates attributes in $o_s$ to an arbitrary constant. Intuitively, we consider the cost of sorting a single *partial sort segment* independently and multiply it by the number of segments. Note that we assume uniform distribution of values for $attrs(o_s)$. Therefore, we estimate $N(e') = \lceil N(e)/D(e, attrs(o_s))\rceil$ and $B(e') = \lceil B(e)/D(e, attrs(o_s))\rceil$. When the actual distribution of values is available, a more accurate cost model that does not rely on the uniform distribution assumption can be used.

## 3.2 Choosing Sort Orders for a Join Tree

Consider a join expression $e = e_1 \bowtie e_2$, where $e_1, e_2$ are input subexpressions and the join predicate is of the form: $(e_1.a_1 = e_2.a_1$ *and* $e_1.a_2 = e_2.a_2 \ldots$ *and* $e_1.a_n = e_2.a_n)$. Note that, *w.l.g.*, we use the same name for attributes being compared from either side and we call the set $\{a_1, a_2, \ldots, a_n\}$ as the join attribute set. In this case, the merge join algorithm has potentially $n!$ interesting sort orders on inputs $e_1$ and $e_2$ [1]. The specific sort order chosen for the merge-join can have significant influence on the plan cost due to the following reasons: (*i*) Clustering and covering indices, indexed materialized views and other operators in the subexpressions $e_1, e_2$ can make one sort order much cheaper to produce than another. (*ii*) The merge-join produces the same order on its output as the one selected for its inputs. Hence, a sort order that helps another operator above the merge-join can help eliminate a sort or just have a partial sort. In this section we show that a special case of the the problem of choosing optimal sort orders for a tree of merge-joins is *NP-Hard* and provide a 1/2 benefit approximation algorithm for the problem[2]. In the next section, we describe our heuristics for a more general setting of the problem in which we make use of the proposed 1/2 benefit approximation algorithm.

### 3.2.1 Finding Optimal is NP-Hard

We now show that the problem of choosing optimal sort orders is NP-Hard by considering a special case of the problem. Let $e = R_1 \bowtie R_2 \bowtie R_3 \ldots \bowtie R_n$ be a join expression with conjunctive join predicates on $n$ relations, where $n$ is a power of 2. Let $T$ be a balanced join order tree for $e$. Figure 3.3 shows an example. For each join node $v$ in $T$, we assign an attribute set $L_v$ (called *representative join attribute set*), which is constructed as follows. If $a_i$ is an attribute involved in the join predicate of $v$ then $\mathcal{H}(a_i) \in L_v$, where $\mathcal{H}(a_i)$ is the *representative* of the attribute equivalence class in the result of $e$. Two attributes $a_i$ and $a_j$ belong to the same attribute equivalence class if they are equated directly or transitively in the join predicates of $e$. The *representative* of an equivalence class is an arbitrarily chosen attribute belonging to the equivalence class. For example, if the predicate $R_1.a_1 = R_2.a_2 \wedge R_2.a_2 = R_3.a_3$ is part of the join predicates of $e$, then $R_1.a_1, R_2.a_2$ and $R_3.a_3$ belong to the same attribute equivalence class, and we will have $\mathcal{H}(R_1.a_1) = \mathcal{H}(R_2.a_2) = \mathcal{H}(R_2.a_2) = R_1.a_1$. In Figure 3.3 we have shown the representative join attribute sets for each join node. For brevity, we omit the relation

[1]We assume merge-join requires sorting on all attributes involved in the join predicate. We do not consider orders on subsets of join attributes since the additional cost incurred at merge-join matches the benefit of sorting a smaller subset of attributes.

[2]The work was carried out in collaboration with Ajit A. Diwan and Ch. Sobhan Babu.

name qualifiers for attributes.

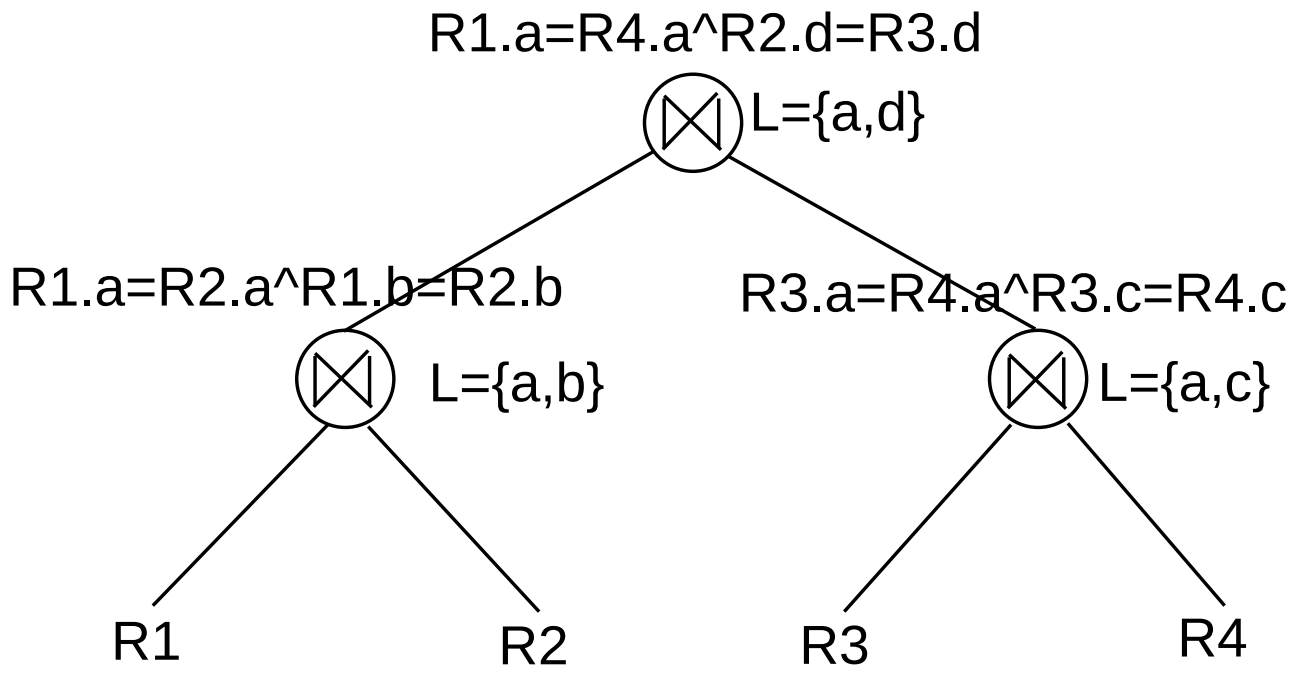

Figure 3.3: A Join Tree with Representative Join Attribute Sets

Now, suppose all the base relations and intermediate results in $T$ are of the same size and no indices are present on the base relations. The problem of choosing optimal sort orders for each join requires us to choose permutations of representative join attribute sets such that we minimize the cost of intermediate sorts. The cost of sorting depends on the sort order already present on the input and the sort order required. In general, the sort cost on any edge $(v_i, v_j)$ of the tree is a monotonically decreasing function of the length of common prefix between attribute permutations chosen for $v_i$ and $v_j$. For example, see our cost model for sort presented in Section 3.1.2. We define the benefit of a solution to be $\sum_{v_i v_j \in E} f(|p_i \wedge p_j|)$, where $E$ is the set of edges in the tree, $p_i, p_j$ are attribute permutations chosen by the solution for vertices $v_i, v_j$ and $f$ is any monotonically increasing function in the length of the common prefix $(|p_i \wedge p_j|)$, with 0 at origin (*i.e.,* $f(0) = 0$). Minimizing the sorting cost requires maximizing the total benefit.

Figure 3.4, shows an example along with a solution, which maximizes the total benefit assuming $f(|p_i \wedge p_j|) = |p_i \wedge p_j|$. The representative join attribute set for each join node is shown in curly braces besides the node. Permutations chosen by the solution are indicated with angle brackets and the number on each edge shows the benefit for that edge. Below we state the problem formally.

**Problem 1** *(Common Prefix Problem) Let $T$ be a tree having $n$ vertices, the vertex set being $V(T)$ and the edge set being $E(T)$. Each vertex $v_i$ $(i = 1, \ldots, n)$ is associated with an attribute set $s_i$. Let $f$ be any non-decreasing function with $f(0) = 0$. Find a set of attribute permutations $p_1, p_2 \ldots, p_n$, where $p_i$ is a permutation of set $s_i$, such that the benefit function $\mathcal{F} = \sum_{\forall v_i v_j \in E(T)} f(|p_i \wedge p_j|)$ is maximized.*

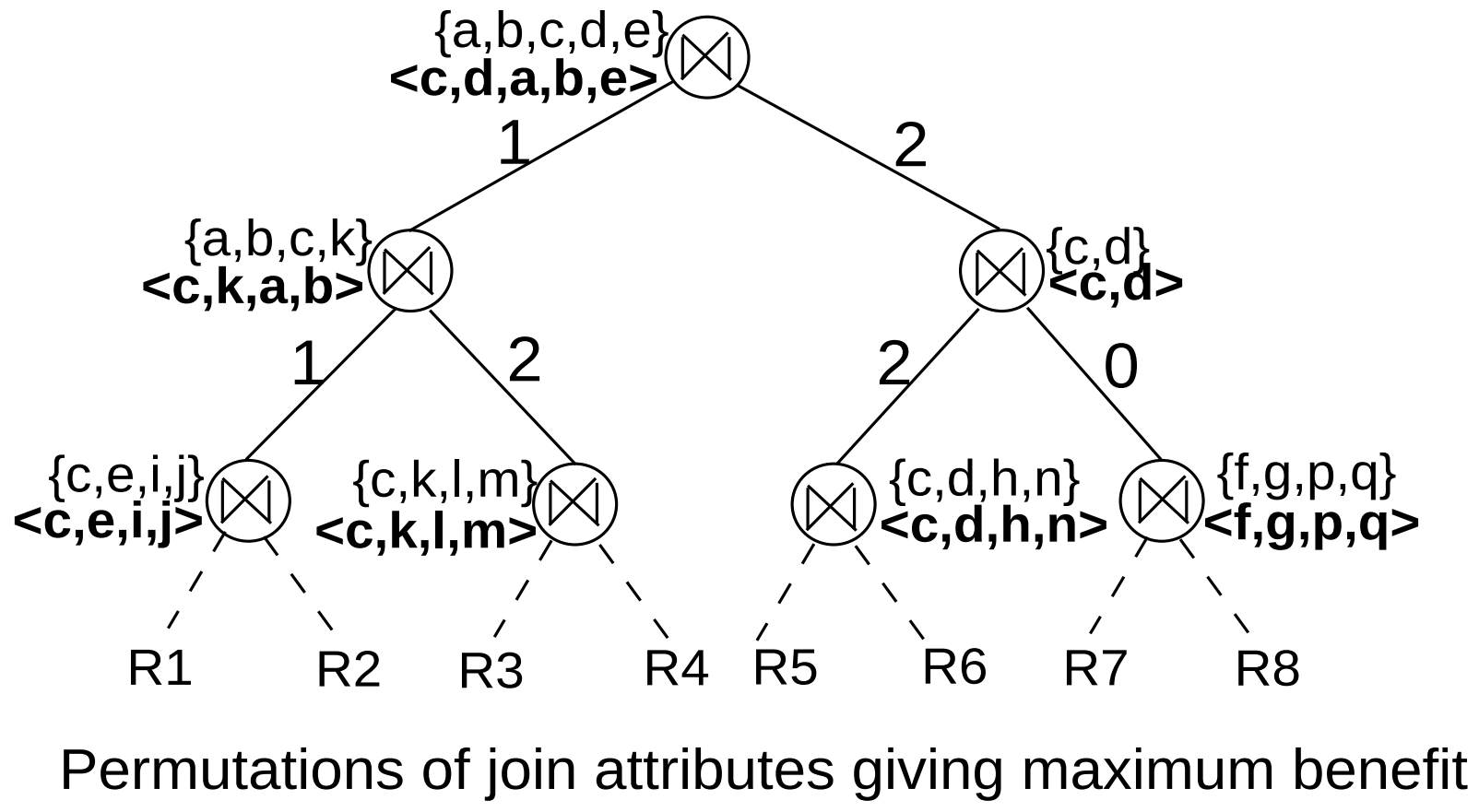

Figure 3.4: A Special Case of Choosing Globally Optimal Sort Order

We prove that Problem 1 is NP-Hard even for binary trees. To do so we consider the special case where $f(|p_i \wedge p_j|) = |p_i \wedge p_j|$.

Before giving a proof for this theorem, we will study some well known NP-Hard problems, which are reducible to the Common Prefix Problem.

**Problem 2** *(Sum-Cut) [11]* Given a graph $G$ with $m$ vertices, number the vertices of $G$ as $1, \ldots, m$ such that $\sum_{1 \leq i \leq m} c_i$ is minimized, where $c_i$ is the number of vertices numbered $\leq i$ that are adjacent to at least one vertex numbered greater than $i$.

The Sum-Cut problem can be rephrased as follows: given a graph $G$ with $m$ vertices, number the vertices of $G$ as $1, \ldots, m$ such that $\sum_{1 \leq i \leq m} \bar{c_i}$ is maximized, where $\bar{c_i}$ is the number of vertices numbered $\leq i$ that are adjacent to no vertex numbered greater than $i$. Let $G'$ be the complement graph of $G$. The complement graph $G'$ of $G$ contains an edge $(v_i, v_j)$ *iff* $(v_i, v_j)$ is not present in $G$. On the complement graph $G'$, it is straight-forward to see that the Sum-Cut problem is equivalent to Problem 3 given below.

**Problem 3** *(Mod-Sum-Cut)* Given a graph $H$ with $m$ vertices, number the vertices of $H$ as $1, 2, \ldots, m$ such that $\sum_{1 \leq i \leq m} q_i$ is maximized, where $q_i$ is the number of vertices that are adjacent to *all* the vertices numbered greater than $i$.

First, we reduce the *Mod-Sum-Cut* problem to the Common Prefix Problem on star trees. A *star tree* or simply a *star* of $n$ vertices is a tree with a root and $n-1$ leaf vertices.

**Lemma 3.1** *The Common Prefix Problem is NP-Hard for stars.*

**Proof**: We reduce the *Mod-Sum-Cut* problem to the Common Prefix Problem on stars, with the function $f$ set to $|p_i \wedge p_j|$ (*i.e.,* the length of the longest common prefix). Let graph $G$ with $m$ vertices be the given instance of Mod-Sum-Cut problem. Let $v_1, \ldots, v_m$ be the vertices of $G$. We construct an instance of the Common Prefix Problem on stars as follows: let $S$ be a star having $m+1$ vertices, with $u_r$ as as its root and $u_1, \ldots, u_m$ as its leaves. The attribute set of root $u_r$ is chosen to be the set of all vertices in $G$ (*i.e.,* $\{v_1, \ldots, v_m\}$), and the attribute set of each leaf $u_i$ is chosen to be *adj*($v_i$), where *adj*($v_i$) is the set of all vertices adjacent to $v_i$ in graph $G$. A pictorial illustration of the construction is shown in Figure 3.5.

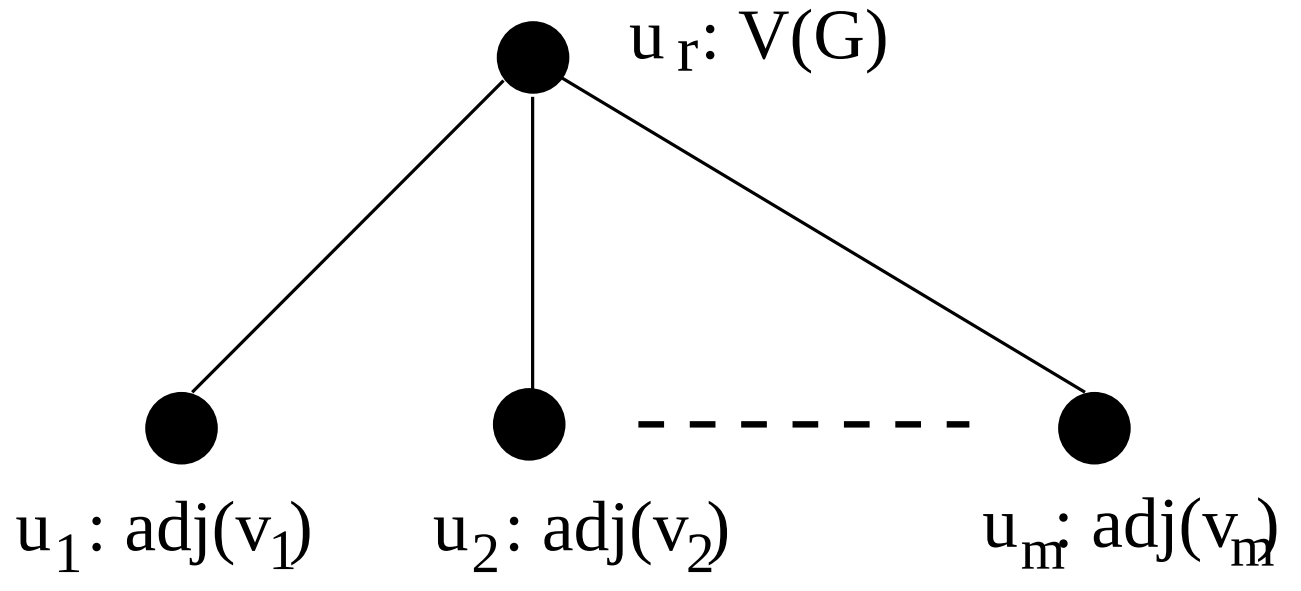

Figure 3.5: Reducing Mod-Sum-Cut to Common Prefix on Star

Now, we show that there exists a solution of value $k$ for Mod-Sum-Cut on $G$ *iff* there exists a solution of value $k$ for the Common Prefix Problem on $S$.

Suppose there exists a solution of value $k$ for Mod-Sum-Cut on $G$. Let the order of vertices in the solution be $v_{g(1)}, v_{g(2)}, \ldots, v_{g(m)}$, where $g$ is a permutation on $1, \ldots, m$ (*i.e.,* a one-to-one function from $\{1, \ldots, m\}$ to $\{1, \ldots, m\}$). We construct the solution for the corresponding Common Prefix Problem (for star $S$) as follows: for the root vertex $u_r$, we choose the attribute permutation to be $o_r = v_{g(m)}, v_{g(m-1)}, \ldots, v_{g(1)}$. For each leaf vertex $u_i$, we choose a permutation $o_i$ of its attribute set *adj*($v_i$) such that the length of the longest common prefix $|o_i \wedge o_r|$ is maximum.

In the solution ordering for Mod-Sum-Cut, let $l_i$ be the smallest integer such that $v_i$ is adjacent to all vertices in the set $v_{g(l_i+1)}, \ldots, v_{g(m)}$. This implies the following: *(i)* in the solution value for Mod-Sum-Cut, vertex $v_i$ will be counted $m - l_i$ times, *i.e.,* $k = \sum_{v_i}(m - l_i)$, and *(ii)* in the corresponding Common Prefix Problem, there exists a common prefix of length $m - l_i$ between the permutations chosen for $u_i$ and the root $u_r$. This shows there exists a solution of value $k$ for the corresponding Common Prefix

Problem.

Now, suppose there exists a solution of value $k$ for the Common Prefix Problem on $S$. In the solution, let the attribute permutation chosen for the root vertex $u_r$ be $o_r = v_{h(1)}, v_{h(2)}, \ldots, v_{h(m)}$, where $h$ is a permutation on $1, \ldots, m$. Now, we construct the solution for Mod-Sum-Cut on $G$ by reversing the order of attributes in $o_r$, *i.e.,* by ordering the vertices of $G$ as $v_{h(m)}, v_{h(m-1)}, \ldots, v_{h(1)}$.

In the solution for the Common Prefix Problem on $S$, let $o_i$ be the permutation (of set $adj(v_i)$) chosen for leaf $u_i$. Let $l_i$ denote the length of the longest common prefix between $o_i$ and $o_r$, *i.e.,* $l_i = |o_i \wedge o_r|$. We observe that, the solution value $k = \sum_{1 \leq i \leq m}(l_i)$. In the corresponding solution for Mod-Sum-Cut on $G$, $l_i$ will be the smallest integer such that vertex $v_i$ is adjacent to all vertices in the set $v_{h(m)}, \ldots, v_{h(m-l_i+1)}$. Hence, in the solution value for Mod-Sum-Cut on $G$, vertex $v_i$ will be counted $l_i$ times. Therefore, the solution for Mod-Sum-Cut will have a value of $\sum_{1 \leq i \leq m}(l_i) = k$. $\Box$

**Theorem 3.2** *Problem 1 is NP-Hard even for binary trees.*

**Proof**: We reduce the Common Prefix Problem on stars to the Common Prefix Problem on binary trees. Let $S$ be a star with $u_r$ as its root and $u_1, \ldots, u_m$ as its leaves. Let $a_r$ denote the set of attributes associated with $u_r$ and $a_1, a_2, \ldots, a_m$ denote the set of attributes associated with vertices $u_1, u_2, \ldots, u_m$ respectively. We now construct an instance of the Common Prefix Problem on binary trees as follows: let $T$ be a binary tree with $2m$ vertices, with $r_1, r_2, \ldots, r_m$ as its internal vertices and $w_1, w_2, \ldots, w_m$ as its leaves. Let the edge set $E(T)$ be $\{r_i r_{i+1} : 1 \leq i < m\} \cup \{r_i w_i : 1 \leq i \leq m\}$. Each internal vertex $r_i$ is assigned the attribute set $\mathcal{A} = a_r \cup \mathcal{L}$, where $\mathcal{L}$ is an arbitrarily chosen set of attributes of size $> m \times |a_r|$ and is disjoint from $a_r \cup a_1 \cup \ldots \cup a_m$. Each leaf vertex $w_i$ is assigned the attribute set $a_i$. Figure 3.6 pictorially illustrates the construction. In the figure, the attribute sets $a_r$ and $a_1, \ldots, a_m$ for the star are assumed to be as in Figure 3.5. Let $\mathcal{Z} = (m-1) \times |\mathcal{A}|$.

First, we show that if there exists a solution of value $k$ for the Common Prefix Problem on $S$ then there exists a solution of value $k + \mathcal{Z}$ for the Common Prefix Problem on $T$.

Suppose there exists a solution of value $k$ for the Common Prefix Problem on $S$. Let the $o_r$ be the attribute permutation assigned for $u_r$ and $o_i$ be the attribute permutation assigned for each $u_i$, $1 \leq i \leq m$. We construct a solution for $T$ as follows: for each intern vertex $r_i$, we assign the permutation $o_r + o_l$, where $o_l$ is a fixed permutation of $\mathcal{L}$, chosen arbitrarily. For each leaf vertex $w_i$, we assign the permutation $o_i$. Since the same permutation is chosen for all the internal vertices, each of the $(m-1)$ pairs of adjacent internal vertices will have a common prefix of length $|\mathcal{A}|$. Further, each of the $m$ pairs of internal and leaf vertices that are adjacent to each other will have a common prefix of length $|o_i \wedge o_r|$. As $k = \sum_{1 \leq i \leq m}(|o_i \wedge o_r|)$, we conclude the solution value for $T$ is $k + \mathcal{Z}$.

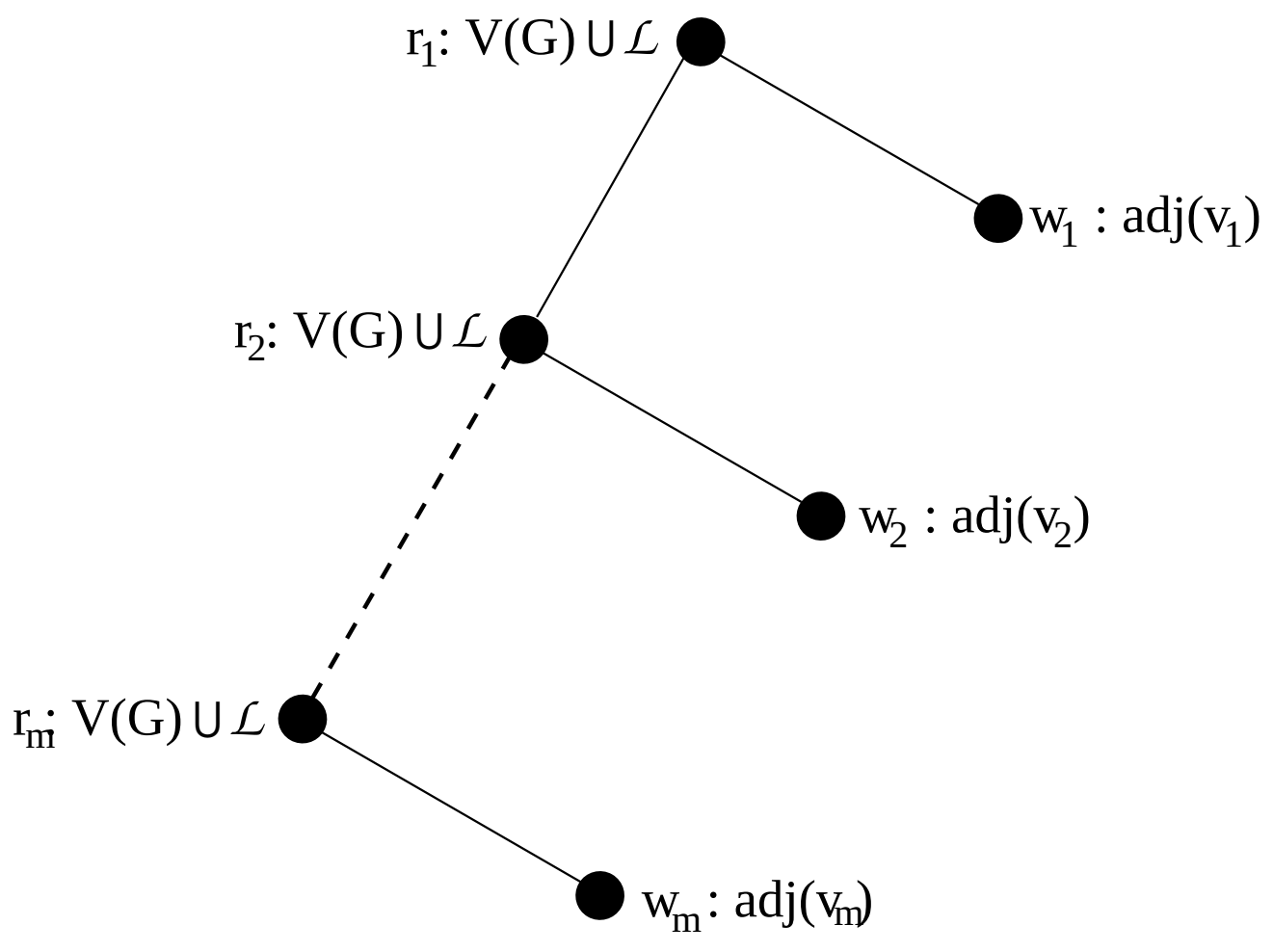

Figure 3.6: Reducing the Common Prefix Problem on Stars to the Common Prefix Problem on Binary Trees

Next, we show that if there exists a solution of value $k$ for the Common Prefix Problem on $T$ then there exists a solution of value $k - \mathcal{Z}$ for the Common Prefix Problem on $S$. To do so, we make use two supporting lemmas, Lemma 3.3 and Lemma 3.4. Below we state and prove these lemmas and then continue with the proof of Theorem 3.2.

**Lemma 3.3** *In any optimal solution for the Common Prefix Problem on $T$, all the internal vertices are assigned an identical permutation $p$.*

**Proof**: Let $T_{opt}$ be an optimal solution for $T$. In the optimal solution, let $p_1, \ldots, p_m$ be the permutations assigned to internal vertices $r_1, \ldots, r_m$ respectively. We prove that $|p_i \wedge p_{i+1}| = |\mathcal{A}|$ for $1 \leq i < m$, which essentially proves this lemma.

**Case 1:** Suppose $|p_i \wedge p_{i+1}| < |a_r|$ for some $i$, $1 \leq i < m$.

This implies, the total benefit of $T_{opt}$, $Ben(T_{opt}) < (m-2) \times |\mathcal{A}| + (m+1) \times |a_r|$. Since $|\mathcal{L}| > m \times |a_r|$, we have $|\mathcal{A}| > (m+1) \times |a_r|$. Therefore, $Ben(T_{opt}) < (m-1) \times |\mathcal{A}|$. However, we know that there exists a solution for $T$ with benefit of at least $(m-1) \times |\mathcal{A}|$.

This is because each of the internal vertices have the same attribute set of size $|\mathcal{A}|$. This contradicts the given fact that $T_{opt}$ is optimal. Therefore, we conclude that our assumption: $|p_i \wedge p_{i+1}| < |a_r|$ for some $i$, $1 \leq i < m$, cannot be true.

**Case 2:** Suppose $|p_i \wedge p_{i+1}| \geq |a_r|$ for all $1 \leq i < m$, but $|p_i \wedge p_{i+1}| < |\mathcal{A}|$ for some $i$, $1 \leq i < m$.

Given a permutation $p$, we use the notation $p[j]$ to denote the attribute at the $j^{th}$ position, where $1 \leq j \leq |p|$. The condition for Case 2 implies the following:

(a) $p_1[j] = p_2[j] = \ldots = p_m[j]$ for $1 \leq j \leq |a_r|$.

(b) in $T_{opt}$, the total benefit of edges incident between internal vertices $\sum_{1 \leq i < m}(|p_i \wedge p_{i+1}|)$ must be less than $|\mathcal{A}| \times (m-1)$.

Now, consider a solution $T'_{opt}$ for $T$ in which each leaf vertex is assigned the same permutation as in $T_{opt}$ and all the internal vertices are assigned an identical permutation $p$ constructed as follows: the first $|a_r|$ attributes of $p$ are chosen in the same order as the first $|a_r|$ attributes in $p_i$ for any $1 \leq i \leq m$ (*i.e.*, $p[j] = p_i[j], 1 \leq j \leq |a_r|$), and the next $|\mathcal{A} - a_r|$ attributes are chosen an in an arbitrary order.

We observe that in both $T'_{opt}$ and $T_{opt}$ the total benefit of edges incident from internal vertices to leaf vertices remains the same. However, in $T'_{opt}$, the total benefit of edges incident between internal vertices will be $|\mathcal{A}| \times (m-1)$ (this is because all internal vertices are assigned an identical permutation). This implies, the total benefit of $T'_{opt}$ is larger than that of $T_{opt}$, which contradicts the given fact that $T_{opt}$ is optimal. Therefore, we conclude the assumption made for Case 2 cannot be true.

We thus conclude in every optimal solution $T_{opt}$, all the internal vertices are assigned an identical permutation, completing the proof of Lemma 3.3. □

Next, we state and prove our second supporting lemma.

**Lemma 3.4** *There exists an optimal solution for $T$ such that, in the permutation $p$ chosen for the internal vertices, every attribute in set $a_r$ occurs before any attribute in set $\mathcal{L}$ occurs.*

**Proof**: Let $T_{opt}$ be an optimal solution for $T$. In $T_{opt}$, let $p_1, \ldots, p_m$ be the permutations assigned to the internal vertices $r_1, \ldots, r_m$ respectively. From Lemma 3.3 we know that all the internal vertices are assigned an identical permutation; let $p_1 = p_2 = \ldots = p_m = p$.

Suppose there exist $x, y$ such that $x < y$, $p[x] \in \mathcal{L}$ and $p[y] \in a_r$. We now modify $p_1, \ldots, p_m$ as follows: in each $p_i$, we swap $p_i[x]$ with $p_i[y]$. Since there is no attribute common to the set $\mathcal{L}$ and the attribute sets associated with the leaf vertices, this modification cannot decrease the total benefit of $T_{opt}$. This modification can be repeated until all the attributes in $a_r$ appear before the attributes in $\mathcal{L}$ in the permutation $p$. $\square$

From Lemmas 3.3 and 3.4, we can make the following statement: if there exists a solution of value $k$ for the Common Prefix Problem on $T$, then there exists a solution $T_{opt}$ of value at least $k$, in which, all internal vertices are assigned an identical permutation $p$ and $|p \wedge a_r| = |a_r|$.

We now construct a solution for the star $S$ as follows: for the root vertex $u_r$ we assign the permutation $p \wedge a_r$. For each leaf vertex $u_i$, we assign the permutation chosen for the corresponding leaf $w_i$ in the solution $T_{opt}$. In $T_{opt}$, the maximum benefit which can be contributed by edges incident between internal vertices of $T$ is $\mathcal{Z}$. Therefore we conclude the corresponding solution on $S$ should have a benefit of at least $k - \mathcal{Z}$. $\square$

### 3.2.2 A Polynomial Time Algorithm for Paths

We now present an efficient algorithm for solving the Common Prefix Problem, when the tree is a path. The algorithm employs dynamic programming. Note that left-deep and right-deep join plans result in problem instances on paths.

**Theorem 3.5** *Let $v_1, v_2, \ldots, v_n$ be a path, where each vertex $v_i$ is associated with an attribute set $s_i$. The optimal solution of Common Prefix Problem for any segment $(i, j)$ of the path, $OPT(i,j) = max\{\ OPT(i,k) + OPT(k+1,j) + f(c(i,j))\ \}$ over all $i \leq k < j$, where $c(i,j)$ is the number of common attributes for the segment $(i,j)$.*

**Proof**:

**Case 1:** Let $c(i,j) = 0$, *i.e.*, there exists no attribute common to all vertices $v_i, v_{i+1}, \ldots, v_j$. Consider an optimal solution for the path $v_i, \ldots, v_j$. Let $p_x$ be the attribute permutation assigned by the optimal solution to vertex $v_x$, $i \leq x \leq j$. The optimal solution must contain two vertices $v_k, v_{k+1}$ such that the benefit for the edge $(v_k, v_{k+1})$ is 0, *i.e.*, $|p_k \wedge p_{k+1}| = 0$. This directly follows from the assumption of Case 1, $c(i,j) = 0$. Now, the problem can be independently solved for the two segments $(v_i, v_k)$ and $(v_{k+1}, v_j)$ and $OPT(i,j) = OPT(i,k) + OPT(k+1,j)$.

**Case 2:** Let $c(i, j) \neq 0$. Let $s(i, j)$ be the set of attributes common to all vertices $v_i, \ldots, v_j$. Note that the cardinality $|s(i, j)| = c(i, j)$. Let $o_s$ be an arbitrarily chosen permutation of set $s(i, j)$. We claim that there exists an optimal solution such that, for every vertex $v_x$ ($i \leq x \leq j$) the attribute permutation $p_x$ chosen by the optimal solution has $o_s$ as its prefix. To see this, consider an optimal solution in which $o_s$ is not a prefix of some $p_x$. We can then reorder the permutations assigned to the vertices, without a decrease in the total benefit $OPT(i, j)$, so as to have $o_s$ as the prefix of each $p_x, i \leq x \leq j$.

Let $v'_i, \ldots, v'_j$ be a path where each vertex $v'_x$ is associated with the attribute set $s_x - s(i, j)$, *i.e.*, $v'_x$ has all the attributes of $v_x$ except those in $s(i, j)$. We can see that the value of the optimal solution $OPT(v_i, v_j)$ is given by Equation 3.1.

$$OPT(v_i, v_j) = OPT(v'_i, v'_j) + (j - i) \times f(c(i, j)) \tag{3.1}$$

Now, consider the path $v'_i, \ldots, v'_j$. From our construction of the path, we know there are no attributes common to all the vertices of $v'_i, \ldots, v'_j$. Therefore, we have:

$$OPT(v'_i, v'_j) = OPT(v'_i, k') + OPT(v'_{k+1}, v'_j) \text{ for some } k', 0 \leq k' \leq n. \tag{3.2}$$

From the construction of the path $v'_i, \ldots, v'_j$, we have:

$$OPT(v'_i, v'_k) = OPT(v_i, v_k) - (k - i) \times f(c(i, j)) \tag{3.3}$$

$$OPT(v'_{k+1}, v'_j) = OPT(v_{k+1}, v_j) - (j - k - 1) \times f(c(i, j)) \tag{3.4}$$

Substituting from Equations 3.3 and 3.4 in Equation 3.2, we get:

$$OPT(v'_i, v'_j) = OPT(v_i, v_k) + OPT(v_{k+1}, v_j) - (j - i - 1) \times f(c(i, j)) \text{ for some } k', 0 \leq k' \leq n. \tag{3.5}$$

Substituting Equation 3.5 in Equation 3.1, we get:

$$OPT(v_i, v_j) = OPT(v_i, v_k) + OPT(v_{k+1}, v_j) + f(c(i, j)) \text{ for some } k', 0 \leq k' \leq n. \tag{3.6}$$

Hence the proof. □

Procedure *PathOrder* in Figure 3.7 computes optimal attribute permutations for any path $(1, n)$, where each vertex $i$, $1 \leq i \leq n$, is associated with an attribute set $s[i]$. The procedure uses dynamic programming and computes solutions bottom up starting from path segments of length 0 (single vertices). The procedure begins by assigning a benefit of 0 to all path segments $(i, i)$, $1 \leq i \leq n$. It then constructs solutions for paths of increasing length. For each path $(i, i+j)$, of length $j$, $1 \leq j \leq n-1$, the value of $k$, which maximizes *ben*=*benefit*$(i, k)$+*benefit*$(k+1, i+j)$ is identified. The values of $k$, *ben* and the attributes common to all the vertices of path $(i, i+j)$ are remembered (memoized). Finally, the sub-procedure *MakePermutation* is used to construct the attribute permutations $p[1], \ldots, p[n]$ using the memo structure. The first call to procedure *MakePermutation* is made with parameter $i$ set to 1 (the first vertex on the path) and $j$ set to $n$ (the last vertex on the path), and each of the attribute permutations $p[1], \ldots, p[n]$ initialized with an empty sort order. Procedure *MakePermutation* constructs the attribute permutation for each of the vertices $i, i+1, \ldots, j$ as follows: a permutation *cp* of *commons(i, j)* (*i.e.,* the set of attributes common to all the vertices from $i, \ldots, j$) is chosen at arbitrary. *cp* is then appended to each of $p[i], \ldots, p[j]$. The common attributes for segment $(i, j)$ are then removed from the common attributes of all subpaths of $(i, j)$. The optimal split point $m$ for the path segment $(i, j)$ is read from the memo structure, and the construction of the permutations continues recursively on subpaths $(i, m)$ and $(m+1, j)$, until $i = j$ (i.e., a single vertex). The overall time complexity of procedure *PathOrder* is $O(n^3)$.

### 3.2.3 A 1/2 Benefit Approximation Algorithm for Binary Trees

For binary trees we propose an approximation with benefit at least half that of an optimal solution. Note that our approximation guarantee implies at least half the best possible *improvement* over the worst case sort cost. This however, does not imply a 2-approximation on the total cost.

We split the tree into two sets of paths, $P_o$ and $P_e$. $P_o$ has the paths formed by edges incident from odd levels and $P_e$ has those formed by edges incident from even levels, Figure 3.8 shows an example. We then find an optimal solutions for each of the two sets of paths. Note that this gives us two solutions for the complete tree, because each set of paths covers all the vertices of the tree (for any left over vertices at the leaf level or the root, we choose an arbitrary permutation). Let the optimal solutions for the two sets of paths be $S_o$ and $S_e$ and the corresponding benefits be *ben*($S_o$) and *ben*($S_e$). Let the set of edges included in $P_o$ and $P_e$ be denoted by $E_o$ and $E_e$ respectively. Consider an optimal solution $S_T$ for the whole tree. In the optimal solution, let the sum of benefits of all edges in $E_o$ be *odd-ben*($S_T$) and that of edges in $E_e$ be *even-ben*($S_T$). Note that *ben*($S_o$) $\geq$

```
Procedure PathOrder
Input: s[n] : array of attribute sets
Output: p[n] : array of permutations or sort orders
Data Structures:
    benefit[n][n], split[n][n] : arrays of integers
    commons[n][n] : array of attribute sets
    apermute(s) : Function that returns an arbitrary permutation of attribute set s
BEGIN
    for i=1 to n
        benefit[i][i] = 0; p[i] = ε; commons[i][i] = s[i]; split[i][i] = -1;
    for j=1 to n-1                          // Consider path segments of length j
        for i = 1 to n-j                    // Consider path segment (i, i+j)
            Let k be the index such that i ≤ k < (i+j) and benefit[i][k]+benefit[k+1][i+j]
            is maximum
            commons[i][i+j] = commons[i][k] ∩ commons[k+1][i+j];
            benefit[i][i+j] = benefit[i][k] + benefit[k+1][i+j] + f(|commons[i][i+j]|);
            split[i][i+j] = k;

    Call MakePermutation(1, n);             // Form the attribute permutations
END PROC

// Procedure to construct attribute permutations from the memo structure, in which
// the optimal split point and common attributes are remembered.
Procedure MakePermutation(i, j)
BEGIN
    Let ca = commons[i][j]; // Attributes common to all the vertices from i to j
    Let cp = apermute(ca); // An arbitrarily chosen permutation of ca.
    for k=i to j
        Append cp to p[k];
    if (i = j)
        return;
    // Remove the common attributes from all subpaths of (i,j), so that the
    // attributes do not repeat.
    For all i', j' s.t. i' ≥ i and j' ≤ j
        commons[i'][j'] = commons[i'][j'] − ca;
    // Construct the permutations of remaining attributes for the two subpaths,
    // to the left and right of the the split point.
    m = split[i][j];
    MakePermutation(i, m);
    MakePermutation(m+1, j);
END PROC
```

Figure 3.7: Optimal Benefit Sort Orders for a Path

*odd-ben($S_T$)* and *ben($S_e$)* $\geq$ *even-ben($S_T$)*. Since the total benefit of the optimal solution *ben($S_T$)* = *odd-ben($S_T$)* + *even-ben($S_T$)*, we have *ben($S_o$)* + *ben($S_e$)* $\geq$ *ben($S_T$)*. Hence at least one of *ben($S_o$)* or *ben($S_e$)* is $\geq 1/2$ *ben($S_T$)*. There may be vertices not included in the chosen solution, e.g., the even level split in Figure 3.8 does not include the root and leaf nodes. For these left over vertices arbitrary permutations can be chosen. Kenkre and Vishwanathan [28] have subsequently improved upon our result and have given a $\frac{log\ log\ n}{1\ +\ log\ log\ n}$ factor approximation.

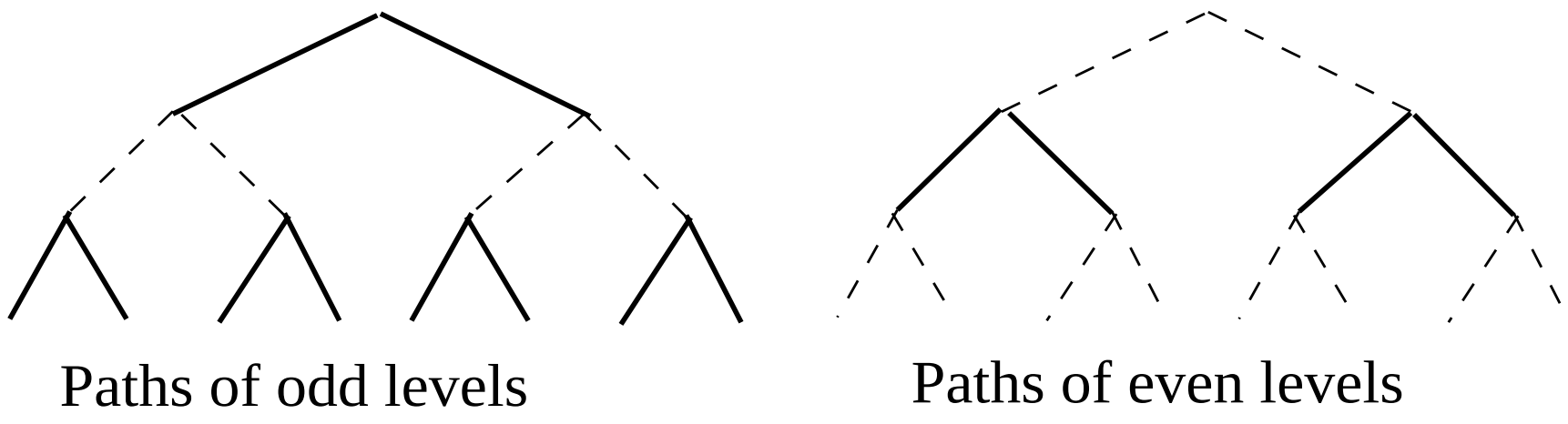

Figure 3.8: A 1/2 Benefit Approximation for Binary Trees

## 3.3 Optimization with Favorable Orders

The benefit model we presented in the previous section, does not take into account factors such as the physical sort order of a relation, available indices and size of base relations and intermediate results. Moreover, we assumed that the join order is fixed. In this section, we present a two phase approach to address the more general problem. In phase-1, which occurs during plan generation, we use the information about available indices and properties of physical operators to efficiently compute a small set of promising sort orders to try. We formalize this idea through the notion of *favorable orders*. Phase-2, is a plan refinement step and occurs after the optimizer makes its choice of the best plan. In phase-2, the sort orders chosen by the optimizer are refined further to reap extra benefit from the attributes common to multiple joins. Phase-2 uses the 1/2 benefit approximation algorithm of Section 3.2.3

### 3.3.1 Favorable Orders

Given an expression $e$, we expect some sort orders (on the result of $e$) to be producible at much lesser cost than other sort orders. Available indices, indexed materialized views, specific rewriting of the expression and choice of physical operators determine what sort orders are easy to produce. To account for such orders, we introduce the notion of *favorable* orders. In the discussion that follows, we use the following notations:

- $cbp(e, o)$ : Cost of the best plan for expression $e$ with $o$ being the required output sort order
- $o_R$ : The clustering order of relation $R$
- $idx(R)$ : Set of all indices over $R$
- $o(I)$ : Order (key) of the index $I$
- $\langle s \rangle$ : An arbitrarily chosen permutation of set $s$
- $P(s)$ : Set of all permutations of set $s$
- $schema(e)$ : The set of attributes in the output of $e$

We first define the ***benefit*** of a sort order $o$ *w.r.t.* an expression $e$ as follows:

$benefit(o, e) = cbp(e, \epsilon) + coe(e, \epsilon, o) - cbp(e, o)$

Intuitively, a positive benefit implies the sort order can be obtained with lesser cost than a full sort of unordered result. For instance, the clustering order of a relation $r$ will have a positive *benefit* for the expression $\sigma_p(r)$. Similarly, query covering secondary indices and indexed materialized views can yield orders with positive benefit. We call the set of all orders, on $schema(e)$, having a positive benefit *w.r.t.* $e$ as the *favorable order set* of $e$ and denoted it as *ford(e)*.

*ford(e)*= { *o*: *benefit(o, e)*> 0 }

### Minimal Favorable Orders

The number of favorable orders for an expression can be very large. For instance, every sort order having the clustering order as its prefix is a favorable order. We call a sort order $o \in$ *ford(e)* as a *minimal favorable order* if the following two conditions hold.

1. $\nexists\ o' \in$ *ford(e)* such that $o' \leq o$ and $cbp(e, o') + coe(e, o', o) = cbp(e, o)$. Intuitively, sort order $o$ is minimal only if there does not exists a sort order $o'$ such that the cost of obtaining order $o$ equals the cost of obtaining sort order $o'$ followed by an explicit sort to obtain order $o$.

2. $\nexists\ o'' \in$ *ford(e)* such that $o \leq o''$ and $cbp(e, o'') = cbp(e, o)$. Intuitively, sort order $o$ is minimal only if there does not exists a sort order $o''$ subsuming order $o$ and available at the same cost as $o$.

We call the set of all minimal favorable orders of an expression $e$ as the *minimal favorable order set* of e and denote it by *ford-min(e)*. Conditions 1 and 2 above, ensure that when a relation has an index that provides sort order $o$ efficiently, orders that are prefixes of $o$ and orders that have $o$ as their prefix are not minimal favorable orders.
We define favorable orders of an expression *w.r.t.* a set of attributes $s$ as: *ford(e, s)*= $\{o \wedge s$: $o \in$ *ford(e)*$\}$. Intuitively, *ford(e, s)* is the set of orders on $s$ or a subset of $s$ that can be obtained efficiently. Similarly, the *ford-min* of an expression *w.r.t.* a set of attributes $s$ is defined as: *ford-min(e, s)*= $\{o \wedge s : o \in$ *ford-min(e)*$\}$

### Heuristics for Favorable Orders

Note that the definition of favorable orders uses the cost of the best plan for the expression. However, we need to compute the favorable orders of an expression **before** the expression is optimized and without requiring to expand the physical plan space. Further, the size of the exact *ford-min* of an expression can be prohibitively large in the worst case. In this section, we describe a method of computing approximate *ford-min*, denoted as *afm*, for *SPJG* expressions. We compute the *afm* of an expression bottom-up. For any expression $e$, *afm(e)* is computable after the *afm* is computed for all of $e$'s inputs.

1. $e = R$, where $R$ is a base relation or materialized view. We include the clustering order of $R$ and all secondary index orders such that the index covers the query.
   $afm(R) = \{o : o = o_R$ or $o = o(I), I \in idx(R)$ and $I$ covers the query$\}$

2. $e = \sigma_p(e_1)$, where $e_1$ is an arbitrary expression.
   $afm(e) = \{o : o \in afm(e_1)\ \}$

3. $e = \Pi_L(e_1)$, where $e_1$ is any expression. We include longest prefixes of input favorable orders such that the prefix has only the projected attributes.
   $afm(e) = \{o : \exists o' \in afm(e_1)$ and $o = o' \wedge L\}$

4. $e = e_1 \bowtie e_2$ with join attribute set $S = \{a_1, a_2, \ldots, a_n\}$. Noting that nested loops joins propagate the sort order of one of the inputs (outer) and merge join propagates the sort order chosen for join attributes, we compute the *afm* as follows: first, we include all sort orders in the input *afms*, next, we consider the longest prefix of each input favorable order having attributes only from the join attribute set and extend it to include an arbitrary permutation of the remaining join attributes.

Let $\mathcal{T} = afm(e_1) \cup afm(e_2)$

Then, $afm(e_1 \bowtie e_2) = \mathcal{T} \cup \{o : o' \in \mathcal{T} \cup \{\epsilon\}$ *and* $o = o' \wedge S + \langle S - attrs(o' \wedge S)\rangle\}$

Note that, for the join attributes not involved in an input favorable order prefix (*i.e.,* $S - attrs(o' \wedge S)$), we take an arbitrary permutation. An exact *ford-min* would require us to include all permutations of such attributes. In the post-optimization phase, we refine the choice made here using the benefit model and algorithm of Section 3.2.3.

5. $e =_L \mathcal{G}_F(e_1)$

   $afm(e) = \{o : o' \in afm(e_1) \cup \{\epsilon\}$ *and* $o = o' \wedge L + \langle L - attrs(o' \wedge L)\rangle\}$

   Intuitively, for each input favorable order we identify the longest prefix with attributes from the projected group-by columns and extend the prefix with an arbitrary permutation of the remaining attributes.

### 3.3.2 Optimizer Extensions

We make use of the approximate favorable orders during plan generation (phase-1) to choose a small set of promising *interesting orders* for sort-based operators. We describe our approach taking merge join as an example but the approach is applicable to other sort based operators. In phase-2, which is a post-optimization phase, we further refine the chosen sort orders.

#### Plan Generation (Phase-1)

Consider an optimization goal of expression $e = e_l \bowtie e_r$ and required output sort order $o$. When we consider merge-join as a candidate algorithm, we need to generate sub-goals for $e_l$ and $e_r$ with the required output sort order being some permutation of the join attributes.

Let $S$ be the set of attributes involved in the join predicate. We consider only conjunctive and equality predicates. We compute the set $\mathcal{I}(e, o)$ of interesting orders as follows.

**Steps to compute $\mathcal{I}(e, o)$:**

1. Collect the favorable orders of inputs plus the required output order
   $\mathcal{T}(e,o) = afm(e_l, S) \cup afm(e_r, S) \ \cup \ o \wedge S$, where $afm(e,S) = \{o' \wedge S : o' \in afm(e)\}$

2. Remove redundant orders
   If $o_1, o_2 \in \mathcal{T}(e,o)$ and $o_1 \leq o_2$, remove $o_1$ from $\mathcal{T}(e,o)$

3. Compute the set $\mathcal{I}(e,o)$ by extending each order in $\mathcal{T}(e,o)$ to the length of $|S|$; the order of extra attributes can be arbitrarily chosen
   $\mathcal{I}(e,o) = \{o : o' \in \mathcal{T}(e,o) \ and \ o = o' \ + \ \langle S - attrs(o')\rangle\}$

We then generate optimization sub-goals for $e_l$ and $e_r$ with each order $o' \in \mathcal{I}(e,o)$ as the required output order and retain the cheapest combination.

**An Example:** Consider Example 3.1 of Section 3.1. For brevity, we refer to the two catalog tables as *ct1* and *ct2*, the rating table as *rt*, and the columns with their starting letters. The *afms* computed as described in Section 3.3.1 are as follows:
$afm(ct1) = \{(y)\}$, $afm(ct2) = \{(m)\}$, $afm(rt) = \{(m)\}$, $afm(ct1 \bowtie ct2) = \{(y), (m), (y, co, c, m), (m, co, c, y)\}$, $afm((ct1 \bowtie ct2) \bowtie rt) = \{(y), (m), (y, co, c, m), (m, co, c, y), (y, m), (m, y)\}$

For $(ct1 \bowtie ct2) \bowtie rt$ we consider two interesting sort orders $\{(y, m), (m, y)\}$ and for $ct1 \bowtie ct2$ we consider four sort orders $\{(y, co, c, m), (m, co, c, y), (y, m, co, c), (m, y, co, c)\}$. As a result the optimizer will consider the plan shown in Figure 3.2.

**A Note on Optimality:** If the set $\mathcal{I}(e,o)$ is computed using the exact set of minimal favorable orders (*ford-min*), then it must contain an optimal sort order, *i.e.*, a sort order, which produces the optimal merge join plan in terms of overall plan cost.

**Theorem 3.6** *The set $\mathcal{I}(e,o)$ computed with exact ford-min contains an optimal sort order $o_p$ for the optimization goal $e = (e_l \bowtie e_r)$ with $(o)$ as the required output sort order, under Assumption A.*

*Assumption A :* If $o_1$, $o_2$ are two sort orders on the same set of attributes (*i.e.,* attrs$(o_1)$ = attrs$(o_2)$), then the CPU cost of sorting the result of an expression $e$ to obtain $o_1$ will be same as that for $o_2$, *i.e.,* cpu-cost$(e, o_1)$ = cpu-cost$(e, o_2)$.

The theorem essentially states the following: to identify an optimal sort order, it is sufficient to consider only the minimal favorable orders and not the full set of favorable orders. Appendix A gives the proof of Theorem 3.6.

### Plan Refinement (Phase-2)

During the plan refinement phase, for each merge-join node in the plan tree, we identify the set of *free attributes*, the attributes which were not part of any of the input favorable orders. Note that for these attributes we had chosen an arbitrary permutation while computing the *afm* (Section 3.3.1). We then make use of the 1/2 benefit approximation algorithm for trees (described in Section 3.2.3) and rework the permutations chosen for the *free attributes.*

Formally, let $p_i$ be the permutation chosen for the join node $v_i$. Let $q_i$ be the order such that $q_i \in$ *afm*($v_i$.*left-input*) $\cup$ *afm*($v_i$.*right-input*) and $|p_i \wedge q_i|$ is the maximum. Intuitively, $q_i$ is the input favorable order sharing the longest common prefix with $p_i$. Let $f_i =$*attrs*$(p_i - (p_i \wedge q_i))$; $f_i$ is the set of *free attributes* for $v_i$.

We now construct a binary tree, where each node $n_i$ corresponding to join-node $v_i$ is associated with the attribute set $f_i$. The attribute permutations for the nodes are chosen using the 1/2 benefit approximation algorithm; the chosen sort order for free attributes is then appended to the sort order chosen during plan generation (*i.e.,* $p_i \wedge q_i$) to get a complete order.

The reworking of the sort orders will be useful only if the adjacent nodes share the same prefix, i.e., $p_i \wedge q_i$ was the same for adjacent nodes. This condition however certainly holds when the inputs for joins have no favorable orders.

Figure 3.9 illustrates the post-optimization phase. Assume all relations involved $(R_1, \ldots, R_4)$ are clustered on attribute $a$ and no other favorable orders exist. *i.e., afm*$(R_i) = \{(a)\}$, for $i = 1$ to 4. The orders chosen by the plan generation phase are shown besides the join nodes with *free attributes* being in *italics.* The reworked orders after the post-optimization phase are shown underlined.

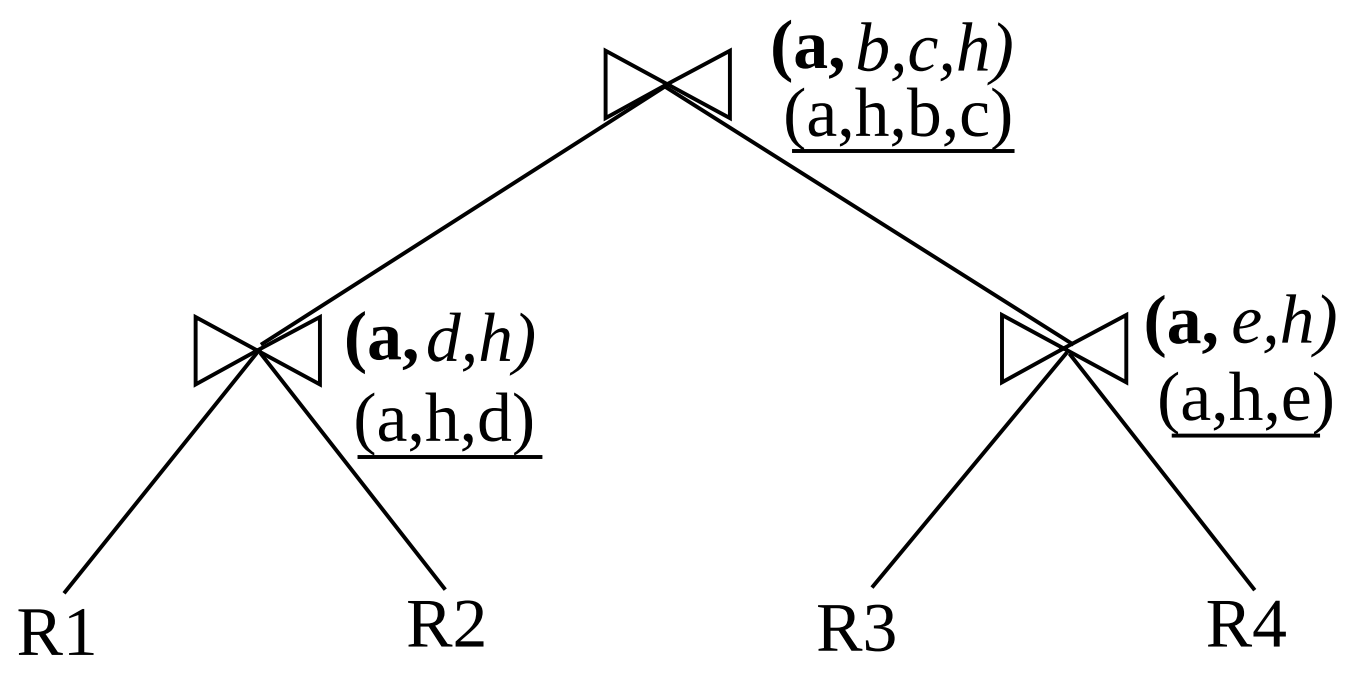

Figure 3.9: Post-Optimization Phase

## 3.4 Application to Nested Queries

The problem of choosing interesting sort orders on sub-query parameters, discussed in Section 2.3, is similar to the problem of choosing sort orders on inputs. We now briefly describe how the techniques presented in this chapter can be used to compute interesting sort orders on sub-query parameters.

Let $e_2$ be an expression nested below expression $e_1$ and $S$ be be the set of parameters bound by $e_1$ that are used inside $e_2$. First, we compute the approximate favorable orders of appropriate sub-expressions in $e_2$ and deduce the corresponding sort order on parameters. As an example, consider a sub-expression $\sigma_p(e_3)$ in $e_2$. We compute $afm(e_3)$ and then deduce the corresponding sort orders on parameters by mapping each attribute to the parameters it is equated to in predicate $p$. Next, we compute the favorable orders for the outer expression *w.r.t.* to the set of parameters it binds, $afm(e_1, S)$. The final set of interesting parameter sort orders is formed by taking the union of interesting parameter sort orders obtained from all the inner sub-expressions and the favorable orders of the outer sub-expression.

Intuitively, our approach considers plans that sort the outer tuples to match a sort order favorable to the inner expression(s), and also plans that sort the inner relations to match a sort order favorable to the outer expression.

## 3.5 Experimental Results

We performed experiments to evaluate the benefits due to the proposed ideas. For comparison, we use PostgreSQL (version 8.1.3) and two widely used commercial database systems (we call them SYS1 and SYS2). All tests were run on an Intel P4 (HT) PC with 512 MB of RAM. We used TPC-H 1GB dataset and additional tables as specified in the individual test cases. For each table, a clustering index was built on the primary key. Additional secondary indices built are specified along with the test cases. All relevant statistics were built and the optimization level for one of the systems, which supports multiple levels of optimization, was set to the highest.

### 3.5.1 Modified Replacement Selection

The first set of experiments evaluate the benefits of exploiting partial sort orders. External sort in PostgreSQL employs the standard replacement selection (SRS) algorithm [32] suitably adapted for variable length records. We modified this implementation to exploit partial sort orders available on the input (as described in Section 3.1), and we call it Modified Replacement Selection (MRS). We now present experiments comparing the performance of MRS with SRS.

#### Experiment A1

The first experiment consists of a simple ORDER BY of the TPC-H *lineitem* table on two columns *(l_suppkey, l_partkey)*.

***Example 3.2*** *An Order-By Query on Two Columns (Query 1)*

*SELECT l_suppkey, l_partkey*
*FROM lineitem*
*ORDER BY l_suppkey, l_partkey;*

A secondary index on *l_suppkey* was available that covered the query (included the *l_partkey* column)[3]. On all three systems, the order by on *(l_suppkey, l_partkey)* took almost the same time as an order by on *(l_partkey, l_suppkey)* showing that the sort operator of these systems did not exploit partial sort orders effectively. We then compared the running times with our implementation that exploited partial sort order *(l_suppkey)* and the results are shown in Figure 3.10.

[3]On systems not supporting indexes with included columns, we used a table with only the desired two columns, clustered on *l_suppkey*

For SYS1 and SYS2, as we did not have access to their source code, we simulated the partial sorting using a correlated rank query. The subquery sorted the index entries matching a given *l_suppkey* on *l_partkey* and the subquery was invoked with all *suppkey* values so as to obtain the desired sort order of *(l_suppkey, l_partkey)*.

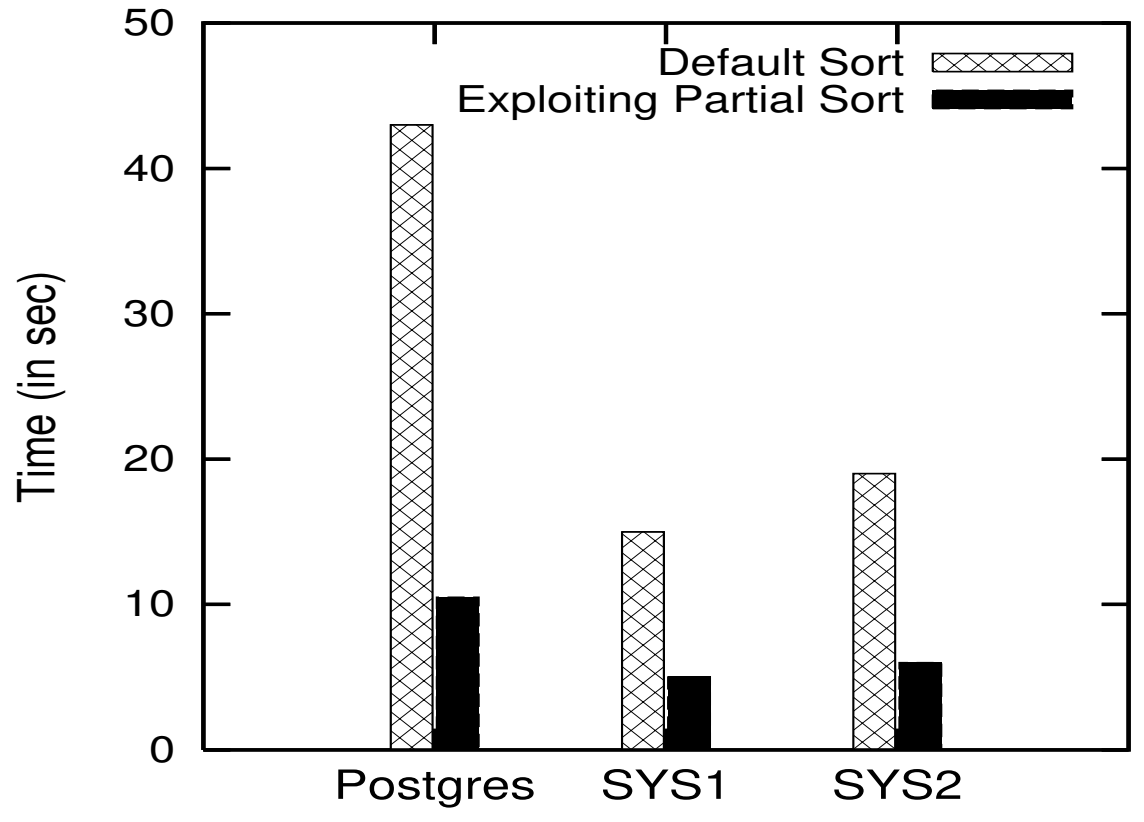

Figure 3.10: Performance Results for Experiment A1

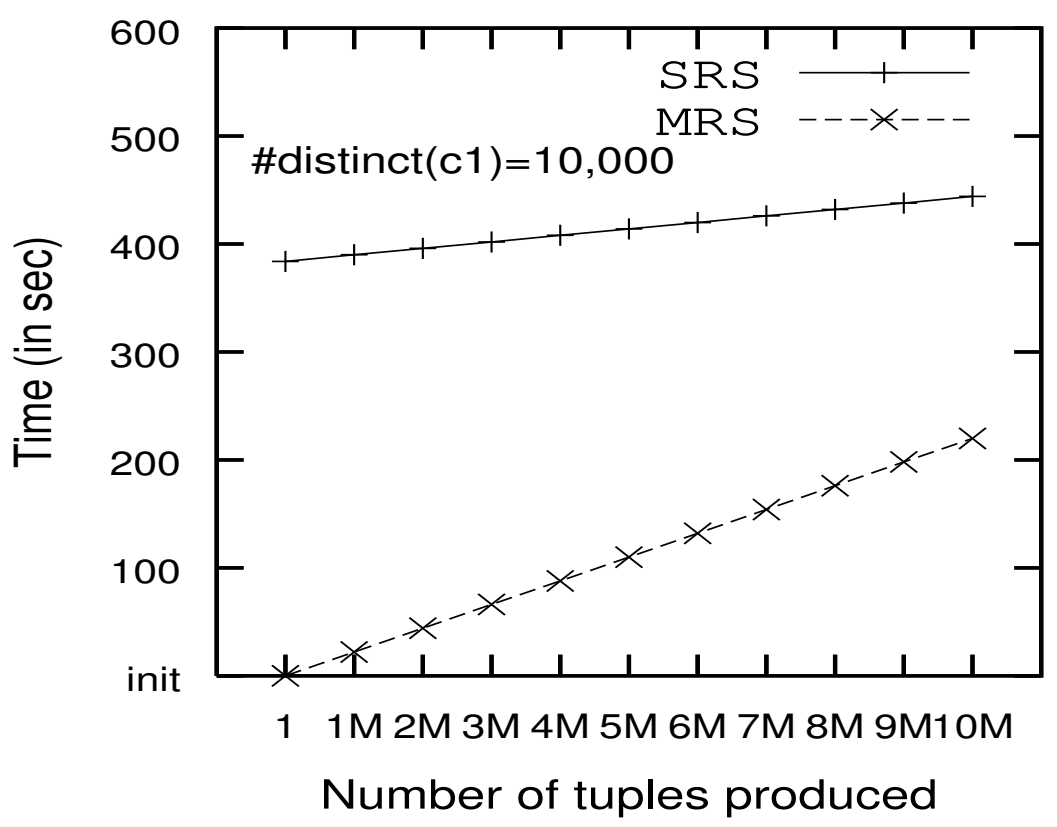

Figure 3.11: Rate of Output

By avoiding run generation I/O and making reduced comparisons, MRS performs 3-4 times better than SRS.

**Experiment A2**

The second experiment shows how MRS is superior in terms of its ability to produce records early and uniformly. Table $R_3$ having 3 columns $(c1, c2, c3)$ was populated with 10 million records and was clustered on $(c1)$. The query asked an order by on $(c1, c2)$. Figure 3.11 shows the plot of number of tuples produced vs. time with cardinality of $c1 = 10,000$.

MRS starts producing the tuples without any delay after the operator initialization where as SRS produces its first output tuple only after seeing all input tuples. By producing tuples early, MRS speeds up the pipeline significantly. Such early output behavior is highly desirable for Top-K queries.

**Experiment A3**

The third experiment shows the effect of *partial sort segment size* on sorting. 8 tables $R_0, \ldots, R_7$, with identical schema of 3 columns $(c1, c2, c3)$ were each populated with 10 million records and average record size of 200 bytes. Each table was clustered on $(c1)$.

Table $R_i$ had $10^i$ tuples for each distinct value of $c1$ (*i.e.,* uniform distribution over $10^{7-i}$ distinct values of $c1$), resulting in a partial sort segment size of $200 \times 10^i$ bytes. Thus $R_0$ had a sort segment of size 1 tuple or 200 bytes, and $R_7$ had a sort segment of size 10 million tuples or 2GB. The query had an order by on $(c1, c2)$. The running times with default and modified replacement selection on PostgreSQL are shown in Figure 3.12.

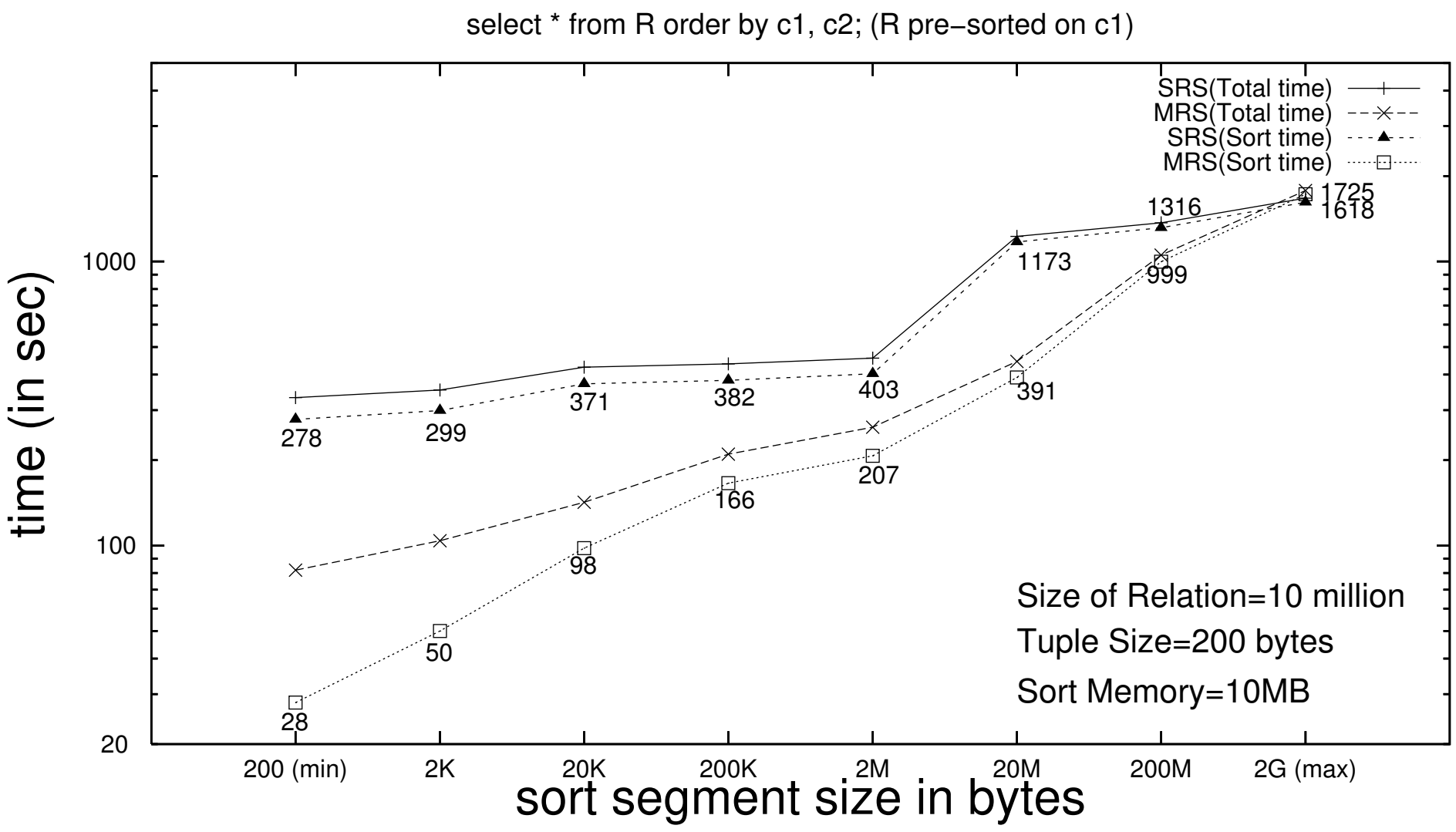

Figure 3.12: Effect of Partial Sort Segment Size

When the partial sort segment size is small enough to fit in memory (up to 10MB or 50K records), SRS produces a single sorted run on disk and does not involve merging of runs. The modified replacement selection (MRS) gets the benefit of avoiding I/O and reduced number of comparisons. When the partial sort segment size becomes too large to fit in memory, we see a sudden rise in the time taken by SRS. This is because replacement selection will have to deal with merging several runs. MRS however deals with merging smaller number of runs initially as each partial sort segment is sorted separately. As the partial sort segment size increases, the running time of MRS rises and becomes same as that of SRS at the extreme point where all records have the same value for $c1$.

**Experiment A4**

To see the influence of MRS on a query with joins and aggregates, we considered the query shown in Example 3.3. The query finds the number of lineitems and available quantity for each supplier, part pair. The supplier and part key columns were common to the join, group-by and the order-by clauses. Two indices, *lineitem(l_suppkey)* and

*partsupp(ps_suppkey)*, were present and covered the query. The indices were thus useful to obtain part of the desired sort order *(suppkey, partkey)*.

***Example 3.3*** *Number of lineitems for each (supplier, part) pair (Query 2)*

```
SELECT   ps_suppkey, ps_partkey, ps_availqty, count(l_partkey)
FROM     partsupp, lineitem
WHERE    ps_suppkey=l_suppkey AND ps_partkey=l_partkey
GROUP BY ps_suppkey, ps_partkey, ps_availqty
ORDER BY ps_suppkey, ps_partkey;
```

On PostgreSQL the query took 63 seconds to execute with SRS, and 25 seconds with MRS. The query plan used in both cases was the same: a merge join of the two relations on *(suppkey, partkey)* followed by aggregation.

### 3.5.2 Choice of Interesting Orders

We extended our Volcano-style cost based optimizer, which we call PYRO, to consider partial sort orders and to use the proposed method for choosing sort orders for merge joins and aggregation. We compare the plans produced by the extended implementation, which we call PYRO-O, with those of PostgreSQL, SYS1 and SYS2.

#### Experiment B1

For this experiment we used the query shown in Example 3.4 below, which lists parts for which the outstanding order quantity is more than the stock available at the supplier.

***Example 3.4*** *Parts Running Out of Stock (Query 3)*

```
SELECT   ps_suppkey, ps_partkey, ps_availqty, sum(l_quantity)
FROM     partsupp, lineitem
WHERE    ps_suppkey=l_suppkey AND ps_partkey=l_partkey AND l_linestatus='O'
GROUP BY ps_availqty, ps_partkey, ps_suppkey
HAVING sum(l_quantity) > ps_availqty ORDER BY ps_partkey;
```

Table *partsupp* had clustering index on its primary key *(ps_partkey, ps_suppkey)*. Two secondary indices, one on *ps_suppkey* and the other on *l_suppkey* were also built

on the *partsupp* and *lineitem* tables respectively. The two secondary indices covered all attributes needed for the query.

The experiment shows the need for cost-based choice of interesting orders. The choice of interesting orders for the join and aggregate are not obvious in this case for the following reasons:

1. The order-by clause favors the choice of a sort order where *partkey* appears first.

2. The clustering index on *partsupp* favors the choice of *(partkey, suppkey)*.

3. The secondary indices favor the choice of *(suppkey, partkey)* that can be obtained by using a low cost partial sort. Note that this option can be much cheaper due to the size of the *lineitem* relation.

Therefore, the optimizer must make a cost-based decision on the sort order to use. Figures 3.13 and 3.14 show the plans chosen by PostgreSQL, PYRO-O, SYS1 and SYS2.

All plans except the hash-join plan of SYS1 and the plan produced by PYRO-O use an expensive full sort of 6 million lineitem index entries on *(l_partkey, l_suppkey)*. Further, PostgreSQL uses a hash aggregate where a sort-based aggregate would have been much cheaper as the required sort order for the group-by was available from the output of merge-join. Note that the sort order *(ps_partkey, ps_suppkey, ps_availability)*, required by the group-by, can be inferred from the sort order *(ps_partkey, ps_suppkey)*, available on the result of merge-join, due to the presence of the functional dependency $\{ps_partkey, ps_suppkey\} \rightarrow \{ps_availqty\}$. On SYS1, it was possible to force the use of a merge-join instead of hash-join and the plan chosen is shown in Figure 3.14(b).

We compared the actual running time of PYRO-O's plan with those of PostgreSQL and SYS1 by forcing our plan on the respective systems. Figures 3.15 and 3.16 show the execution times. It was not possible for us to force our plan on SYS2 and make a fair comparison and hence we omit the same. The only surprising result was the default plan chosen by SYS1 performed slightly poorer than the forced merge-join plan on SYS1. In all cases, the forced PYRO-O plan performed significantly better than the other plans. The main reason for the improvement was the use of a partial sort of *lineitem* index entries as against a full sort. The final sort on *partkey* was not very expensive as only a few tuples needed to be sorted.

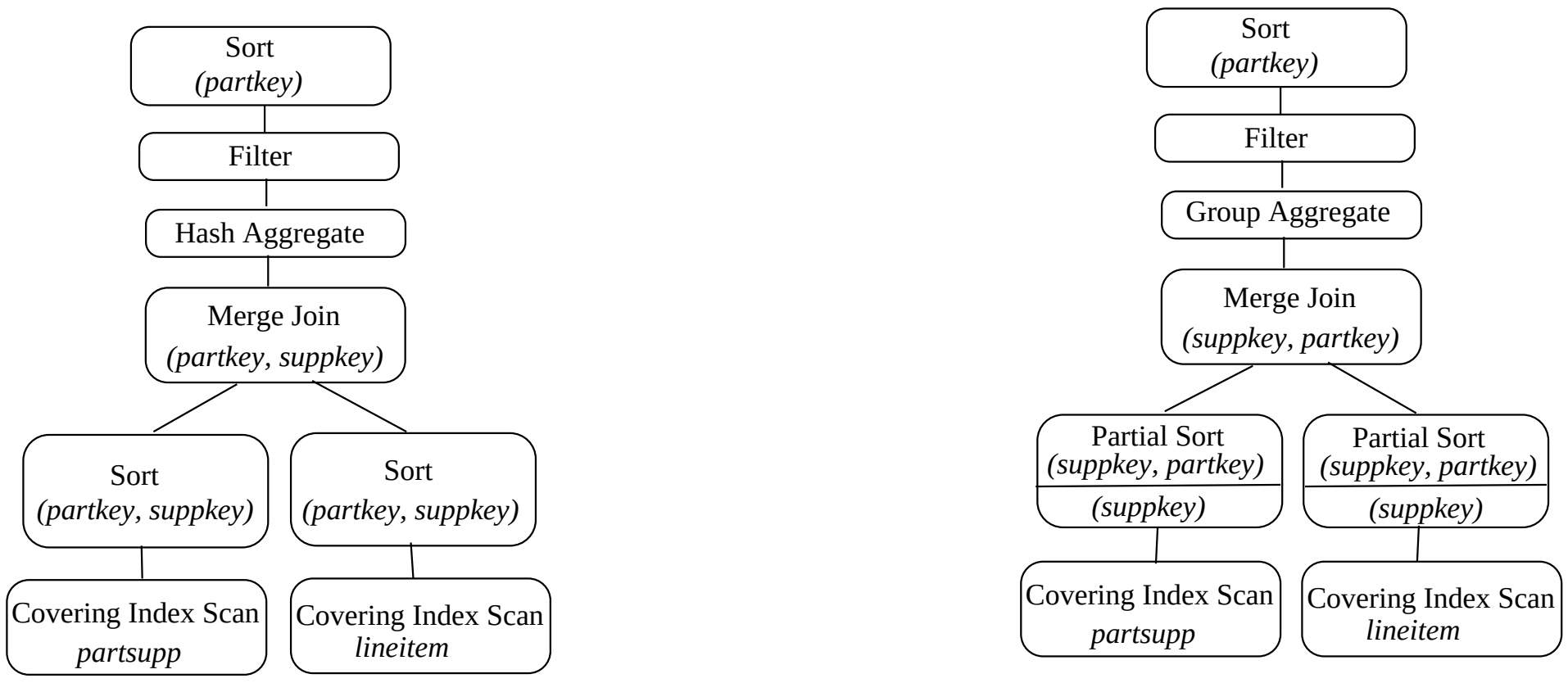

(a) Default Plan (Postgres)

(a) Plan Generated by PYRO−O

Figure 3.13: Plans for Query 3 (PostgreSQL and PYRO-O)

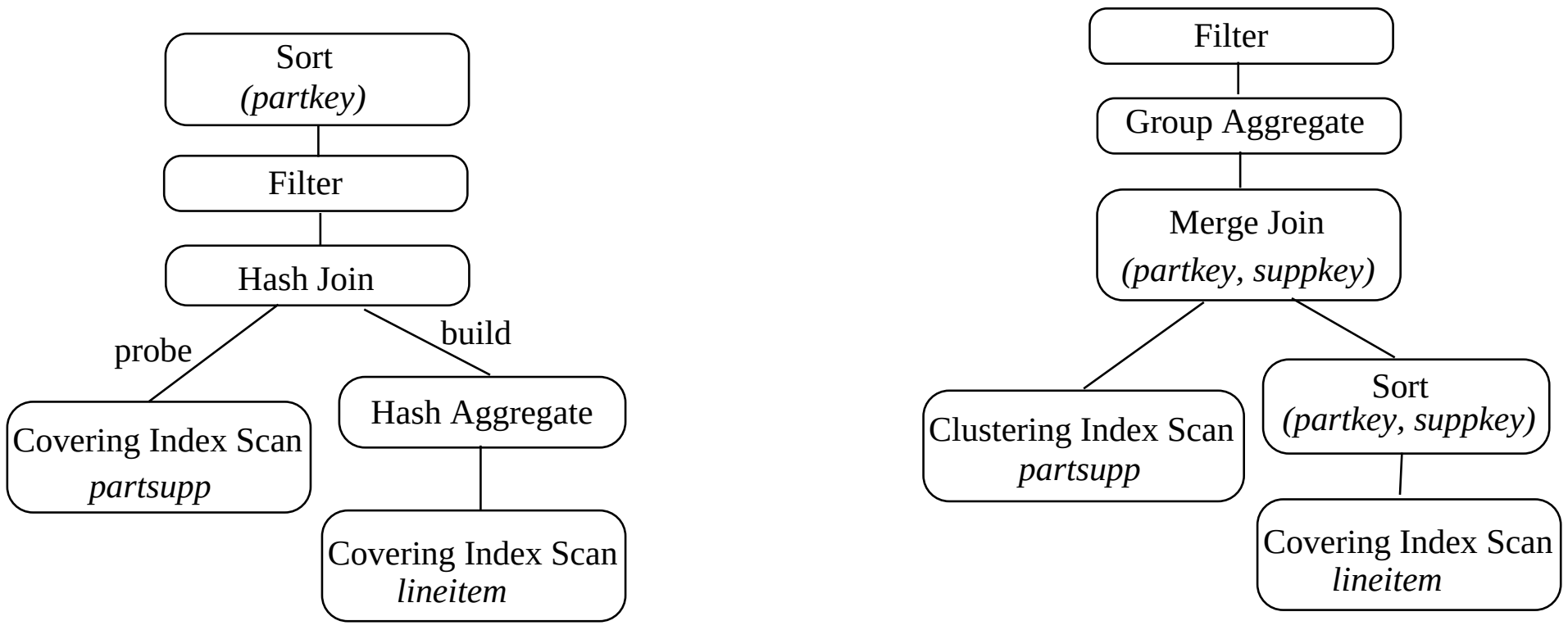

(a) Default plan on SYS1

(b) Forced merge−join plan on SYS1 and default plan on SYS2

Figure 3.14: Plans for Query 3 (SYS1 and SYS2)

For Query 3 (Example 3.4) the plan generation phase (phase-1) was sufficient to select the sort orders and phase-2 does not make any changes. We shall now see a case for which phase-1 cannot make a good choice and the sort orders get refined by phase-2.

### Experiment B2

This experiment uses the query shown in Example 3.5, which has two full outer joins with two common attributes between the joins. We performed this experiment to see whether the systems we compare with are designed to exploit attributes common to multiple sort-based operators.

The tables R1, R2 and R3 were identical and each populated with 100,000 records. No indexes were built. As shown in Figure 3.17(a), both SYS1 and PostgreSQL chose sort orders that do not share any common prefix. The plan chosen by PYRO-O is shown

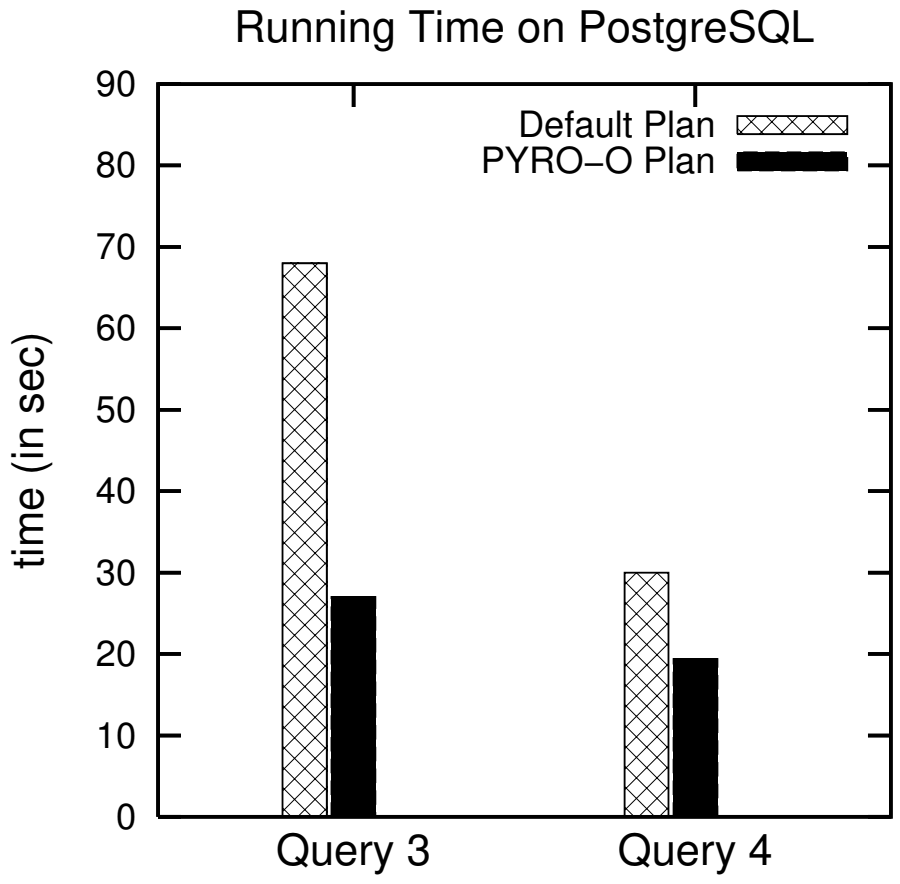

Figure 3.15: Performance on PostgreSQL

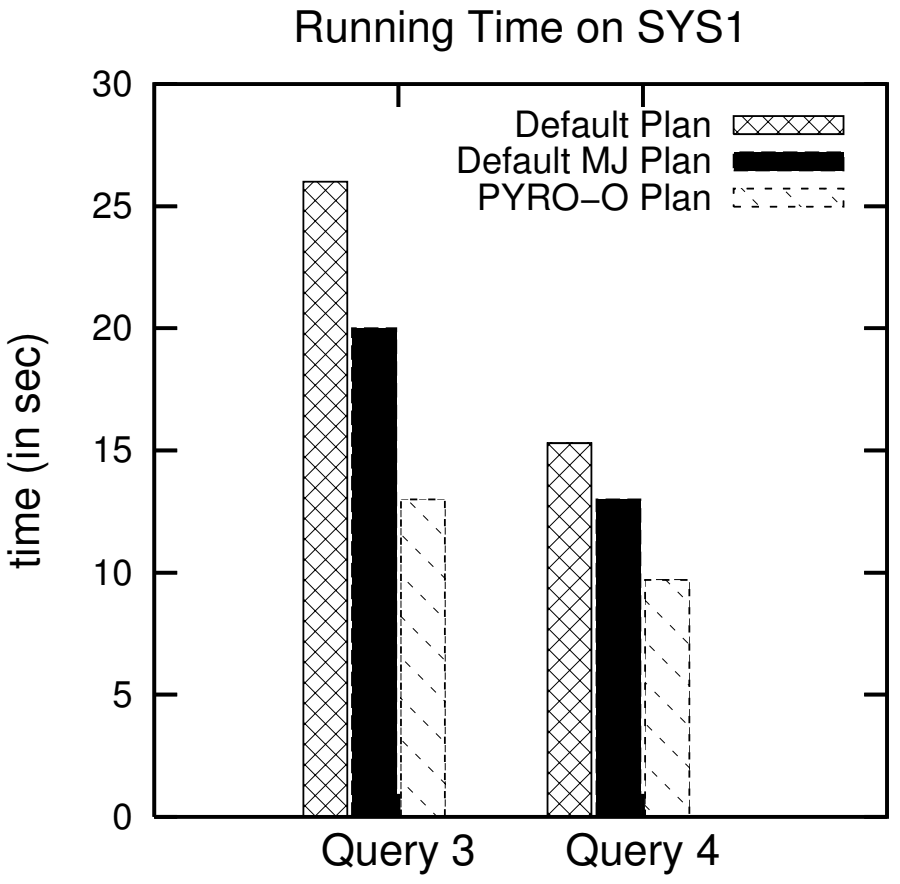

Figure 3.16: Performance on SYS1

***Example 3.5*** *A query with attributes common to multiple joins (Query 4)*

```
SELECT * FROM R1
        FULL OUTER JOIN R2
                ON (R1.c5=R2.c5 AND R1.c4=R2.c4 AND R1.c3=R2.c3)
        FULL OUTER JOIN R3
                ON (R3.c1=R1.c1 AND R3.c4=R1.c4 AND R3.c5=R1.c5);
```

in Figure 3.17(b). In the plan chosen by PYRO-O, the two joins share a common prefix of *(c4, c5)*, and thus the sorting effort is expected to be significantly less. SYS2, not having an implementation of full outer join, chose a union of two left outer joins. The two left outer joins used to get a full outer join used different sort orders making the union expensive, illustrating a need for coordinated choice of sort orders. The execution timings for Query 4 on PostgreSQL and SYS1 are shown in Figures 3.15 and 3.16 respectively.

### Experiment B3

In this experiment we compare our approach of choosing orders, PYRO-O, with the exhaustive approach, and a heuristic used by PostgreSQL. PostgreSQL uses the following heuristic: for each of the $n$ attributes involved in the join condition, a sort order beginning with that attribute is chosen; in each order, the remaining $n-1$ attributes are ordered arbitrarily. We implemented PostgreSQL's heuristic in PYRO along with the extensions to exploit partial sort orders and we call it PYRO-P. The exhaustive approach, called PYRO-E, enumerates all $n!$ permutations and considers partial sort orders. In addition, we also compare with baseline PYRO, which chooses an arbitrary sort order, and a variation of

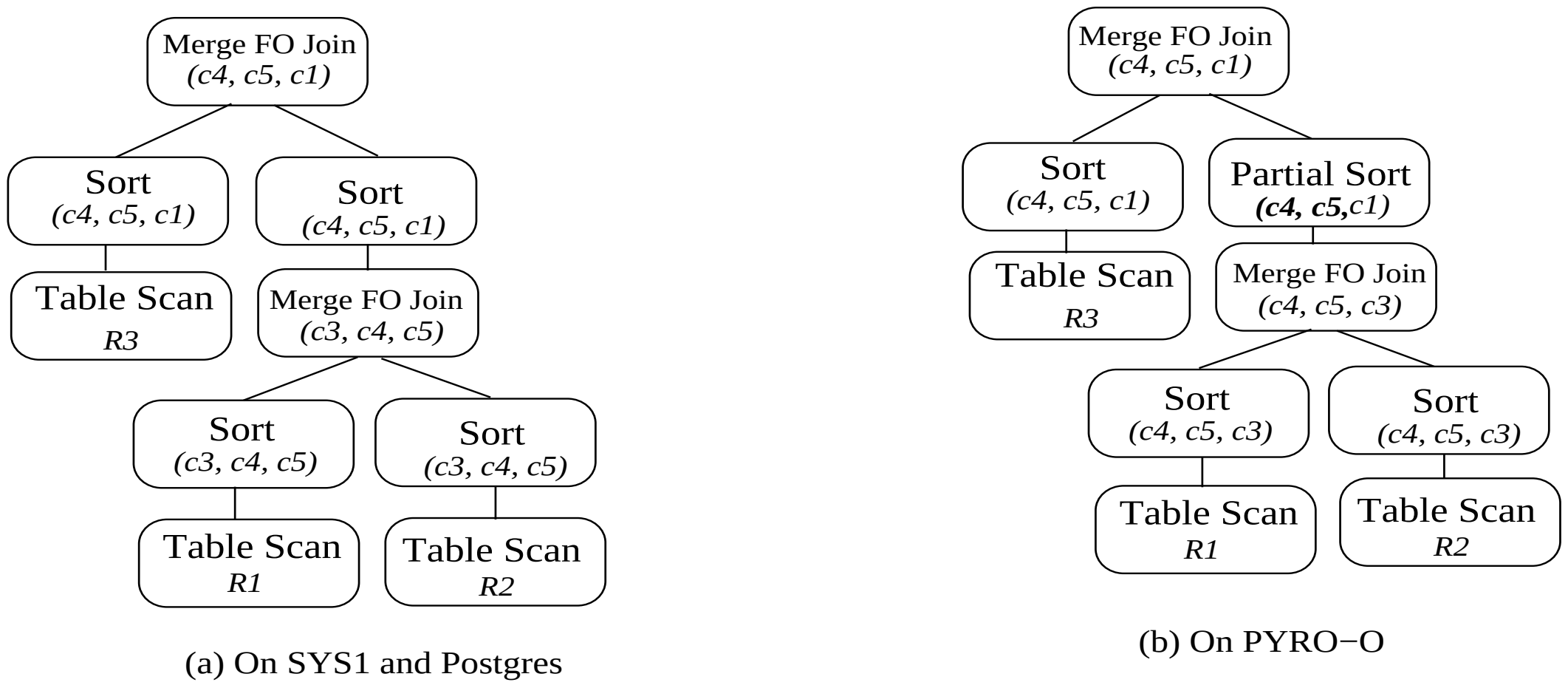

Figure 3.17: Plans for Query 4

PYRO-O, called PYRO-O$^{-}$ that considers only exact favorable orders (no partial sort). Figure 3.18 shows the estimated plan costs. Note the logscale for y-axis. The plan costs are normalized taking the plan cost with exhaustive approach to be 100. In the figure, Q3 and Q4 stand for Query 3 (Example 3.4) and Query 4 (Example 3.5) of Experiments B1 and B2. Q5 and Q6 stand for Query 5 and Query 6, and are shown below as Examples 3.6 and 3.7 respectively. For Q3 and Q4, as very few attributes were involved in the join condition, PostgreSQL's heuristic along with extensions to exploit partial sort orders, produced plans which were close to optimal. However, for more complex queries the heuristic does not perform well since it makes an arbitrary choice for secondary orders.

---

***Example 3.6*** *Total value executed for a given order (Query 5)*

---

*SELECT T1.UserId, T1.BasketId, T1.ParentOrderId, T1.WaveId, T1.ChildOrderId,*
*(T1.Quantity * T1.Price) as OrderValue,*
*SUM(T2.Quantity * T2.Price) as ExecValue*
*FROM TRAN T1, TRAN T2*
*WHERE T1.UserId=T2.UserId AND T1.ParentOrderId=T2.ParentOrderId AND*
*T1.BasketId=T2.BasketId AND T1.WaveId=T2.WaveId AND*
*T1.ChildOrderId=T2.ChildOrderId AND T1.TranType='New' AND*
*T2.TranType='Executed'*
*GROUP BY T1.UserId, T1.BasketId, T1.ParentOrderId, T1.WaveId, T1.ChildOrderId;*

---

**Example 3.7** *Basket Analytics (Query 6)*

*SELECT * FROM BASKET B, ANALYTICS A*
*WHERE B.ProdType = A.ProdType AND B.Symbol = A.Symbol AND*
*B.Exchange = A.Exchange;*

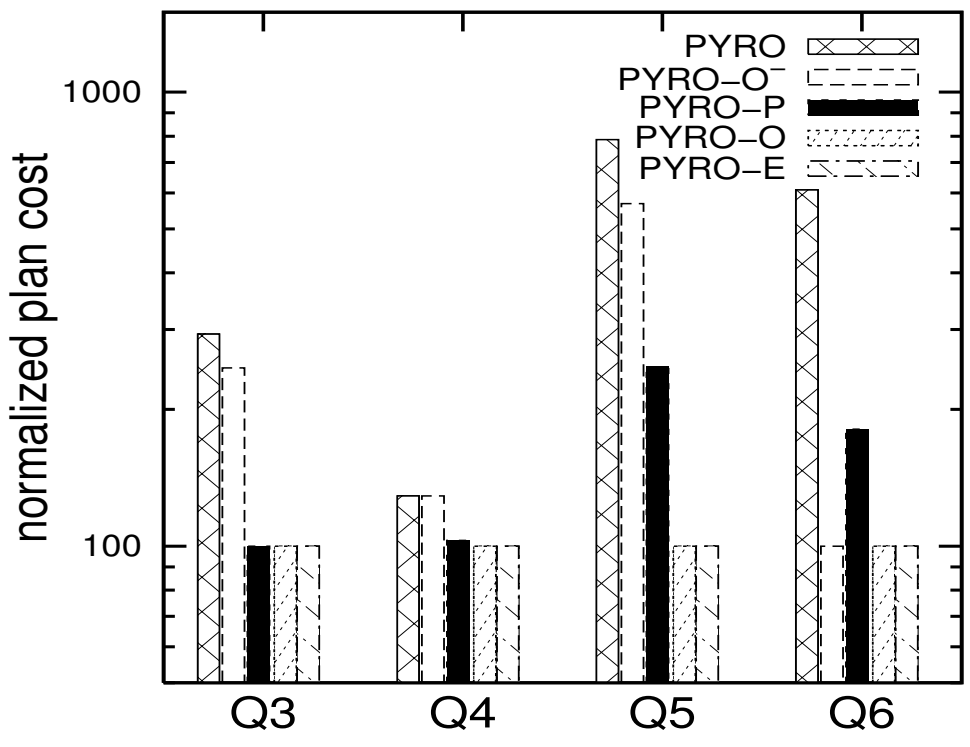

Figure 3.18: Evaluating Different Heuristics

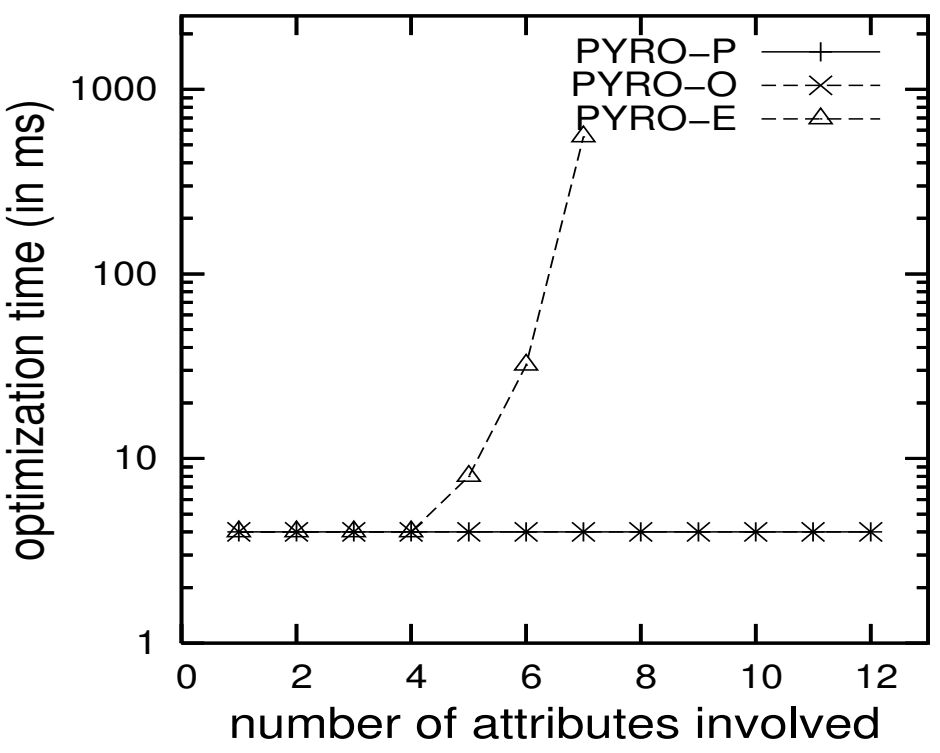

Figure 3.19: Optimizer Scalability

### 3.5.3 Optimization Overheads

The optimization overheads due to the proposed extensions were negligible. During plan generation, the number of sort orders we try at each join or aggregate node is of the order of the number of indices that are useful for answering the query, which in most practical case is expected to be small. Figure 3.19 shows the scalability of the three heuristics. For this experiment a query that joined two relations on varying number of attributes was used. Though PYRO-P and PYRO-O take the same amount of time in this experiment, in most cases, the number of favorable orders is much less than the total number of attributes involved and hence PYRO-O generates fewer interesting orders than PYRO-P.

The plan-refinement algorithm presented in Section 3.2.3 was tested with trees up to 31 nodes (joins) and 10 attributes per node. The time taken was negligible in each case. The execution of plan refinement phase took less than 6 ms even for the tree with 31 nodes.

Both the optimizer extensions and the extension to external-sorting (MRS) were straight forward to implement. The optimizer extensions neatly integrated into our existing Volcano style optimizer.

## 3.6 Related Work

Both System R [48] and Volcano [23] optimizers consider plans that could be locally sub-optimal but provide a sort order of interest to other operators, and thus yield a better plan overall. However, both System R and Volcano assume that operators have one or few *exact* sort orders of interest. This is not true of operators like merge-join, merge-union, grouping and duplicate elimination, which have a factorial number of interesting orders. Heuristics such as the PostgreSQL heuristic, are commonly used by optimizers. Details of the heuristics are publicly available only for PostgreSQL. Further, System R and Volcano optimizers consider only those sort orders as useful that completely meet an order requirement. Plans that partially satisfy a sort order requirement are not handled. In this chapter we addressed these two issues.

The seminal work by Simmen et.al. [51] describes techniques to infer sort orders from functional dependencies and predicates applied, and thereby avoids redundant sort enforcers in the plan. Simmen et.al. [51] briefly mention the problem of non-exact sort order requirements and mentions an approach of propagating an order specification that allows any permutation on the attributes involved. Though such an approach is possible for single input operators like group-by, it cannot be used for operators such as merge-join and merge-union for which the order guaranteed by both inputs must match. Moreover, the paper does not make it clear how the flexible order requirements are combined at other joins and group-by operators. Simmen et.al. [51] also note that the approach of carrying a flexible order specification increases the complexity of the code significantly. Our techniques do not use flexible order specifications and hence can be incorporated into an existing optimizer with minimal changes. Further, our techniques work uniformly across all types of operators that have a flexible order requirement.

Claussen et.al. [7] explore early-sorting as a means to reduce sorting cost in query plans. The key idea is to avoid sorting of large intermediate join results by pushing sorting to base relations, and using order preserving hash joins. There has been significant work on avoiding redundant sorting by inferring sort orders and groupings using functional dependencies [51, 56, 39, 40]. These techniques are complementary to our work.

## 3.7 Summary

In this chapter we addressed the problem of choosing efficient sort orders for sort-based operators such as merge-join and sort-based grouping. We showed that even a simplified version of the problem is *NP-Hard*, and proposed principled heuristics for choosing interesting orders. Our heuristics are guided by the notion of favorable orders. We take into account important issues such as partially matching sort orders and attributes common to multiple operators. We then explained how the solution can be used for choosing efficient parameter sort orders for nested queries. We presented a detailed experimental study on widely used database systems, and the results showed significant performance improvements due to the proposed techniques for several queries.

# Chapter 4

# Rewriting Procedures for Batched Bindings

In this chapter we consider parameter batching as a means of improving performance of iteratively invoked database procedures. Several data retrieval and update tasks need more expressive power than what standard SQL offers. Therefore, many database applications perform queries and updates from within procedural code that encodes business logic. Stored procedures and user-defined functions written using procedural extensions to SQL (*e.g.*, PL/SQL [42], PL/pgSQL [41]) are widely used. Other paradigms that allow mixing of procedural constructs with database access are SQL extensions to procedural languages, *e.g.*, SQLJ [53] and Microsoft's language integrated query (LINQ) [35], and application programming interfaces (API) for database access, *e.g.*, JDBC, ODBC. Such procedures/programs can run either inside a database system, as stored procedures or user-defined functions (UDF), or outside the database system. As queries in SQL can in turn invoke UDFs (as part of WHERE/SELECT clauses), control can alternate between the SQL execution engine and the procedural code.

We have earlier seen an example (Example 1.3 in Chapter 1) of a query that invokes a simple UDF in its WHERE clause. In general, UDFs/procedures can be more complex with arbitrary control-flow and looping. Such a UDF is shown in Example 4.1, where the UDF counts the number of items in a given category and all its sub-categories.

Parameter batching is an important technique to speedup iterative execution of queries and updates [22, 18]. Parameter batching allows the choice of efficient set-oriented plans for queries and updates, thus reducing random IO. For inserts and updates batching

**Example 4.1** *Query with a Complex UDF - Items in a Category and Sub-Categories*

*SELECT * FROM category WHERE count_items(category_id) > f(level);*

```
         INT count_items(INT catid)
         DECLARE
             INT totalcount; INT curcat; INT catitems; INT subcat;
             INT stack[100]; INT top; RECORD catrec;
         BEGIN
s1:          totalcount := 0;
s2:          top := 0;
s3:          stack[top] = catid;
s4:          top := top + 1;
s5:          WHILE top > 0 LOOP
s6:              top := top − 1;
s7:              curcat := stack[top];
s8:              catitems := SELECT count(item_id) FROM item WHERE category_id = curcat;
s9:              totalcount := totalcount + catitems;

                 // Now push all the subcategory ids onto the stack
s10:             FOR catrec IN SELECT category_id FROM category
                              WHERE parent_category=curcat LOOP
s11:                 stack[top] := catrec.category_id;
s12:                 top := top + 1;
                 END LOOP;
             END LOOP;
s13:         RETURN totalcount;
         END;
```

allows efficient integrity checks and index maintenance. When the calls span across a network, batching also helps in reducing network round-trip delays.
In this chapter we present our work on automatic rewrite of iterative programs to fulfill the following needs:

(a) Rewrite UDFs and stored procedures to accept batched bindings

(b) Rewrite programs that repeatedly execute a parameterized query or stored procedure to use batched invocation when possible

The rest of this chapter is organized as follows. Section 4.1 defines the problem, introduces the notion of *batched forms* and *batch-safe* operations. We identify the program transformation goal of pulling expensive operations out of loops as a key for addressing the two needs mentioned above. Section 4.2 describes the background needed for our approach. In Section 4.3 we present the set of program transformation rules, which together achieve the required program transformation, when the program satisfies certain conditions. The program transformation rules rely on the *data dependence graph* of the given program and can work with complex procedures such as the ones shown in Example 1.3 and Example 4.1. We present the results of our experiments in Section 4.6. The experiments are based on real-life examples of performance problems. We discuss related work in Section 4.7 and summarize in Section 4.8.

## 4.1 Rewriting for Batched Bindings

In order to exploit the benefits of batching, we must have an efficient *batched form* of the operation being invoked, and a way of using the *batched form* in place of repeated invocations of the operation. In this section, we formally define *batched forms* of operations and introduce the problem of rewriting loops so as to make use of the *batched forms* of expensive operations invoked within them. We also consider the issues in batching invocations of operations that have side-effects. Finally, we introduce the problem of generating *batched forms* of complex procedures.

### 4.1.1 Batched Forms of Operations

Informally, the batched form of an operation takes a batch (or set) of parameters at once and processes them. Batched forms of operations are typically more efficient than iterative invocation of the corresponding non-batched forms. For example, a database bulk load operation can be thought of as a batched form of the *insert* operation, assuming logging can be ignored. Similarly, a relational *join* can be thought of as a batched form of relational *selection* with a parameterized predicate. Note that we model a batch as a *set* and not a *sequence*. This is due to the fact that most batched forms do not guarantee the order in which the elements in the batch are processed, and this is an important reason for their efficiency. We now define the batched forms formally.

#### Batched Forms of Pure (Side-Effect Free) Functions

Let $f : D \rightarrow R$ be a side-effect free function, where $D$ is the domain and $R$ is the range of $f$. A function $fb : BD \rightarrow BR$ is considered a batch form of $f$ if the following are true.

1. The domain $BD$ is the power set of $D$
2. The range $BR$ is the power set of $D \times R$
3. $\forall b \in BD, fb(b) = \bigcup_{b_i \in b} \{(b_i, f(b_i))\}$

**Example:** Consider the square function defined as $sq(x) = x^2$. The corresponding batched form can be defined as $sqb(sx) = \{(x, x^2) : x \in sx\}$

Intuitively, the batched form of a function takes a set of parameters and returns a set comprising of all the results. To establish the correlation between a parameter and the corresponding result we require the batched form to return the parameter value along with the result.

### Batched Forms of Parameterized Relational Queries

Relational queries are pure functions that return (multi)sets of tuples. Though we can use the above definition of batched forms for queries, it makes the return type of batched queries violate the first normal form (1NF) as queries may have set-valued return type. We desire the first normal form on batched queries so as to be able to make our techniques easily implementable in existing relational query processing systems. Hence we use a slightly modified definition for the batched form of a query.

Let $q(p_1, p_2, \ldots, p_n)$ be a query with $n$ parameters. Let $v_1, v_2, \ldots, v_m$ be the attributes in the result-set that $q$ returns. The batched form $qb$ of $q$ takes a set $p$ of *n-tuples* as its parameter (each *n-tuple* gives a binding for the parameters $p_1, p_2, \ldots, p_n$). The result-set of $qb$ contains the union of $q$'s results for all the parameter tuples in $p$. Each tuple in *qb's* result-set contains $n + m$ attributes $p_1, p_2, \ldots, p_n, v_1, v_2, \ldots, v_m$. Often only a subset of the parameters are sufficient to establish the correlation with the corresponding results. However, for simplicity, we assume the batched form returns all the parameter values along with the results. Formally,

$$qb(p) = \bigcup_{p_t \in p} \begin{cases} \{\{p_t\} \times q(p_t)\} & \text{if } q(p_t) \neq \phi \\ \{\{p_t\} \times \{\text{null-m}\}\} & \text{otherwise} \end{cases}$$

When the result of $q$ is an empty set for any parameter binding, the result-set of $qb$ contains a tuple corresponding to the specific parameter binding but the attributes $v_1, v_2, \ldots, v_n$ will be set to null.

**Example:** Consider the following parameterized select-project query:

$$q(custid) = \Pi_{ordrid}(\sigma_{customer\text{-}id=custid}(\text{ORDERS}))$$

The corresponding batched form can be defined as:
$qb(\text{cs}) = \Pi_{(custid,ordrid)}(\text{cs} ⟕_{custid=customer\text{-}id} \text{ORDERS})$,
where *cs* is the parameter relation having the attribute *custid*.

Batched forms of relational queries have been used in the context of query decorrelation [49, 31, 17, 9]. As shown in the above example, batched forms of simple SPJ queries use a join or an outer join. Batched forms of aggregate queries either use grouping followed by join or an outer-join followed by grouping. The details of deriving correct batched forms of SQL queries can be found in the literature on decorrelation. Example 4.2 shows the batched forms of queries q1 and q2 used inside the UDF of Example 1.3 in Chapter 1.

---

***Example 4.2*** *Batched Forms of Queries in Example 1.3*

---

**q1b(r):** *The batched form of query* **q1**, *with* r *as the parameter batch*

*SELECT r.curcode, c.exchrate FROM r JOIN curexch c ON r.curcode=c.ccode;*

**q2b(r):** *The batched form of query* **q2**, *with* r *as the parameter batch*

```
SELECT r.itemcode, r.amount_usd, count(b.itemcode) AS count_offers
FROM r LEFT OUTER JOIN buyoffers b ON b.itemid = r.itemcode AND
        b.price >= r.amount_usd
GROUP BY r.itemcode, r.amount_usd;
```

---

Most database systems also support batched bindings for basic data manipulation operations like insert, delete and update. The *insert into ... select from ...* construct can be used as the batched form of *insert* operation. For updates, we can use the *merge* construct of SQL:2003 (or the *update ... from ...* construct of SQLServer), as the batched form. Batched forms of these operations employ various techniques such as set-oriented index update and set-oriented integrity checks to offer increased efficiency over the corresponding non-batched operations.

**Operations having Side-Effects**

An operation with side-effects, in addition to returning a value, modifies the system state. Further, the return value may be a function of not only the arguments (parameters) but also the system state. We can model such an operation with a pair of functions, $fv : S \times D \rightarrow R$ and $fs : S \times D \rightarrow S$, where $S$ is the set of all possible system states, $D$ is domain of parameters and $R$ is the domain of result values. Since we assume batched forms are free to process the arguments in any order, we can define the batched forms for only a restricted class of side-effect causing operations. We call this restricted class of operations, for which the batched forms are defined, as *batch-safe* operations. We call an operation *batch-safe* if the following conditions hold:
When processing a set of arguments,

1. the operation's return value, for any parameter, is independent of the order in which the parameters are processed.
   $\forall s \in S, \forall x, y \in D, fv(S, x) = fv(fs(S, y), x)$

2. the final system state is independent of the order in which the arguments are processed.
   $\forall s \in S, \forall x, y \in D,\ fs(fs(s, x), y) = fs(fs(s, y), x)$

For example, an INSERT operation on a table that has no constraints defined, is *batch-safe*. However, in the presence of table constraints (*e.g.,* a unique column), the INSERT operation may or may not be *batch-safe* depending on the exact set of parameters. Operations that write to an external file/device or communicate with other systems may or may not be *batch-safe* depending on the specific application. If the programmer has enclosed an operation inside a loop that gives no guarantee of order, *e.g.,* iteration over the result set of a query without order-by clause, we may treat the operation as *batch-safe*. Such analysis for *batch-safety* of an operation can be extended further, but is beyond the scope of this work. Also, in this chapter, we assume that operations with side-effects do not have a return value, they just cause a change in the system state.

### 4.1.2 Rewriting Loops to Use Batched Forms

Consider a statement that invokes operation $q$ inside loop $L$ of a program $P$. We say that the invocation of $q$ is *batchable w.r.t* loop $L$ if it is possible to rewrite $P$ into an equivalent program $P'$ by removing the invocation of $q$ from the body of the loop and making a single invocation of the batched form of $q$ (*i.e.,* $qb$) outside the loop. As an example, consider the query in Example 1.3, which invokes a UDF. A program corresponding to the *iterative plan* for this query is shown in Example 4.3.

---

***Example 4.3*** *An Iterative Program/Plan for the Query in Example 1.3*

```
for each t in select orderid, itemcode, amount, curcode from sellorders where mkt='NSE' loop
    < body of the udf with parameters bound from t >
    if (return value > 0)
        output t.orderid;
end loop
```

---

In this example, both the queries inside the body of the UDF can be batched *w.r.t* the enclosing loop. Example 4.4 shows a rewritten form of the query and the UDF in Example 1.3. The rewritten query makes used of a batched form of the UDF, which is called *count_offers_batched*. The batched form of the UDF, shown in Example 4.4, takes a table-valued parameter. The table-valued parameter (or the batch) is constructed from all the distinct parameter values with which the non-batched form of the function would be invoked by the original query. The UDF *count_offers_batched*, makes use of a temporary table $r2$. Note that all the updates performed by the UDF are only on the temporary table. The UDF *count_offers_batched* in turn makes use of the set-oriented forms the queries *q1* and *q2* . These set-oriented forms called *q1b* and *q2b* were shown in Example 4.2. The UDF uses the SQL MERGE construct to merge the results of the set-oriented queries with the temporary table $r2$, which stores the values of all the local variables and function parameters. In the batched version we still have an iterative loop (the only loop in the body of *count_offers_batched*), but it contains only inexpensive operations. It is possible to pull even such operations out of the loop, we discuss about this in Chapter 5.

Even if an operation is *batch-safe*, it may not always be possible to batch a given invocation of the operation *w.r.t* a loop, because of side effects within the loop. In Section 4.3 we provide a set of program transformation rules that allow batching the invocations of an operation when the program satisfies certain conditions.

***Example 4.4*** *Batched Form of the Query in Example 1.3*

*Let r1 = SELECT DISTINCT itemcode, amount, curcode FROM sellorders WHERE mkt='NSE';*

*Now, the query of Example 1.3 can be written as:*

```
SELECT orderid FROM sellorder so, count_offers_batched(r1) br
WHERE so.mkt='NSE' AND so.itemcode=br.itemcode AND so.amount=br.amount AND
      so.curcode=br.curcode AND br.count_offers > 0;
```

*where,* count_offers_batched *is the batched form of the UDF* count_offers *defined as follows. For brevity, we omit the schema details when it is obvious.*

```
TABLE count_offers_batched(TABLE r1)
DECLARE
    TABLE (itemcode, amount, curcode, cond1, amount_usd, count_offers) r2; // A temporary table
BEGIN
    FOR EACH t1 IN r1 LOOP
        FLOAT amount_usd; BOOLEAN cond1; INT count_offers;
        cond1 := (t1.curcode == "USD");
        cond1 == true? amount_usd := t1.amount;
        // variables below take default values if unassigned
        r2.addRecord((t1.itemcode, t1.amount, t1.curcode, cond1, amount_usd, count_offers));
    END LOOP

    MERGE INTO r2 USING q1b(e1) AS q1b ON (r2.curcode=q1b.curcode)
    WHEN MATCHED THEN UPDATE SET amount_usd = amount * q1b.exchrate;

//  where e1 is SELECT DISTINCT curcode FROM r2 WHERE cond1=false;
//  and q1b, the batched form of query q1, is shown in Example 4.2.

    MERGE INTO r2 USING q2b(e2) AS q2b
    ON (r2.itemcode=q2b.itemcode AND r2.amount_usd=q2b.amount_usd)
    WHEN MATCHED THEN UPDATE SET count_offers = q2b.count_offers;

//  where e2 is SELECT DISTINCT itemcode, amount_usd FROM r2;
//  and q2b, the batched form of query q2, is shown in Example 4.2.

    RETURN (SELECT itemcode, amount, curcode, count_offers FROM r2;)
END
```

**Note:** *MERGE is a SQL:2003 construct.*

### 4.1.3 Generating Batched Forms of Procedures

To speed up applications or queries that make repeated calls to stored procedures or UDFs we need efficient batched forms of these procedures. However, batched forms of complex operations like stored procedures and UDFs are (as far as we know) not available

*fb-trivial(pt)* $\Longleftrightarrow$ *Apply(pt, f)*
where the function *Apply* is defined as :

> ***Apply(pt, f):***
> $r = \{\}$ ;
> for each $t$ in *pt*
>     $<$ *body of f with parameters bound from attributes of t* $>$
>     *rf* = *return value of f*;
>     *r.addRecords*({t} $\times$ *rf*);
> return $r$;

Figure 4.1: Trivial Batched Form of a Procedure

unless implemented by the programmers manually. We therefore consider the problem of automatically generating batched forms of stored procedures and UDFs. Our goal is to generate efficient batched forms by batching the expensive operations within the body of the procedure/UDF.

Given any side-effect free function or *batch-safe* operation (which could be a stored procedure) *f*, we can generate its trivial batched form as shown in Figure 4.1. Batched form of any procedure can thus be generated by enclosing it in a loop that iterates over the parameter set and executes the statements inside the procedure repeatedly. However, such a rewriting is not of significant benefit as *cost(fb-trivial)* for a batch size of $k$ is nearly same as $k \times cost(f)$; Such a rewriting can still be useful in reducing round-trip delays in client server environments. More interesting batched rewrites are the ones that use specialized and efficient strategies for batch processing, *e.g.,* batched *selection* within the procedure can be processed as a *join*, while a query in the procedure which performs a selection followed by an aggregate would have a batched form that employs grouping followed by join.

To generate an efficient batched form of a procedure, we can start with the trivial batched form of the procedure and try to batch each expensive sub-operation in the body of the procedure *w.r.t.* to the enclosing loop. To do so, the sub-operation must be taken out of the enclosing loop and substituted by its batched equivalent. If the sub-operation is a query, its batched form may be known. If the sub-operation is a procedure call, we recursively invoke the method to generate the batched form of the called procedure.

As we can see, generating batched form of a complex procedure reduces to the task of batching selected operations *w.r.t.* a loop (see Section 4.1.2). We address this problem in Section 4.3 after introducing some preliminaries in Section 4.2.

## 4.2 Background for Our Work

In this section, we formally outline the language constructs we support and provide background material on the terminology used in this chapter.

### 4.2.1 Language Constructs

For our illustration, we use a simple procedural language. The language offers expressions, assignment, conditional branching and looping. The supported language constructs are briefly described below.

- *while loops* are of the form *while*`(predicate)` *loop* ... *end loop;*.
- *cursor loops* are of the form *for each* `record` *in* `query/table` *loop* ... *end loop;* An *order by* clause can be present if the iteration is over a table. When present, the *order by* is assumed to be *ascending* by default. Unlike the more general and powerful *while loops*, the cursor loops iterate over the result of a query and hence their iteration space is known once the query is evaluated.
- Branching is possible through *if-then-else* having the syntax *if* `(predicate)` {...} *else* {...}.
- Scalar variables: we consider only scalar variables in our discussion. However, our techniques can be easily extended to handle arrays, records and collection types. Each looping block can have variables local to the block. Statements can access variables local to their block or variables defined in any of the ancestor blocks.
- Result of scalar queries (queries that return exactly one tuple) can be assigned to variables.
  *e.g.,* $v_1, v_2, \ldots, v_n =$ *select* $c_1, c_2, \ldots, c_n$ *from* ...;
  Note that a single assignment can be used to simultaneously assign values to multiple variables. Set-valued queries can be used only in the context of *cursor loops.*

We also use a few additional constructs in the transformed code. We assume these constructs are not available for the end-user and hence cannot be present in the input program.

- The TABLE type is used to construct the parameter batches. The TABLE type can be implemented so as to make use of both main memory and disk.
- Updatable cursor loops are of the form *for each* `record` *by ref in* `table` *loop* ... *end loop;* Any updates to the record modify the underlying table variable.
- The transformed program may use relational operators such as selection, projection and join.

## Assumptions

We make the following assumptions about the program.

1. Unconditional control transfer statements like GOTO, EXIT and CONTINUE are not used.
2. Statements have no hidden side-effects. Information about all reads and writes performed by a statement (either on memory locations or on external resources like files and databases) are captured in the *data dependence graph* (explained in the next section).

### 4.2.2 Data Dependence Graph

The *Data Dependence Graph* (DDG), sometimes referred to as Program Dependence Graph [37, 15], of a program is a directed multi-graph in which program statements are nodes (vertices) and the edges represent data dependencies between the statements. The different types of data dependence edges are explained below.

- A *flow-dependence* edge ($\xrightarrow{FD}$) exists from statement (node) $s_a$ to statement $s_b$ if $s_a$ writes a location that $s_b$ may read, and $s_b$ follows $s_a$ in the forward control-flow.
- An *anti-dependence* edge ($\xrightarrow{AD}$) exists from statement $s_a$ to statement $s_b$ if $s_a$ reads a location that $s_b$ may write, and $s_b$ follows $s_a$ in the forward control flow.
- An *output-dependence* edge ($\xrightarrow{OD}$) exists from statement $s_a$ to $s_b$ if both $s_a$ and $s_b$ may write to the same location, and $s_b$ follows $s_a$ in the forward control flow.

- A *loop-carried flow-dependence* edge ($\xrightarrow{LFD_L}$) exists from $s_a$ to $s_b$ if $s_a$ writes a value in the $i^{th}$ iteration of a loop $L$ and $s_b$ may read the value in a later iteration ($j^{th}$ iteration where $j > i$).
- Similarly, there are loop carried counter parts of *anti* and *output* dependencies and are denoted by ($\xrightarrow{LAD_L}$) and ($\xrightarrow{LOD_L}$) respectively.

The data dependence graph for the sample UDF of Example 4.1 is shown in Figure 4.2.

### External Dependencies

Statements may have dependencies not only through program variables but also through the database and other external resources like files. For example, we have $s_1 \xrightarrow{FD} s_2$ if $s_1$ writes a value to the database, which $s_2$ may read subsequently. Though standard dataflow analysis performed by compilers considers only dependencies through program variables, it is not hard to extend the techniques to consider external dependencies, at least in a conservative manner. For instance, we could model the entire database (or file system) as a single program variable and thereby assume every query/read operation on a database/file to be conflicting with an update/write of the database/file. In practice, it is possible to perform a more accurate analysis on the external writes and reads. When referring to external dependencies explicitly, we use $E$ as a superscript to the corresponding type of dependence edge *e.g.,* $s_1 \xrightarrow{FD^E} s_2$.

## 4.3 Program Transformation

Recall from Section 4.1.2 that an invocation of an operation $q$ inside a loop $L$ is said to be *batchable w.r.t* loop $L$ if it is possible to rewrite the program into an equivalent program where the invocation of $q$ is removed from the body of the loop and a single invocation of the batched form $qb$ is made outside the loop. For such a rewrite, it is necessary that the operation should be *batch-safe*. However, the data and control dependencies between program statements may make it impossible to batch a statement that invokes an operation even if the operation is *batch-safe*. In this section, we present a set of program transformation rules, which enable us to batch a statement *w.r.t.* a loop when the program satisfies certain conditions.

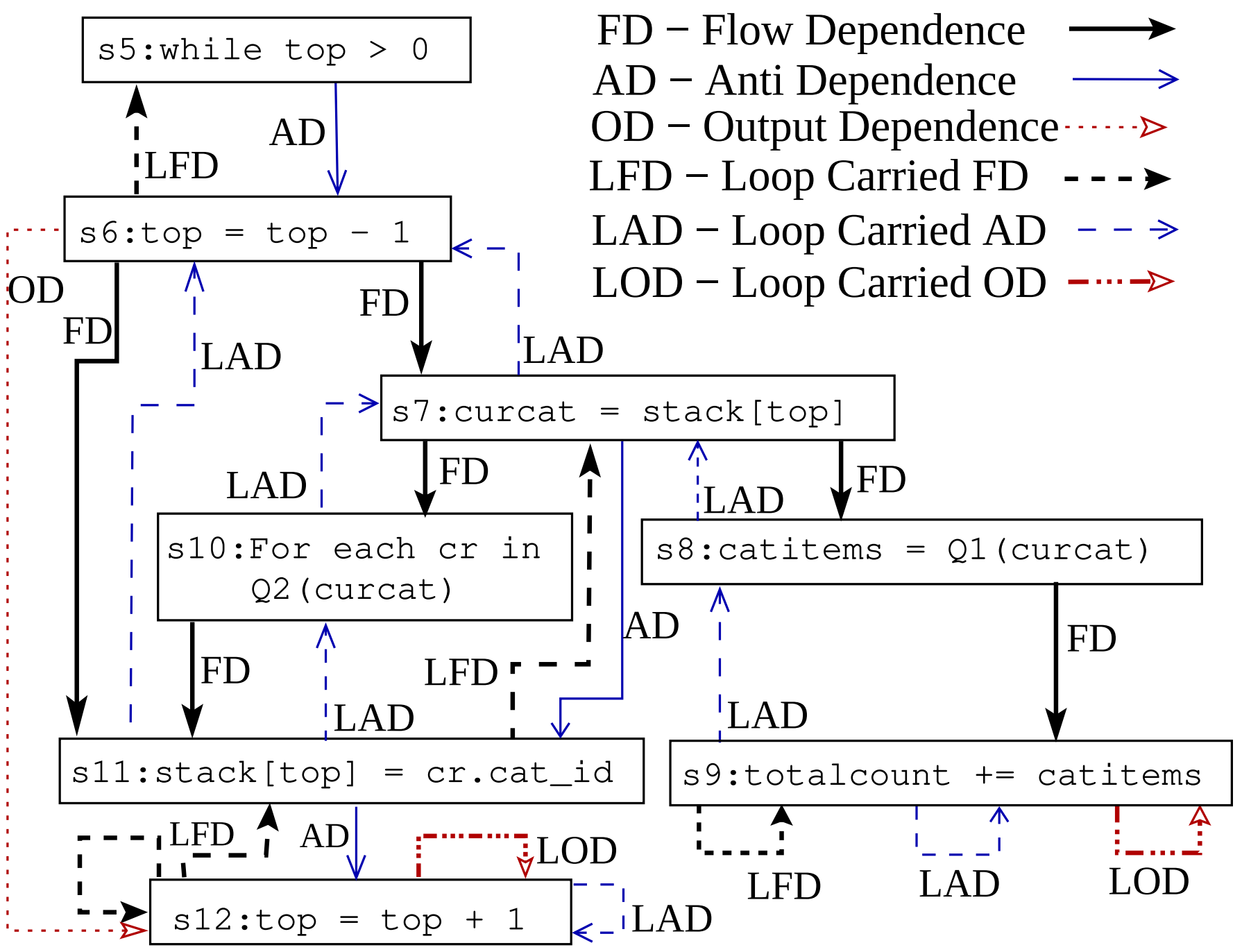

Figure 4.2: A Subgraph of the Data Dependence Graph for the UDF in Example 4.1

The program transformation rules we present, like the equivalence rules of relational algebra, allow us to repeatedly refine a given program. Applying a rule to a program involves substituting a program fragment that matches the antecedent (LHS) of the rule with the program fragment instantiated by the consequent (RHS) of the rule. Some rules facilitate the application of other rules and together achieve the goal of batching a desired statement *w.r.t.* a loop. Applying any rule results in an equivalent program and hence the rule application process can be stopped at any point.

### Notation Used in the Transformation Rules

- $R(s)$ : The read-set of $s$ is the set of variables read by statement or statement sequence $s$.
- $W(s)$ : The write-set of $s$ is the set of variables written by statement or statement sequence $s$.
- $U(s)$ : $R(s) \cup W(s)$. Called the use-set of $s$.
- *pred*? $s$ : Conditional statement. Equivalent to *if (pred) then* $s$.
- *LV(s)* : Set of variables local to the block statement $s$.

- *NLV(s)* : Set of variables accessible but not local to the block statement $s$. These are variables defined in an ancestor block.
- $|ss|$ : Length of the statement sequence $ss$
- $ss[i]$ : Stmt at the $i^{th}$ position in sequence $ss$, $1 \leq i \leq |ss|$.
- $s_1 + s_2$ : Concatenation (of statement sequences or strings).
- $SUBS(s, v, v')$ : Statement obtained by substituting all occurrences of variable $v$ in statement $s$ by variable $v'$.
- $SUBS(s, vs, map)$ : Statement obtained as follows. For each variable $v \in U(s) \cap vs$, substitute all occurrences of $v$ in statement $s$ by $map(v)$.
- $s \cup^* r$ : Disjoint union (UNION ALL) of relations s and r.
- $\Pi^d_{a_1,a_2,\ldots,a_n}(r)$ : Projection without duplicate elimination
- $(a_1, a_2, \ldots, a_n)$ : Tuple constructor
- *type-of(e)* : Data type of expression $e$.

### Predicates on the DDG

- $s_1 \xrightarrow{FD} s_2$ : True only if the DDG contains a flow-dependence edge (either internal or external) from $s_1$ to $s_2$.
- $s_1 \xrightarrow{FD+} s_2$ : True only if the DDG contains a path from $s_1$ to $s_2$ having only FD edges.
- $s_1 \xrightarrow{(FD|LFD)+} s_2$ : True only if the DDG contains a path from $s_1$ to $s_2$ having only FD or LFD edges.
- $indep(s_1, s_2)$ : True only if there are no dependencies between statements $s_1$ and $s_2$.
- Similarly we have predicates for the existence of other types of dependencies.

### Conventions

1. Loops of the form "*for each t by ref in r*" are updatable cursor loops. The underlying set $r$ can be modified by assignments to the tuple's attributes.

2. Suppose a table-valued expression $e$ has arity $n$. The rename operator $\rho_{x(a_1,a_2,\ldots,a_n)}(e)$ returns the result of expression $e$ under the name $x$, and with the attributes renamed to $a_1, a_2, \ldots, a_n$.

3. The merge operator $\mathcal{M}_{a_1=b_1,\ldots,a_n=b_n}(r, s)$ (based on the SQL:2003 merge construct) updates relation $r$ by merging in the records of $s$. For each record in $s$ that matches a record in $r$ on the attributes common to $r$ and $s$, the record in $r$ is updated by assigning the values of attributes $b_1, \ldots, b_n$ from the $s$ tuple to the attributes $a_1, \ldots, a_n$ of the $r$ tuple correspondingly.

4. Projection that removes the specified attributes:
$\Pi_{\bar{a_1},\bar{a_2},\ldots,\bar{a_n}}(r)$ is equivalent to $\Pi_{S-\{a_1,a_2,\ldots,a_n\}}(r)$, where $S$ is the set of all attributes in *schema(r)*

5. Multi-assignment from scalar queries: Let $q$ be a query returning exactly one tuple of arity $n$. The assignment $v_1, v_2, \ldots, v_m = q$ (where $m \leq n$) assigns the values of the first $m$ attributes of the returned tuple to the $m$ variables on the LHS in that order.

In all the rules, unless specified, we assume **q** to be a **batch-safe** operation with **qb** as its batched form.

### 4.3.1 Rewriting Set Iteration Loops (Rule 1)

In the simplest case, the loop contains a single statement that invokes the operation we want to batch. In this rule, we consider cursor update loops - the loop iterates over a set of tuples and the values returned by the operation to be batched are assigned back to the attributes of the tuple associated with current iteration. Rules 1A through 1C are the basic rules that allow replacing a loop by a batched invocation. Rule 1D and all the other rules presented in this section help us transform the program so as to enable the application Rules 1A, 1B or 1C.

In Rule-1A, $q$ can be any batch-safe operation (with or without side-effects). Note the use of projection without duplicate elimination ($\Pi^d$) for constructing the parameter multiset. However, in rules 1B and 1C, we require $q$ to be a *pure* function returning exactly one tuple (*e.g., a scalar query*). In rules 1B and 1C we construct a duplicate-free parameter set using the standard relational projection. This avoids the duplicate record problem while merging back the results of the batched invocation.

For brevity, in rules 1B and 1C we omit the form with loop invariant parameters. In rules 1B and 1C we deal only with assignments to cursor attributes and not variables.

**Rule 1** Batching Simple Set Iteration Loops

Rule 1A: Unconditional invocation with no return value

1A(i) Basic form

for each $t$ in $r$ loop
    $q(t.c_1, t.c_2, \ldots, t.c_m)$; $\iff$ $qb(\Pi^d_{c_1,c_2,\ldots,c_m}(r))$;
end loop;

1A(ii) Form with loop invariant parameters

for each $t$ in $r$ loop
    $q(t.c_1, t.c_2, \ldots, t.c_m, v_1, v_2, \ldots, v_n)$; $\iff$ $qb(\Pi^d_{c_1,c_2,\ldots,c_m}(r) \times \{(v_1, v_2, \ldots, v_n)\})$;
end loop;

Rule 1B: Unconditional invocation with return value

for each $t$ by ref in $r$ loop
    $t.c_{w1}, t.c_{w2}, \ldots, t.c_{wn} = q(t.c_{r1}, t.c_{r2}, \ldots, t.c_{rm})$;
end loop;
where $q$ is a *pure* function.

$\Updownarrow$

$\mathcal{M}_{c_{w1}=c_{w1'},\ldots,c_{wn}=cwn'}(r, e)$
where $e = \rho_{x(c_{r1},\ldots,c_{rm},c_{w1'},\ldots,c_{wn'})} qb(\Pi_{c_{r1},\ldots,c_{rm}}(r))$;

Rule 1C: Conditional Invocation

for each $t$ by ref in $r$ loop
    $(t.cv == true)?\ t.c_{w1}, \ldots, t.c_{wn} = q(t.c_{r1}, \ldots, t.c_{rm})$;
end loop;
where $q$ is a *pure* function.

$\Updownarrow$

$\mathcal{M}_{c_{w1}=c_{w1'},\ldots,c_{wn}=cwn'}(r, e)$, where
$e = \rho_{x(c_{r1},\ldots,c_{rm},c_{w1'},\ldots,c_{wn'})} qb(\Pi_{c_{r1},\ldots,c_{rm}}(\sigma_{cv=true}\ r))$;

Note that in rules 1B and 1C we use duplicate-eliminating projection ($\Pi$) and not ($\Pi^d$).

Rule 1D: Removal of Order-By

| | | |
|---|---|---|
| for each $t$ [by ref] in $r$ order by *cols* loop<br>    *batch-safe-operation(t)*<br>end loop; | $\iff$ | for each $t$ [by ref] in $r$ loop<br>    *batch-safe-operation(t)*<br>end loop; |

The reason for this will be clear when we describe rules 2 and 3. For now, it suffices to know that the later transformations bring the program into a form in which we can apply one of the rules described in this section.

### 4.3.2 Splitting a Loop (Rule 2)

In general, the body of a loop may contain several statements along with the query execution statement(s) we are interested in batching. In such a case, we try to split the loop into multiple loops, such that the statement we are interested in batching appears in a loop by itself. Consider the examples in Figures 4.3 and 4.4. The statements to be batched are shown in bold. As shown in the figures, we split the loop into multiple set-iteration loops. The aim is to have the statement to be batched appear in a loop by itself, a form in which we can apply Rule 1. For example, in the rewritten code of Figure 4.3, the loop containing a single INSERT statement can be replaced by a batched invocation, by first removing the *order by* using Rule 1D and then applying Rule 1A(i).

```
for each r in  SELECT grantid, empid, gnum FROM grantload loop
    int internalid = foo(r.grantid, r.empid);
    INSERT INTO grants VALUES (internalid, r.empid, r.gnum);
    total + = r.gnum;
end loop;
```

$\Downarrow$

```
TABLE(key, empid, gnum, internalid) t;
int loopkey = 0;
for each r in  SELECT grantid, empid, gnum FROM grantload loop
    RECORD(key, empid, gnum, internalid) s;
    s.empid = r.empid;  s.gnum = r.gnum;
    int internalid = foo(r.grantid, r.empid);
    s.internalid = internalid;
    s.key = loopkey++;
    t.addRecord(s);
end loop;

for each s in t loop order by key
    INSERT INTO grants VALUES (s.internalid, s.empid, s.gnum);
end loop;

for each s in t loop order by key
    total += s.gnum;
end loop;
```

Figure 4.3: Splitting a `cursor` loop

If the sequence of statements $ss$ in a loop is made up of two consecutive sub-sequences $ss_1, ss_2$ (*i.e.,* $ss = ss_1 + ss_2$), and if there are no loop-carried flow dependencies from any

```
while (top > 0) loop
      top = top − 1;
      curcat = stack[top];
      catitems = SELECT count(itemid) FROM item WHERE category=curcat;
      totalcount + = catitems;
end loop;
```

⇓

```
TABLE(key, curcat, catitems) t;
int loopkey = 0;
while(top > 0) loop
      RECORD(key, curcat, catitems) r;
      top = top − 1;
      curcat = stack[top];
      r.curcat = curcat;
      r.key = loopkey++;
      t.addRecord(r);
end loop;

for each r by ref in t order by key loop
      t.catitems = SELECT count(itemid) FROM item WHERE category=t.curcat;
end loop;

for each r in t order by key loop
      totalcount + = t.catitems;
end loop;
```

Figure 4.4: Splitting a `while` loop

statement in $ss_2$ to any statement in $ss_1$ or to the loop predicate, then the loop can be split such that $ss_1$ and $ss_2$ appear in separate loops.

Unlike cursor loops, the iteration space for general *while* loops cannot be known upfront [45] and is constructed dynamically. In general, the loop splitting transformation, for the case of *while* loops, can be expressed as Rule 2[1].

We call a variable as *split variable* if it is involved in a loop-carried anti or output dependency that crosses the split boundaries. While splitting a loop we introduce a table-valued variable, which has attributes corresponding to each of the split variables. Each loop in the original program, if required to be split, introduces exactly one table. The table essentially serves to break the loop-carried anti and output dependencies. We call the table associated with loop $L$ in the original program as the $L$-table. Note that after

[1] In the interest of readability, the earlier examples in Figures 4.3 and 4.4 contained minor variations from the construction in Rule 2.

**Rule 2** Splitting a WHILE Loop

**while** $p$ **loop**
    $ss_1$;  $s$;  $ss_2$;
**end loop;**

such that:

(a) No loop-carried flow dependencies (*i.e.,* LCFD edges, external or otherwise) cross the points before and after $s$.
(b) No loop-carried *external* anti or output dependencies cross the points before and after $s$.

$\Updownarrow$

*Table(T)* t;
int loopkey = 0;
**while** $p$ **loop**
    *Record(T)* r;  $ss'_1$;  r.key=loopkey++;  *t.addRecord(r);*
**end loop;**
**for each** r **by ref in** t **order by** t.key **loop**
    $s'$
**end loop;**
**for each** r **by ref in** t **order by** t.key **loop**
    $ss_r$;  $ss_2$
**end loop;**
**delete** t;

where $T$, $ss'_1$, $s'$ and $ss_r$ are constructed as follows.
Let $SV$ (split variables) be the set of variables for which either an LCAD or LCOD edge crosses the split boundaries (the edge is incident from $ss_2$ to $s$ or $ss_1$, or from $s$ to $ss_1$).

1. Table $t$ and record $r$ have attributes corresponding to each variable in $SV$ and a key.
2. $ss'_1$ is same as $ss_1$ but with additional assignment statements to attributes of $r$. Each write to a split variable $v$ is followed by an assignment statement $r.v = v$;. If the write is conditional, then the newly added statement is also conditional on the same guard variable.
3. The statement $s'$ is same as $s$, except that each reference $v$ to a variable in set $SV$ is replaced by $r.v$. Formally, $s' = SUBS(s, SV, map : v \rightarrow r.v)$
4. $ss_r$ is a statement sequence assigning attributes of $r$ to corresponding variables. Each assignment in $ss_r$ is conditional; the assignment is made only if the attribute of $r$ is non-null (*assigned*).

**Note:** If the all operations in the loop are *batch-safe* we can omit the *loopkey* and the *key* attribute.

Rule 2A: Splitting a Cursor Loop

The rule for splitting cursor loops is a minor variant of Rule 2, and we omit the details for brevity.

the split, the newly formed loops that iterate on the $L$-table must have an ORDER BY clause. The ORDER BY clause can then be eliminated using Rule 1D if the statements inside the loop are batch-safe.

Although we succeed in splitting a loop by breaking loop-carried anti and output dependencies, loop-carried flow dependencies prohibit splitting of the loop (see precondition (a) of Rule-2). For example, if a statement in $ss_2$ wrote a variable whose value is read in $ss_1$ or by the loop predicate, then Rule 2 does not apply. However, later in this chapter we shall see that in some cases it is possible to reorder the statements within a loop so as avoid such loop-carried flow dependencies. All types of loop-carried external dependencies prohibit splitting the loop (see preconditions (a) and (b) of Rule-2). Recall that external dependencies are data dependencies through external resources like files. Rule 2 generalizes rule T4 of Lieuwen and DeWitt [34]. We compare our work with [34] in Section 4.7.

### 4.3.3 Separating Batch-Safe Operations (Rule 3)

A program statement may contain the expression we want to batch in combination with other non batch-safe operations. In such a case, we isolate the batch-safe operation by introducing an extra variable. Figure 4.5 shows an example and Rule 3 expresses the transformation formally.

**Rule 3** Isolating Batch-Safe Expressions

Let $e$ be a batch-safe expression in statement *stmt*. Then,

$$stmt; \quad \Leftrightarrow \quad T\ v = e; \quad stmt';$$

where $stmt' = SUBS(stmt, e, v)$ and $T =$*type-of*$(e)$;

Variable assignment is not batch-safe in general, *e.g.,* assignment to a global variable. However, assignments to different locations, *e.g.,* different rows of a cursor loop, can be performed in any order and hence batch-safe in the context of split variables that are converted to attributes of the $L$-table by Rule 2. If the return value of a query is assigned to a non-local variable, applying Rule 3 introduces a new split variable and thus enables batching the query.

### 4.3.4 Control to Flow Dependencies (Rule 4)

Conditional branching (*if-then-else*) and *while* loops lead to control dependencies. If the predicate evaluated at a conditional branching statement $s1$ determines whether or not

**(i) Original Code**

```
while (n > 0) loop
    x = s[−−n];
    // q() is to be batched; print() is not batch-safe
    print(q(x));
end loop;
```

⇔

**(ii) After applying Rule-3**

```
while (n > 0) loop
    x = s[−−n];
    // Let T be type-of(q(...))
    T v = q(x);
    print(v);
end loop;
```

**(iii) After splitting the loop**

```
Table(...) t; int loopkey = 0;
while (n > 0) loop
    Record(. . .) r;
    x = s[−−n];
    r.x = x; r.key = loopkey++; t.addRecord(r);
end loop;

// order-by removed with Rule 1D
for each r by ref in t loop
    r.v = q(r.x);
end loop;

// order-by is required
for each r in t order by t.key loop
    print(r.v);
end loop;
```

Figure 4.5: Separating Batch-Safe Operations

```
for each t by ref in sales loop
  if (t.brcode == 58)
    t.brcode = 1;
    q(t.item, t.qty, t.brcode);
  end if
end loop;
```

⟺

```
for each t by ref in sales loop
  // Using a control variable remember
  // the branching decision
  boolean cv = (t.brcode == 58);
  (cv==true)? t.brcode = 1;
  (cv==true)? q(t.item, t.qty,
                        t.brcode);
end loop;
```

*Note: After this transformation, we can apply Rule-2 and split the loop. The conditional invocation of* **q** *can then be batched using Rule 1C.*

Figure 4.6: Transforming Control-Dependencies to Flow-Dependencies

control reaches statement $s2$, then $s2$ is said to be control dependent on $s1$. During loop split, it may be necessary to convert the control dependencies into flow dependencies [29]. Figure 4.6 shows an example. Rule 4 specifies the transformation formally.

**Rule 4** Converting control-dependencies to flow-dependencies

if $(p)$ { $ss_1$ } else { $ss_2$ }

$\Updownarrow$

boolean $cv = p$;
$ss$

where $ss[i] = (cv == true)?ss_1[i], 1 \leq i \leq |ss_1|$ and
$ss[k+j] = (cv == false)?ss_2[j], 1 \leq j \leq |ss_2|, k = |ss_1|$

### 4.3.5 Reordering Statements (Rule 5)

Consider the example in Figure 4.7. Suppose we want to batch the query invocation **q(category)** in statement s1. We cannot directly split the loop so as to batch s1 because there are loop-carried flow-dependencies from s3 to s1 and to the loop predicate, which violate pre-condition (a) of Rule 2. Statement s3, which appears after s1, writes a value and statement s1 reads it in a subsequent iteration. We therefore reverse the order of statements s1 and s3 before splitting the loop (Figure 4.7). Intuitively, we first collect all the categories in the hierarchy and then perform a batched invocation of the query that computes the item counts for the categories. The basic rules that allow us to reorder statements are specified in Rule 5. To be able to split a loop so as to batch the desired statement, multiple applications of Rule 5 may be needed. It is important that Rule 5 be applied in a carefully chosen sequence so as to achieve the desired reordering. We will return to this issue in Section 4.4, where we give an algorithm to reorder statements such that the pre-conditions for Rule 2 are met.

### 4.3.6 Batching Across Nested Loops (Rule 6)

Loops in a program may be nested within other loops and form a hierarchy. The query or update operation we are interested in batching may lie anywhere in the loop hierarchy. It is often desirable to batch the query or update operation *w.r.t.* as many ancestor loops as possible. The aim here is to make fewest possible calls to the expensive operation, in other words, to make the size of the batch in each invocation as large as possible.

Consider a loop $L_c$ nested under loop $L_p$ and a query $q$ inside $L_c$. When the child loop ($L_c$) is split using Rule 2, a TABLE valued local variable, $L_c$-table, is introduced in the parent loop ($L_p$). With the application of Rule 1, $q$ is pulled out of $L_c$ and is replaced

**Original Program**

```
s0:  while (category != null) loop
s1:        int icount = q(category); // Query to batch
s2:        sum = sum + icount;
s3:        category = getParentCategory(category);
     end loop;
```

⇕

**After Order Reversal**

```
s0:  while (category != null) loop
s1':       int category_stub = category;
s3:        category = getParentCategory(category);
s1:        int icount = q(category_stub);
s2:        sum = sum + icount;
     end loop;
```

⇕

**After Loop Split**

```
TABLE(...) r;
int loopkey = 0;
while (category != null) loop
     RECORD(...) t;
     int category_stub = category;
     t.category_stub = category_stub;
     category = getParentCategory(category);
     t.key = loopkey++;
     r.addRecord(t);
end loop;

for each t by ref in r loop
     t.icount = q(t.category_stub);
end loop;

for each t in r order by key loop
     sum = sum + t.icount;
end loop;
```

Figure 4.7: Reordering Statements to Satisfy Pre-Condition of Rule-2

by $qb$ that lies directly inside $L_p$. In turn, when the parent loop is split, the $L_c$-table (of the child loop) becomes a TABLE valued attribute (nested table) in the parent's $L_p$-table. We now perform a second level batching of $qb$ *w.r.t.* $L_p$ by unnesting the $L_p$-table. Rule 6 enables this transformation. Intuitively, we first try to pull the statement out of the inner most loop enclosing it and then out of the next (higher) level loops. Figure D.5 in Appendix D gives a complete example of batching a statement across nested loops.

In Rule 6 we make use of the *nest*($\nu$) and *unnest*($\mu$) operators of nested relational algebra [8, 3]. Below we give a brief description of these operators and refer to [8] for the

**Rule 5** Basic Rules that Facilitate Reordering of Statements

Rule 5A: Reordering Independent Statements
Two statements can be reordered if there exists no dependency between them.

$s_1; s_2;$ where *indep*$(s_1, s_2)$ $\Longleftrightarrow$ $s_2; s_1;$

Rule 5B: Shifting an Anti-Dependence Edge
An anti-dependence edge between two statements can be shifted by using an extra variable.

$s_1; s_2;$
where $s_1 \xrightarrow{AD_v} s_2$
$\Updownarrow$
$v' = v; s_1'; s_2;$

where $s_1'$ is constructed from $s_1$ by replacing all reads of $v$ by reads of $v'$.

Rule 5C: Shifting an Output-Dependence Edge

$s_1; s_2;$
where $s_1 \xrightarrow{OD_v} s_2$
$\Updownarrow$
$s_1; s_2'; v = v';$

where $s_2'$ is constructed from $s_2$ by replacing all writes of $v$ by write to $v'$.

formal definitions.

**Nest**: The *nest* operator ($\nu_{S\to s}(r)$) groups the tuples of $r$ on attributes *schema(r)* $- S$, then for each group forms a single tuple with a relation valued attribute $s$ containing the $S$ values of the tuples grouped together.

**Unnest**: The *unnest* operator ($\mu_s(r)$), where $s$ is a relation valued attribute of $r$, performs the inverse operation of *nest*.

$\mu_s(r) = \bigcup_{t\in r}(\Pi_R\{t\} \times t.s)$, where $R$ is the set of all attributes in *schema(r)* excluding $s$.

Though we use a nested relational model for the $L$-tables, our techniques are easy to implement on any RDBMS by storing the nested tables separately.

### 4.3.7 Correctness of Transformation Rules

We give here a brief description of our approach for proving the correctness of the program transformation rules. Appendix B gives formal proofs of correctness.

**Rule 6** Batching Across Nested Loops

Let $s$ be a table valued attribute of table $r$, and let $S = schema(r.s)$.

Rule 6A. No Return Value
for each $t$ in $r$ loop
    $qb((t.c_1, \ldots, t.c_n) \times t.s);$
end loop;

$$\Updownarrow$$

$qb(\Pi^d_A(\mu_s(r)))$ where $A = \{c_1, \ldots, c_n\} \cup S$

Rule 6B: With Return Value
for each $t$ in $r$ loop
    $\mathcal{M}_{c1=c'_1,\ldots,c_n=cn'}(t.s, qb(t.s))$
end loop;
where $qb$ is a *pure* function.

$$\Updownarrow$$

$rs = \mu_s(r);$
$\mathcal{M}_{c1=c'_1,\ldots,c_n=cn'}(rs, qb(\Pi_S(rs)));$
$r = \nu_{S \to s}(rs);$

Let $P_L$ be a program fragment that matches the LHS of a rule and $P_R$ be the program fragment instantiated by the corresponding RHS. Let $p$ be the program position at which $P_L$ begins. Let $(G, S)$ be the pair of *any* valid program and system states at $p$. The program state $G$ comprises of values for all variables accessible at the program position $p$ and the system state $S$ comprises of the state of all external resources like database and file system. To prove the correctness of a transformation rule, we must show the following. If the execution of $P_L$ on $(G, S)$ results in the state $(G', S')$ then the execution of $P_R$ on $(G, S)$ will also result in the state $(G', S')$. Note that we assume intermediate program and system states are not observable. This is a valid assumption in many practical applications. The correctness of many rules directly follows from the definition of *batch-safe* operation and that of *batched forms*. In some cases, we need to show the multiset equivalence of the $L$-tables, which is being updated, at the end of the execution of $P_L$ and $P_R$. The proof of correctness of *Rule 2* uses an argument on the values seen by each statement in the $i^{th}$ iteration $(1 \leq i \leq n)$, where $n$ is the number of loop iterations.

## 4.4 Control Algorithm for Rule Application

Rewriting a program for set-orientation involves the following steps: *(i)* identify iteratively invoked query execution statements, *(ii)* decide whether it is beneficial to batch the query execution and the ancestor loop (in the hierarchy of loops enclosing the statement) with respect to which the statement must be batched and *(iii)* rewrite the program by *systematically* applying the transformation rules presented in the previous section.

```
procedure batch(Stmt s, Loop l)
Inputs:
    s:  The query execution statement to be batched
    l:  A program loop w.r.t. which the query must be batched (i.e., s must be pulled out of l).
        s may be present directly inside l or within a descendant loop of l.

    Note: We assume the above inputs are provided manually; cost-based decision
          is a future work.
Goal:
    Rewrite the program to batch the query execution in statement s w.r.t. loop l.

begin
    Let lp be the loop which directly encloses s.
    // Batch s w.r.t. lp. Let the batched statement be sb.
    sb = do-batching(s, lp);
    if (lp != l)
        Let lpp be the parent loop of lp
        batch(sb, lpp);
end;

procedure do-batching(Stmt s, Loop l)
begin
    Rewrite s using Rule-3 so that s is a simple assignment with only the query invocation
    expression on its RHS.

    If s is control dependent on any statement inside loop l (other than the loop predicate),
    convert the control-dependency to flow dependency (using Rule-4).

    Reorder the statements in l to satisfy pre-conditions for Rule-2 (making use of Rule 5).
    reorder(s, l.block);      // Procedure reorder is described later (Figures 4.9 and 4.11)

    Applying Rule-2 split the loop l at the program points before and after s.

    Batch the query execution using Rule 1 or Rule 6.

    Return the reference to statement sb, the batched form of s.
end;
```

Figure 4.8: Control Algorithm for Rule Application

Identifying the query execution statements in a loop is usually straight forward. However, the decision on whether a statement should be batched and the level in the loop hierarchy up to which the statement must be batched, requires a cost-based analysis. Cost-based analysis is a future direction for our work and some of the parameters needed for cost-based analysis are discussed in Chapter 5. In this section, we assume these two inputs (the query invocation to be batched and the ancestor loop up to which the query must be batched) are available from the user. Given a query execution statement and a loop with respect to which the statement must be batched, the transformation rules presented in this chapter can be used to rewrite the program. However, it is important to apply the rules in a systematic way so as to achieve the goal of batching the given statement.

In Figure 4.8 we give the procedure for applying the rules so as to batch a given query execution statement *w.r.t.* a given ancestor loop enclosing it. The procedure *batch* recursively pulls out the given statement, starting from the inner most loop enclosing it. Procedure *do-batching* performs the actual task of rewriting by applying the rules. First, we apply Rule-3 on the statement and ensure the RHS contains only the query execution expression. We then convert all the control-dependencies to flow-dependencies by applying Rule-4. This allows us to treat the entire body of the loop as a *basic block* (a straight-line sequence of statements with no branches into or out of the sequence). We perform a reordering of the statements (if needed) to satisfy the pre-conditions for Rule-2, and then use Rule 2 to split the loop at the program points which immediately precede and succeed the query execution statement. This leaves the query execution statement in a loop by itself - a form in which we can apply Rule-1 or Rule-6 and make use of the batched form. Rule-1 gets applied for the inner most loop enclosing the statement and Rule-6 gets applied for the higher level loops.

## Reordering Statements to Enable Loop Splitting

As described in Section 4.3.5, to facilitate splitting of a loop, it may be necessary to reorder the statements within a loop. A loop can be split using Rule-2 only if there are no loop-carried flow dependencies that cross the split boundaries. Section 4.3.5 gives an example and specifies the basic transformations (Rule 5) that enable reordering of statements. The basic transformations specified in Rule 5 must be applied in an appropriate sequence to

```
procedure reorder(BasicBlock b, Stmt s_q)
// Goal: Reorder the statements within b, such that no LCFD edges cross
// the program points immediately preceding and succeeding s_q.
// Assumption: s_q does not lie on a true-dependence cycle in the subgraph
// of the DDG induced by statements in b.
begin
    while there exists an LCFD edge crossing the split boundaries for s_q
        Pick an LCFD edge (v1, v2) crossing the split boundaries.

        if there exists a true-dependence path from v1 to s_q
        /* No true-dependence path from s_q to v1 */
            stmtToMove = s_q;
            targetStmt = v1;
        else
        /* No true-dependence path from v1 to s_q, which implies no true-dependence
           path from v2 to s_q as there exists an LCFD edge from v1 to v2 */
            stmtToMove = v2;
            targetStmt = s_q;

        // Move stmtToMove past the targetStmt
        Compute srcDeps, the set of all statements between stmtToMove and targetStmt,
        which have a flow dependence path from stmtToMove.

        while srcDeps is not empty
            Let v be the statement in srcDeps closest to targetStmt
            moveAfter(v, targetStmt);     // see Figure 4.11

        moveAfter(stmtToMove, targetStmt);
end;
```

Figure 4.9: Procedure *reorder*

achieve the desired reordering of statements in a loop. We now give an algorithm to do so. The goal is to reorder the statements such that no loop-carried flow dependencies cross the desired split boundaries. We make use the following definition in the description to follow.

**Definition 4.1** *A true-dependence path (or cycle) in a data dependence graph is a directed path (or cycle) where each edge represents either a flow-dependence (FD) or a loop-carried flow-dependence (LCFD).*

Note that a true-dependence path excludes anti, output, loop-carried anti and loop-carried output dependence edges.

The algorithm *reorder*, shown in Figure 4.9, works as follows. For each loop-carried flow dependence edge that crosses the split boundaries (the two program points in the basic block that immediately precede and succeed the statement to batch), the algorithm decides the statement to move, and its target position. There are four cases to consider while deciding the statement to move and its target position. The cases are shown in Figure 4.10. The way in which we choose the *stmtToMove* and *targetStmt* ensures the following. If $s_q$, the statement to be batched, does not lie on a true-dependency cycle, then there exists no true-dependence path from the *stmtToMove* to the *targetStmt*. We then compute two sets *srcDeps* and *tgtDeps*. *srcDeps* is the set of statements present between *stmtToMove* and *targetStmt*, and which have a path of flow-dependence edges from *stmtToMove*. *tgtDeps* is the set of statements present between *stmtToMove* and *targetStmt*, and from which there exists a path of flow-dependence edges to the *targetStmt*. Note that the two sets will be disjoint. Each statement in *srcDeps* is then moved past the *targetStmt* using the *moveAfter* procedure. The procedure *moveAfter* (shown in Figure 4.11) performs the required reordering by swapping pairs of adjacent statements. While doing so, the procedure resolves any anti and output dependencies by creating stub statements, which make use of temporary variables.

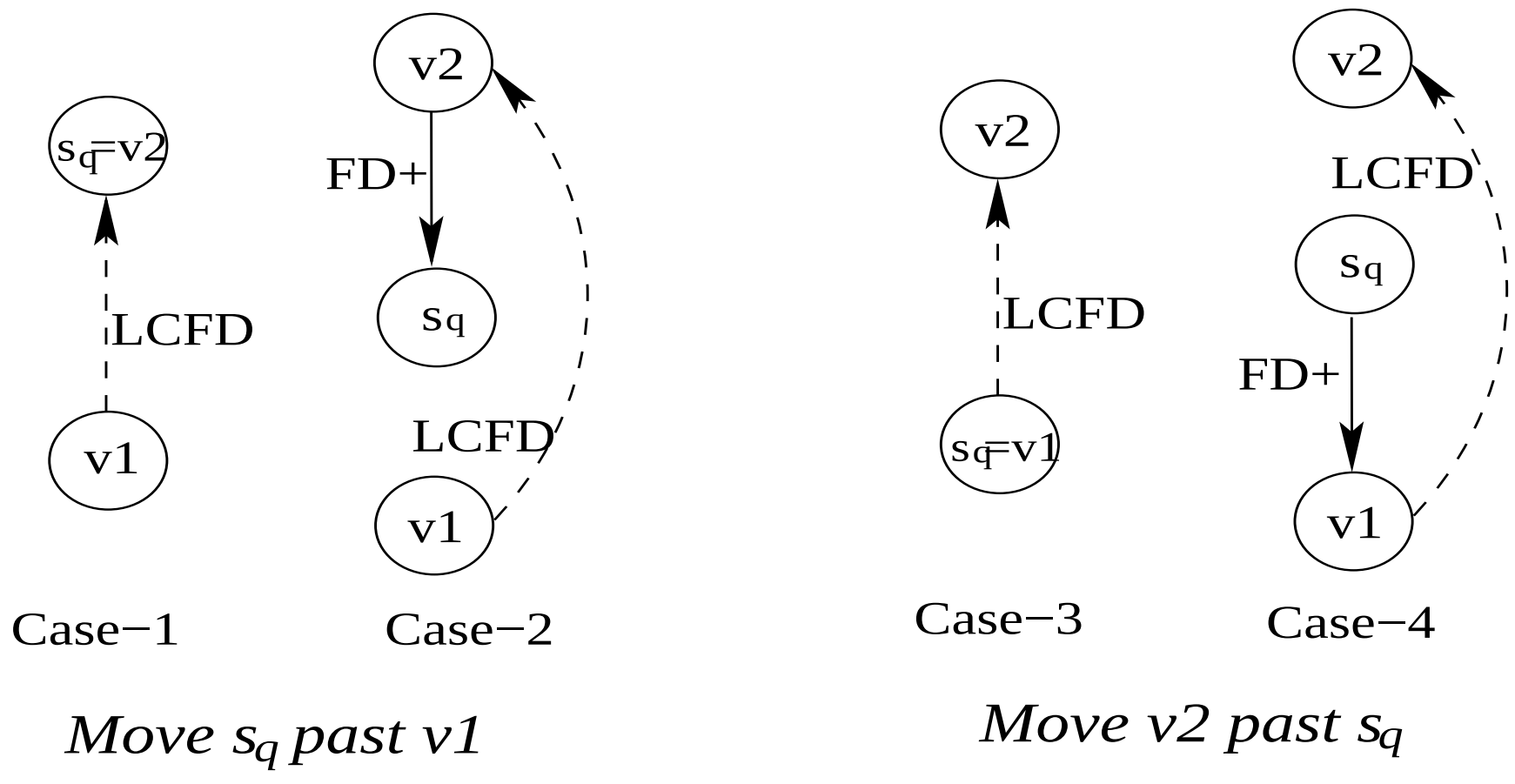

Figure 4.10: Cases for Reordering Statements

Figures 4.12 through 4.14 show examples illustrating statement reordering. Figure 4.15 shows the data dependence graph for the original and reordered code of Figure 4.14. In Figure 4.15, for each flow-dependence (FD) edge from $x$ to $y$, there exists a corresponding loop-carried anti-dependence (LCAD) edge from $y$ to $x$ but these edges are not shown. Similarly, AD and OD edges have corresponding LCFD and LCOD edges respectively, which are not shown. In this example, $s1$ is the statement containing the query to batch. The LCFD edge from $s4$ to $s1$ crosses the split boundary and hence $s1$ must be moved past $s4$. As can be seen in Figure 4.15, after the reordering, no LCFD

```
procedure moveAfter(Stmt s, Stmt t)
External variables used:
    List srcDeps, Stmt s_q      // Constructed by procedure reorder in Figure 4.9
begin
    if s succeeds t in the basic block
        return;
    Stmt next = successor(s);
    do {
        if no flow/anti/output dependence edges between s and next
            /* Reorder the statements by applying Rule-5A */
            swap s and next;
        else {
            for each OD_v edge from s to next { // OD_v: output dependence on variable v
                /* Shift the OD edge by applying Rule-5C */
                Replace writes to v in next by writes a new variable v';
                Insert a new statement as'_v that assigns v' to v immediately after next;
                moveAfter(as'_v, t);
            }

            for each AD_v edge from s to next { // AD_v: anti-dependence on variable v
                /* Shift the AD edge by applying Rule-5B */
                if there exists an AD_v edge from s_q to next      // Use a reader stub
                    Insert a new statement as'_v that assigns v to a new temp variable v'
                    immediately before s;
                    Replace all read references to v in s by v';
                else        // Use writer stub
                    Replace write of v in next by write to a new temp var v';
                    Insert a new statement as_v that assigns v' to v immediately after next;
                    moveAfter(as_v, t);
            }
            swap s and next;
        }
        lastStmt = next;
        if (lastStmt ! = t)
            next = successor(s);
    }
    while(lastStmt ! = t) ;
end
```

Figure 4.11: Procedure *moveAfter*

| | | |
|---|---|---|
| **while**(category != null) **loop**<br>(s1) icount = **q(category)**;<br>(s2) sum = sum + icount;<br>(s3) category = getParent(category);<br>**end loop**; | Move s1 past s3<br>$\Longrightarrow$ | **while**(category != null) **loop**<br>(ts1) category1 = category;<br>(s3) category = getParent(category);<br>(s1) icount = **q(category1)**;<br>(s2) sum = sum + icount;<br>**end loop**; |

Figure 4.12: Example-1 of Statement Reordering

```
while(top > 0 ) loop
(s6) top = top-1;
(s7) curcat = stack[top];
(s8) catitems = q(curcat);
(s9) totalcount = totalcount + catitems;
(s10') stack, top = block(curcat, top);
end loop;
```

Move s8 past s10' ⟹

```
while(top > 0 ) loop
(s6) top = top-1;
(s7) curcat = stack[top];
(s10') stack, top = block(curcat, top);
(s8) catitems = q(curcat);
(s9) totalcount = totalcount + catitems;
end loop;
```

Figure 4.13: Example-2 of Statement Reordering (from the UDF in Example 4.1)

```
while(pred(c)) loop
(s1) cv1? a = q(b);
(s2) cv2? a,c = f(x);
(s3) d = g(a, b);
(s4) cv3? a,b = h(c);
end loop;
```

Move s1 past s4 ⟹

```
while(pred(c)) loop
(s2) cv2? a3,c = f(x);
(n1) b2 = b;
(n2) b5 = b;
(s4) cv3? a1,b = h(c);
(s1) cv1? a = q(b5);
(n3) cv2? a = a3;
(s3) d = g(a, b2);
(n4) cv3? a = a1;
end loop;
```

Figure 4.14: Example-3 of Statement Reordering

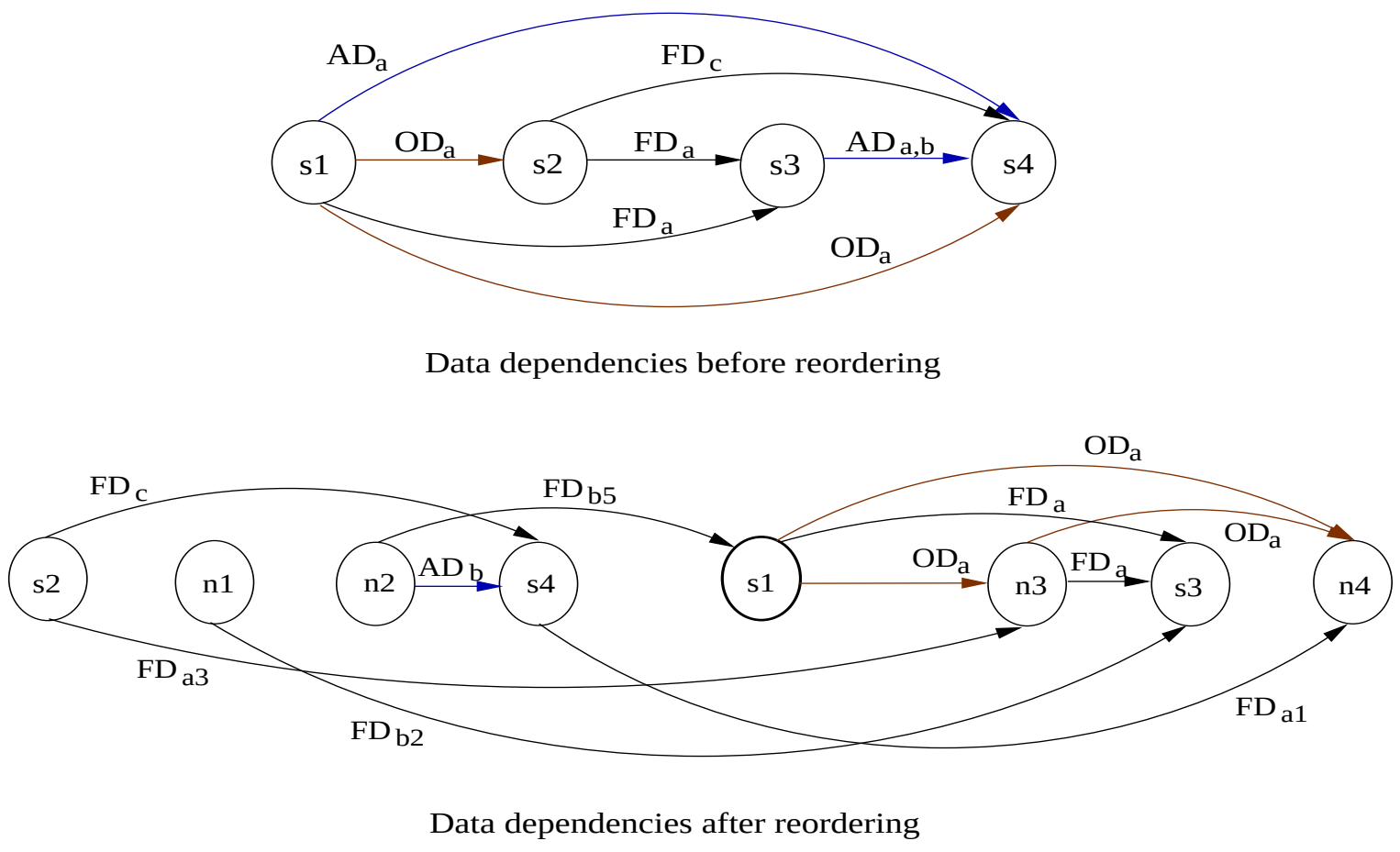

Figure 4.15: Data Dependence Graphs for the Example of Figure 4.14

edges cross the split boundary. The LCAD edges crossing the split boundary do not prohibit splitting of the loop, because these will be removed by the introduction of the *loop local table* while splitting the loop.

```
Table(key, a1, ..., b5, cv1, ..., cv3) t;
int loopkey = 0;
while(pred(c)) loop
      Record r;
(s2)  cv2? a3,c = f(x);
(s2') cv2? r.a3 = a3;
(n1)  b2 = b;  (n1') r.b2 = b2;
(n2)  b5 = b;  (n2') r.b5 = b5;
(s4)  cv3? a1,b = h(c);  (s4') cv3? r.a1 = a1;
      r.key = loopkey++;
      t.addRecord(r);
end loop

for each r by ref in t order by key loop
(s1)  r.cv1? r.a = q(r.b5);
end loop;

for each r by ref in t order by key loop
      boolean cv1, cv2, cv3;
      int a1, a3, b2;
      assigned(r.cv1)? cv1 = r.cv1;
      ...
      assigned(r.b5)? b5 = r.b5;

(n3)  cv2? a = a3;
(s3)  d = g(a, b2);
(n4)  cv3? a = a1;
end loop;
```

Figure 4.16: Loop Splitting Applied to Reordered Code in Figure 4.14

## 4.5 Applicability of Transformation Rules

Our program transformation algorithm succeeds in rewriting fairly complex programs for batched bindings. However, it may not be always be possible to rewrite a program to batch the invocation of a specific operation. As an example, consider the program and its DDG shown in Figure 4.17. We can batch the query invocation in statement `s2` but not the one in statement `s1`. The query invocation in statement `s1` lies on the true-dependence cycle `s1` $\xrightarrow{FD}$ `s4` $\xrightarrow{LFD}$ `s1` and hence we cannot reorder the statements so as to satisfy the pre-conditions of Rule-2. Similarly, in the DDG of Figure 4.2, the query invocation in statement `s8` is batchable, whereas the one in statement `s10` is not batchable.

Flow dependencies that result from control-dependencies (Rule 4) must be taken into account while checking for the presence of a true-dependence cycle. For instance,

```
s0: while(eid ! = NULL) loop
s1:     mgr =SELECT manager
                FROM emp WHERE empid=eid;
s2:     idx = SELECT perfindex FROM rating
                WHERE reviewer=mgr and reviewed=eid;
s3:     sumidx += idx;
s4:     eid = mgr;
    end loop;
```

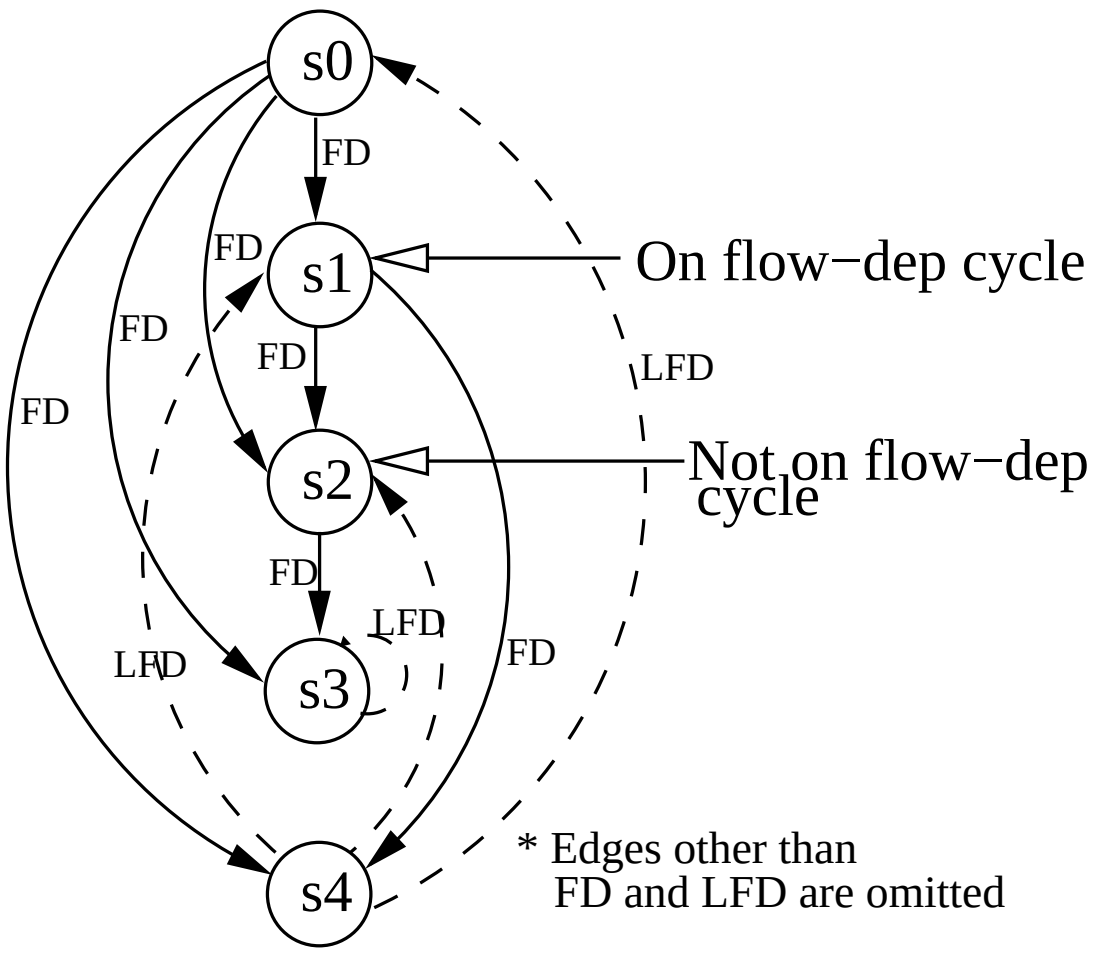

Figure 4.17: Cyclic True-Dependencies

statement `s2` in the code fragment of Figure 4.18 lies on true-dependence cycle after converting the control dependence of `s2` on `s0` into a flow dependence. Intuitively, an operation cannot be batched if the execution of the operation in any iteration depends on the value it returned in a previous iteration.

```
s0:  while(x < n) loop
s1:       y = y − 1;
s2:       x = q(y);
     end loop;
```

Figure 4.18: True-Dependence Cycle Created Due to Control-Dependency

## Condition for Batching

In this sub-section, we state and prove the condition under which procedure *reorder* can reorder the statements in a loop such that a given statement $s_q$ can be batched. Before presenting the theorem and its proof, we introduce some useful terms and prove a supporting lemma.

Let $b$ be the basic block of statements containing $s_q$, the statement to be batched. Let $C$ be the set of LCFD edges that cross the split boundaries of $s_q$ in the given basic block $b$. Let $T$ be the set of LCFD edges $(v_1, v_2)$ such that both $v_1$ and $v_2$ succeed $s_q$ in $b$. For example, in Figure 4.19, $C = \{(s_2, s_1), (s_3, s_q)\}$ and $T = \{(s_3, s_2)\}$.

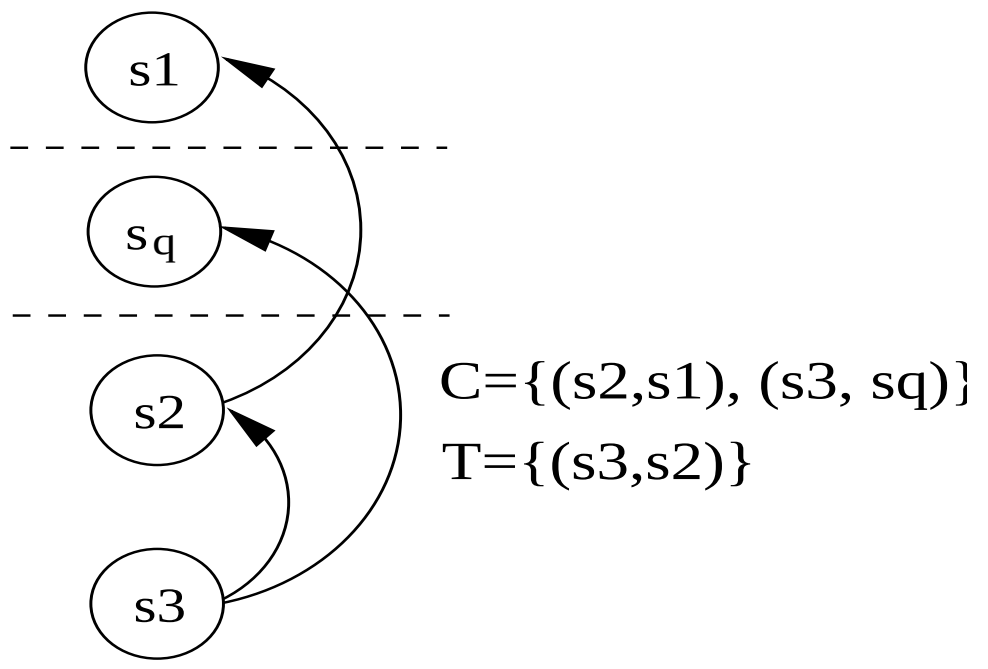

Figure 4.19: Example of Dependence Edges in the $C$ and $T$ Sets

We now state and prove a lemma, which characterizes the behavior of procedure *moveAfter*.

**Lemma 4.2** *Each call to procedure moveAfter(s, t), made from procedure reorder, satisfies the following pre-conditions and post-conditions.*

**Pre-conditions for procedure moveAfter(s, t)**

*(R.1) $t$ = targetStmt (the original target statement assigned in procedure reorder)*

*(R.2) No true-dependence path from $s$ to $t$ exists.*

*(R.3) No statement between $s$ and $t$ has a true-dependence path from $s$.*

**Post-conditions for procedure moveAfter(s, t)**

*(T.1) $|srcDeps|$ decreases by at least 1 if $s$ was in srcDeps.*

*(T.2) $|C| + |T|$ does not increase when $s \neq stmtToMove$, and $|C| + |T|$ decreases by at least 1 when $s = stmtToMove$ (here, stmtToMove is the original statement to be moved past the targetStmt, and is assigned in procedure reorder)*

*(T.3) Program correctness is preserved.*

*(T.4) The procedure terminates, and $s$ is moved past $t$.*

**Proof**: First, we prove that all the calls to procedure *moveAfter*, made from procedure *reorder* satisfy the three pre-conditions R.1, R.2 and R.3. We then prove that procedure *moveAfter* ensures the post-conditions whenever the pre-conditions hold.

(R.1) All calls made to procedure *moveAfter* from procedure *reorder* have the *targetStmt* as parameter $t$. All recursive calls within procedure *moveAfter* do not change parameter $t$. Hence, pre-condition R.1 holds.

(R.2) *(i)* If $s_q$ does not lie on a true-dependence cycle, then procedure *reorder* picks *stmtToMove* and *targetStmt* such that there exists no true-dependence path from *stmtToMove* to *targetStmt*. This can be observed in Figure 4.10.

*(ii)* Each statement $v$ in *srcDeps* has a true-dependence path from *stmtToMove* and hence there cannot be a true-dependence path from $v$ to *targetStmt* (otherwise, *stmtToMove* would be on a true-dependence cycle).

*(iii)* Observations *(i)* and *(ii)* above prove that pre-condition R.2 holds for all calls to procedure *moveAfter* made from procedure *reorder*.

(R.3) For all the calls to procedure *moveAfter*, which are made from procedure *reorder*, this condition is seen to hold from the way we pick statement $v$ (passed as argument $s$ for *moveAfter*).

**Post-conditions**

We now show the following: If the pre-conditions for procedure *moveAfter* are met then, the post-conditions and pre-conditions for all recursive calls will be met.

**Case I**: $s \neq stmtToMove$

*Case IA*: $t$ immediately follows $s$.

IA.1 *No flow/anti/output dependencies exist from $s$ to $t$.*

Procedure *moveAfter* swaps $s$ and $t$. Post-condition *(T.1)* holds because $s$ no longer lies between *stmtToMove* and $t$ and is accordingly removed from *srcDeps*. Post-condition *(T.2)* holds because no changes happen to $C$ and $T$ (because $s$ is a statement that lies after *stmtToMove*). The program correctness is preserved (Post-condition *(T.3)* holds) because no flow/anti/output dependencies exist between $s$ and $t$; the *do* loop (and hence the procedure) terminates after the first iteration and $s$ is moved past $t$ (Post-condition *(T.4)* holds).

IA.2 *A single output-dependence edge ($OD_v$) exists from $s$ to $t$.*

Transformations made by procedure *moveAfter* are shown pictorially in Figure 4.20. We first note that the recursive call to *moveAfter*$(as'_v, t)$ satisfies the pre-conditions for *moveAfter*. $as'_v$ appears after $t$ and hence the recursive call terminates. We then swap $s$ and $t$.

Post-conditions *(T.1)* and *(T.2)* hold for the same reasons stated for Case IA.1. Rewriting the LHS of $t$ with $v'$ makes $s$ and $t$ independent and hence the swap preserves correctness. The newly introduced copy-back statement, which assigns $v'$ to $v$, ensures later statements see the correct value of $v$. Hence, post-condition *(T.3)* (correctness) holds. The *do* loop (and hence the procedure) terminates after the first iteration and $s$ is moved past $t$ (Post-condition *(T.4)* holds).

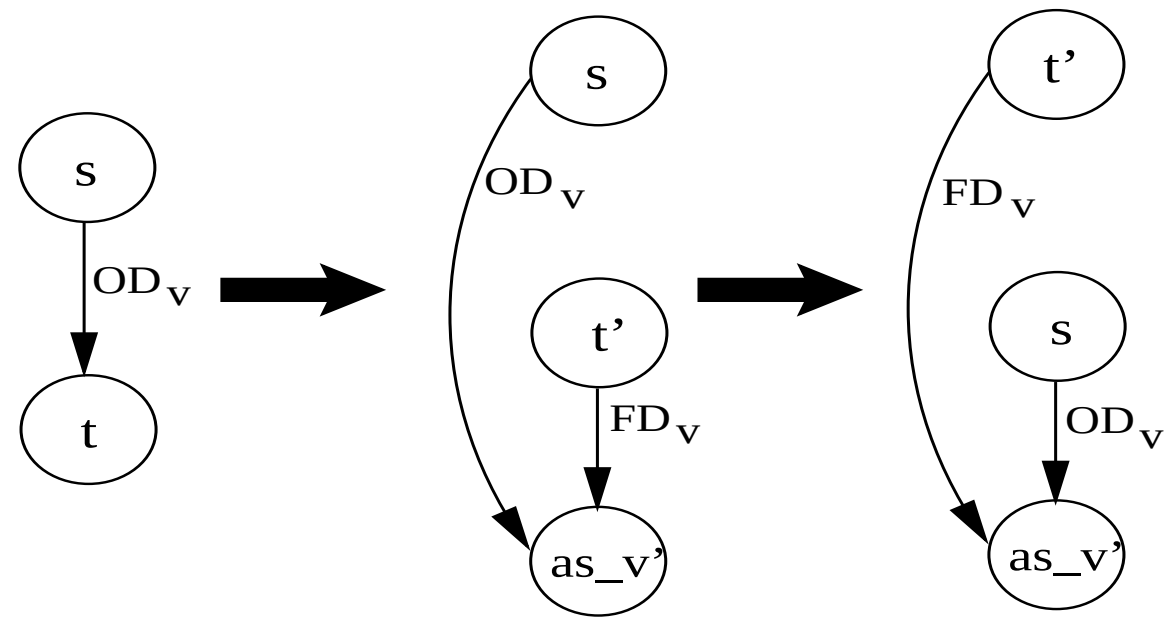

Figure 4.20: Pictorial Depiction of Reordering with OD Edge

IA.3 *A single anti-dependence edge ($AD_v$) exists from $s$ to $t$.*

Assume there exists an $AD_v$ edge from $s_q$ to *next*. Transformations that procedure *moveAfter* makes are shown pictorially in Figure 4.21. The newly introduced statement $as'_v$, which preserves the value of $v$ in $v'$, cannot be part of *srcDeps* for the following reasons. Assume $as'_v$ has a true-dependence path from *stmtToMove*, which in this case must be $s_q$. This implies either $s_q$ or some statement having a true-dependence path from $s_q$ writes $v$. Let this statement be $p$. This implies a true-dependence cycle involving $s_q$ and $p$ since $s_q$ reads $v$ and $p$ writes $v$. Since it is given that $s_q$ does not lie on a true-dependence cycle, we conclude that $as'_v$ cannot have a true-dependence path from $s_q$ and hence not part of *srcDeps*. Now, it is straight-forward to see that the post conditions hold.

Assume there exists no $AD_v$ edge from $s_q$ to *next*. Transformations that procedure *moveAfter* makes are shown pictorially in Figure 4.22. It is straight-forward to see that the pre-conditions for the recursive call hold and that the subsequent swap of $s$ and *next* will be correct and the transformation satisfies the post-conditions.

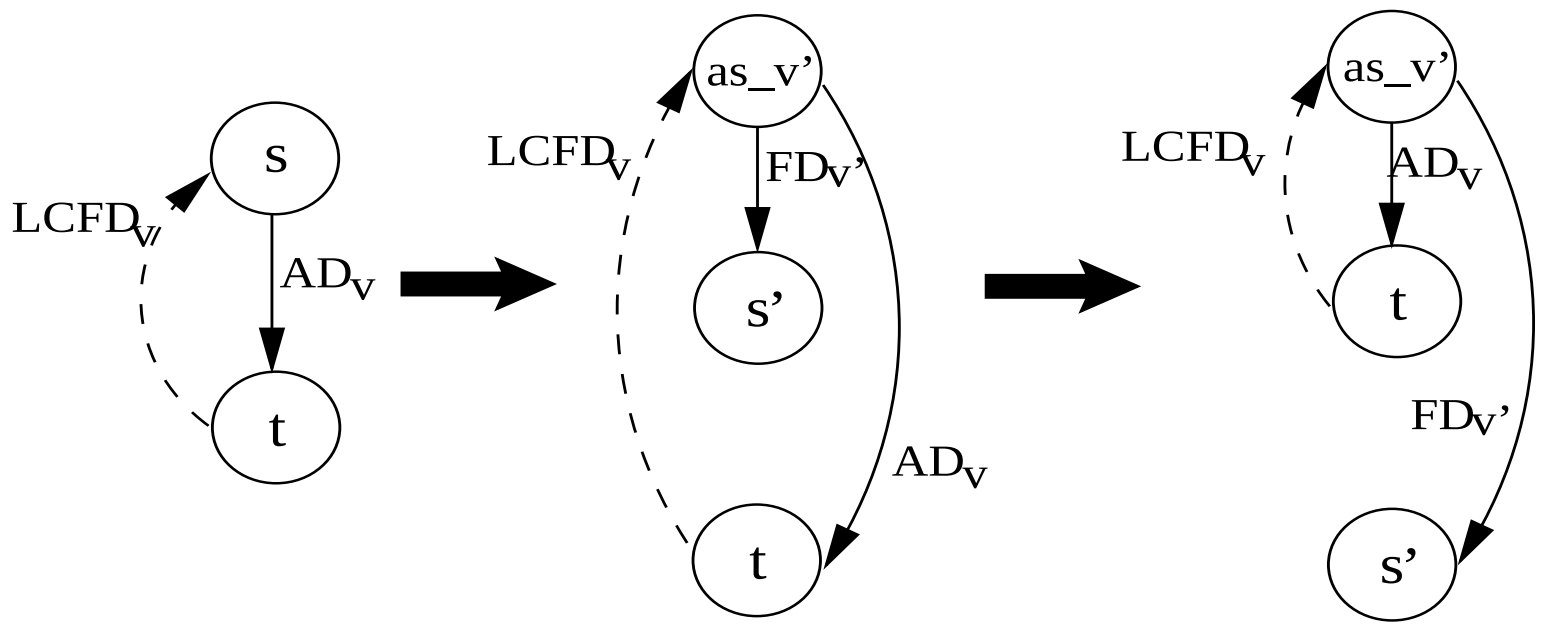

Figure 4.21: Pictorial Depiction of Reordering with AD Edge (R-Stub)

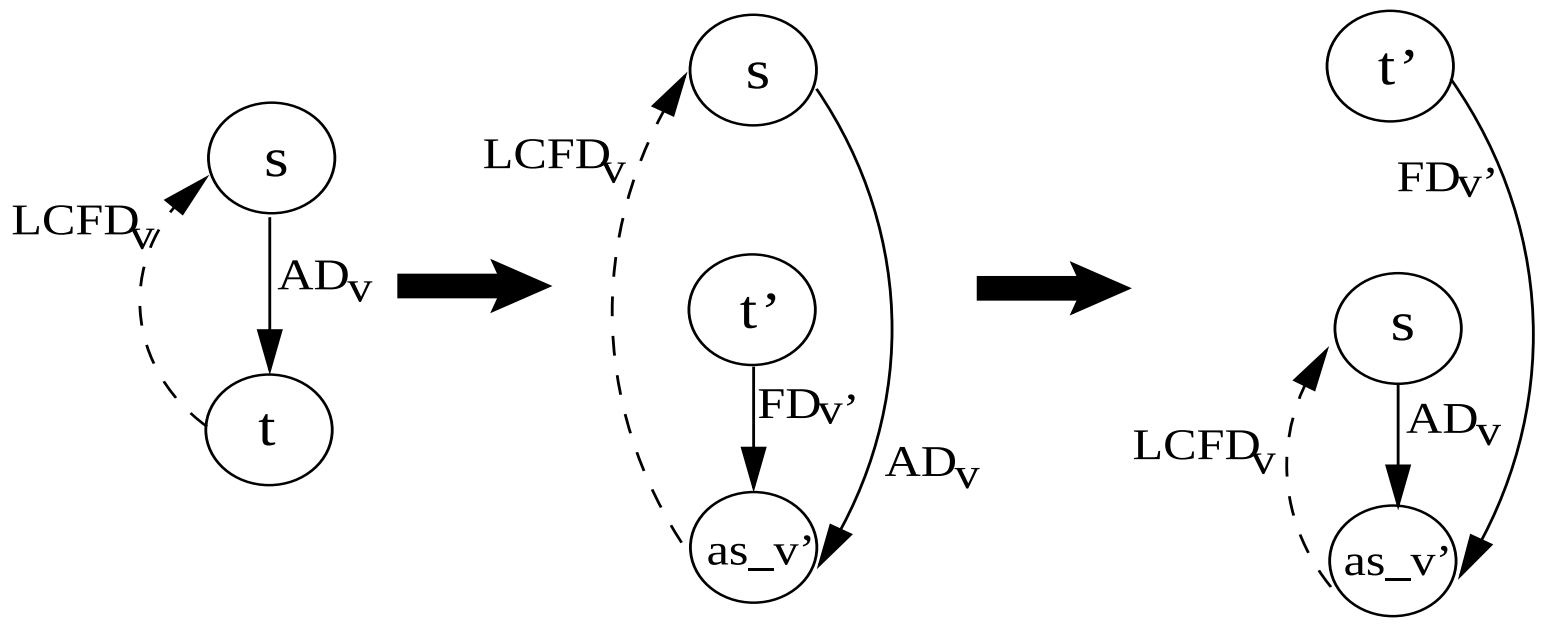

Figure 4.22: Pictorial Depiction of Reordering with AD Edge (W-Stub)

IA.4 *A flow-dependence edge exists from $s$ to $t$.*

This case is ruled out from pre-condition *R.2*.

IA.5 *Multiple AD and OD edges exist from $s$ to $t$.*

Each of the AD and OD edges can be treated independently as is done in procedure *moveAfter*. And we can see that the post-conditions continue to hold.

*Case IB*: $t$ does not immediately follow $s$ (there exist other statements between $s$ and $t$).

From pre-condition *R.3*, the only possible dependencies between $s$ and any statement $u$ between $s$ and $t$ are AD and OD. We can keep swapping $s$ and its successor till $s$ is moved past $t$. Each swap preserves the pre and post conditions as argued above.

**Case II**: $s = stmtToMove$

*Case IIA*: $t$ immediately follows $s$.

IIA.1 *No flow/anti/output dependencies exist from $s$ to $t$.*

Procedure *moveAfter* swaps $s$ and $t$. Let $k$ be the number of LCFD edges incident on $t$. Now, the change in the number of edges in sets $C$ and $T$ is as follows (see also Figure 4.10).

If *stmtToMove*=$s_q$: $|C_{new}| \leq |C_{old}| + k - 1$ and $|T_{new}| \leq |T_{old}| - k$

If *stmtToMove* $\neq s_q$: $|C_{new}| \leq |C_{old}| - k$ and $|T_{new}| \leq |T_{old}| + k - 1$

In both the above cases, $|C|+|T|$ decreases by at least 1. Hence we can see that post-condition *(T.2)* holds. Post-condition *(T.1)* is satisfied since *stmtToMove* is not in *srcDeps*. The swap preserves the correctness since there are no flow/anti/output dependencies from $s$ to $t$ (post-condition *(T.3)* holds). Post-condition *(T.4)* holds as argued in IA.1.

IIA.2 *A single output-dependence edge ($OD_v$) exists from $s$ to $t$.*

The resulting dependencies after the transformation are pictorially shown in Figure 4.20. It can be seen that, $|C| + |T|$ decreases after the transformation.

IIA.3 *A single anti-dependence edge ($AD_v$) exists from $s$ to $t$.*

The AD edge is moved using a reader stub if $s = s_q$ and is moved using a writer side stub otherwise. The dependencies after this reordering are shown in Figures 4.21 and 4.22. Again, it can be seen that $|C| + |T|$ decreases after the transformation.

IIA.4 *A flow-dependence edge exists from $s$ to $t$.*

This case is ruled out from pre-condition *R.2*.

IIA.5 *Multiple AD and OD edges exist from $s$ to $t$.*

Each of the AD and OD edges can be treated independently as is done in procedure *moveAfter*. And we can see that the post-conditions continue to hold.

*Case IIB*: $t$ does not immediately follow $s$ (there exist other statements between $s$ and $t$).

This case is similar to *Case IB*, except that when $s$ is eventually moved past $t$, $|C| + |T|$ strictly decreases. Hence the post-conditions hold.

This completes the proof of Lemma 4.2. □

We now state and prove our main theorem.

**Theorem 4.3** *Given a basic block of code $b$ and statement $s_q$ in $b$ such that $s_q$ does not lie on a true-dependence cycle in the DDG, procedure reorder terminates, reordering the statements of $b$ such that:*

*(a) No LCFD edges cross the program points that immediately precede and succeed $s_q$.*

*(b) Program correctness is preserved (*i.e.*, the reordered block is equivalent to the original)*

**Proof**: First, we prove that procedure *reorder* terminates.

1. The inner *while* loop of procedure *reorder* executes until the set *srcDeps* is empty. In each iteration, a call to procedure *moveAfter* is made, with an element of *srcDeps* being passed as the parameter $s$. From the post-condition *(T.1)* of Lemma 4.2 we know that, after each call to *moveAfter*, the size of *srcDeps* decreases by at least 1. Hence, the inner *while* loop of procedure *reorder* must terminate.

2. The outer *while* loop of procedure *reorder* executes until $|C| = 0$. In each iteration of the outer *while* loop, the last call to procedure *moveAfter* is made with $s$ being set to $stmtToMove$. Hence, from the post-condition *(T.2)* of Lemma 4.2 we know that $|C| + |T|$ decreases by at least 1 at the end of each iteration of the outer *while* loop. None of the other calls to the procedure *moveAfter* cause any increase in $|C| + |T|$ (follows from the post-condition *(T.2)* of Lemma 4.2). Hence, the outer *while* of procedure *reorder* must terminate.

3. Each call to procedure *moveAfter*, made from procedure *reorder*, terminates (post-condition *(T.4)* in Lemma 4.2).

From 1, 2 and 3 above, it is straight-forward to observe that procedure *reorder* terminates.

Any reordering of statements within the basic block $b$ are performed by procedure *moveAfter* and not directly by procedure *reorder.* Hence, the post-condition *(T.3)* of Lemma 4.2 suffices to prove that procedure *reorder* preserves program correctness.

When procedure *reorder* terminates, no LCFD edges cross the program points that immediately precede and succeed $s_q$. This directly follows from the termination condition, $|C| = 0$, of the outer *while* loop of procedure reorder.

This completes the proof of Theorem 4.3 □

## 4.6 Experimental Results

Our rewrite rules can conceptually be used with any language. However, to implement the rules we need to perform dataflow analysis of a program and build the data dependence graph. For our implementation, we chose Java, since tools for its dataflow analysis are available in public domain. Our implementation uses the *SOOT* optimization framework [52]. *SOOT* uses an intermediate code representation called *Jimple* and provides dependency information on *Jimple* statements. Our implementation transforms the *Jimple* code using the dependence information. Finally, the *Jimple* code is translated back into a Java program.

Our current implementation requires that queries and updates be performed using our API layer built on top of JDBC. During rewrite we recognize these calls and transform them for batched bindings when possible. We have not yet implemented query rewriting to get batched forms and this step is done manually. The techniques for deriving batched forms of queries are well known and we expect the implementation to be straight-forward. Tables (batches) used in the rewritten procedures are constructed in-memory and transferred to the database before evaluating the batched queries. Nest/unnest and merge operations are performed on these in-memory tables.

There are no benchmarks for procedural SQL that we could use for our experiments. However, we had seen three real-world applications which were facing serious performance problems due to non-set-oriented execution, which were affecting their usability. We use these scenarios for our experiments. Our current implementation does not support all the transformation rules presented in this chapter. Hence, in some cases part of the rewriting was performed manually in accordance to the transformation rules. We do not have access to the actual data used in these applications and hence we used synthetic data. In one case we used TPC-H data as it matched the scenario. As we cannot report timings on the actual application code, we used independent programs having only the code required for the specific scenarios. The experiments were performed on a widely used commercial database system (we call it SYS1) running on an Intel P4 (HT) PC with 1GB of RAM.

### Experiment 1: Traversal of Category Hierarchy

For this experiment, we used a program, which is a slight variant of the UDF in Example 4.1. The program finds the item (part) with maximum size under a given category (including all its sub-categories) by performing a DFS of the category hierarchy. For each node (category) visited, the program queries the *item* table. The TPC-H *part* table, augmented with a new column *category-id* and populated with 2 million rows, was used as the *item* table. The *category* table had 1000 rows - 900 leaf level, 90 middle level and 10 top level categories (approximately). A clustering index was present on the *category-id* column of the *category* table and a secondary index was present on the *category-id* column of the *item* table. All relevant statistics were built. Figure 4.23 shows the performance of the program before and after rewrite.

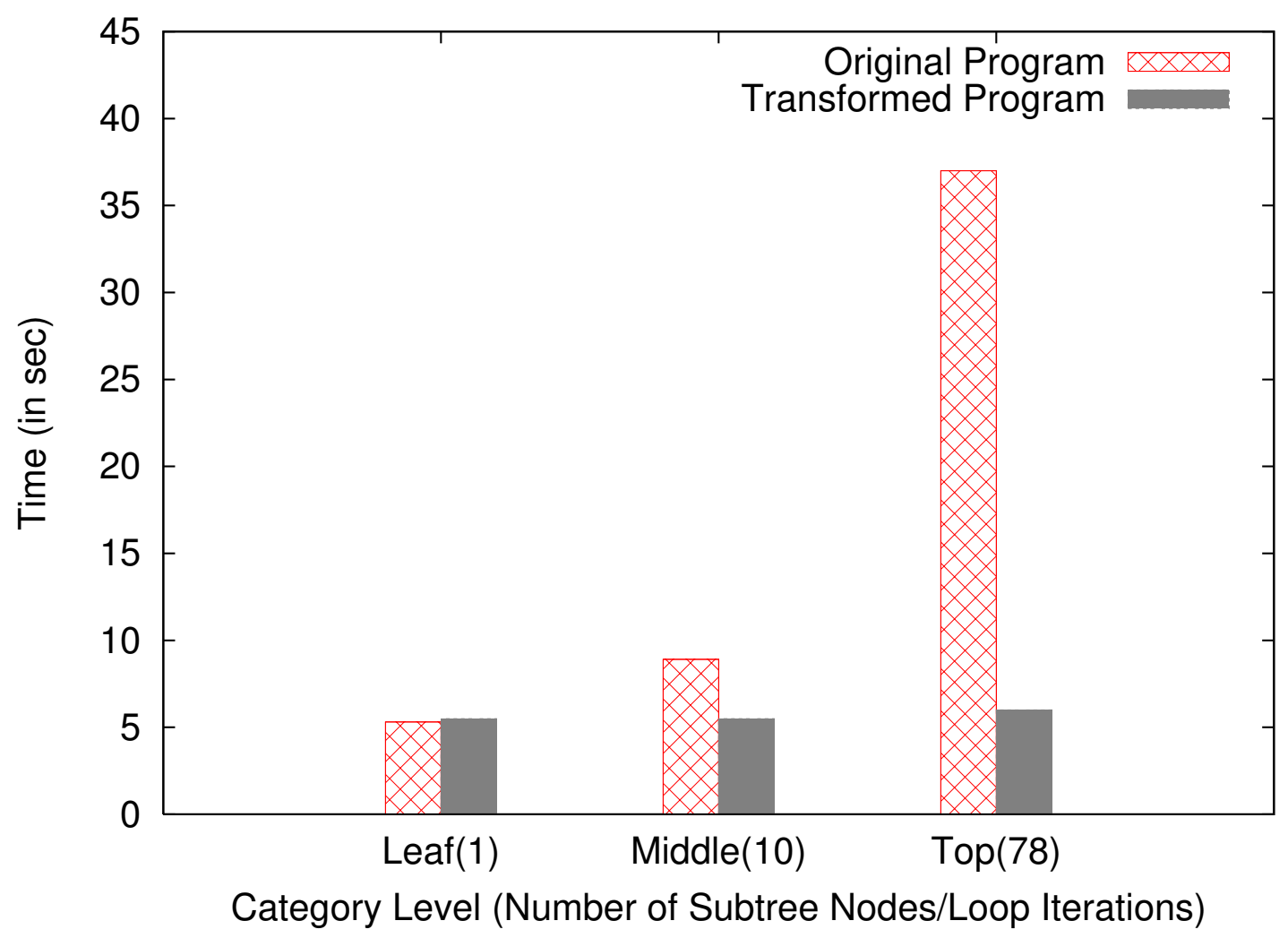

Figure 4.23: Performance Results for Experiment 1

For the non-batched query on the *item* table, SYS1's default choice was to use the secondary index. This plan results in a lot of random IO, and we found an alternative plan, which performs a sequential scan takes less time since the entire relation is brought into memory on the first invocation, and there is no IO on subsequent invocations. Since this plan was found to be cheaper, we enforced it using optimizer hints. Figure 4.23 compares the time taken by the best plan for the original program with the batched version. The batched version, in this case, performed a *group-by* followed by a join, whereas the original program repeatedly executed a query that performed selection followed by group by; as a result the batched version showed much better performance. In this experiment, the performance of the batched form was almost independent of the batch size because the group-by query computed the results for all the parameters in each case.

In transforming the program, Rule-5 (reordering), Rule-2 (loop splitting) and Rule-1 (batching) were applied in that order. There was a 12.5% increase in the program size (lines of code) due to the transformation.

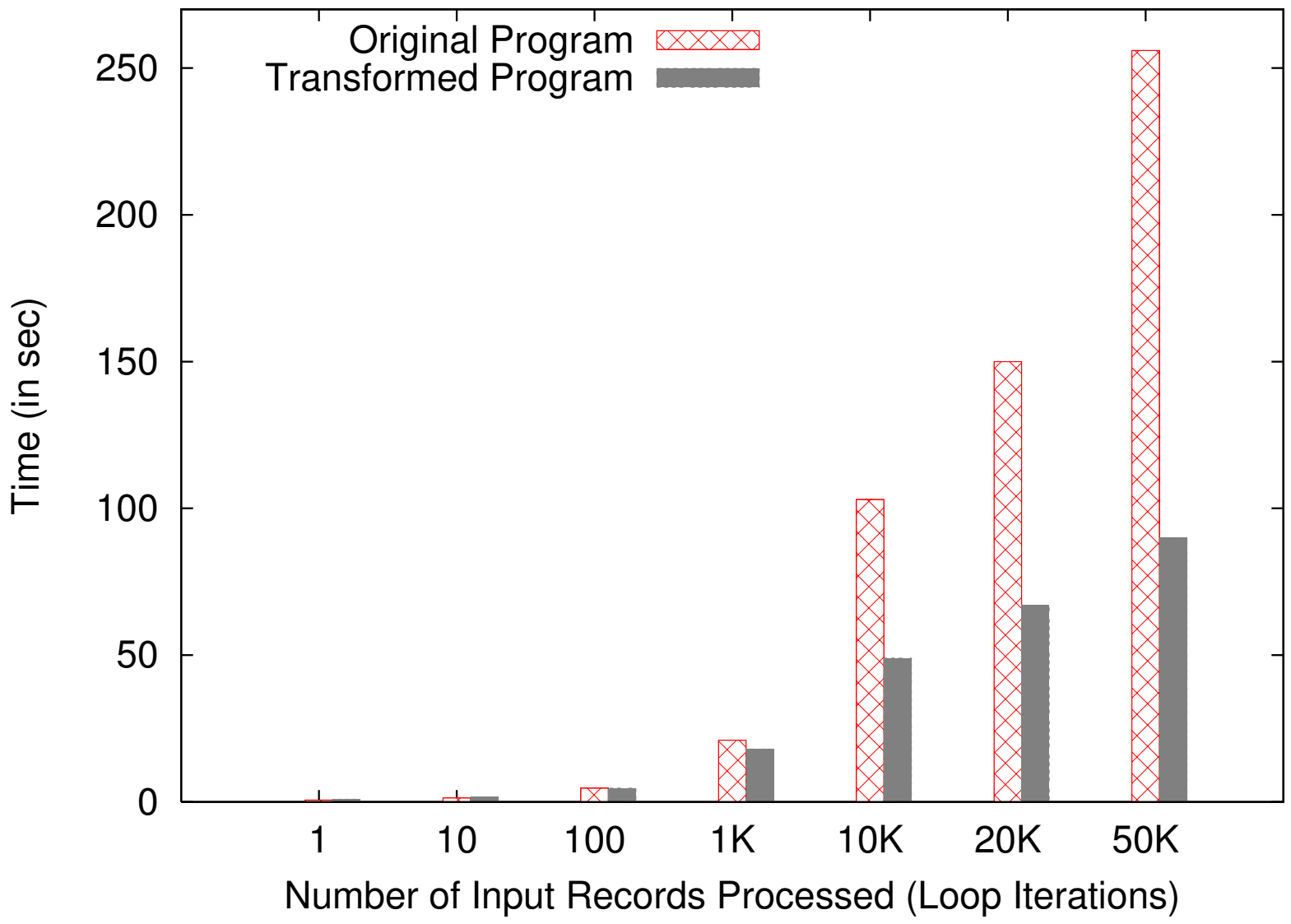

Figure 4.24: Performance Results for Experiment 2

## Experiment 2: ESOP Management Application

For this experiment, an application used for managing stock option grants of multiple organizations was considered. During each *upload* operation, a large number of records from a delimited file were processed. The application performed a mix of queries, inserts and updates. A brief outline of the program logic is given below to indicate the complexity of control-flow involved.

For each record read from the input file, the program performs validation and pre-processing of the fields and then queries the *options* table to check if a record for the person already exists. The query predicate is parameterized on the values read from the input record. If a record is present, the old values of the various fields and the *internal-emp-id* are obtained as part of the same query. Further, the *contactinfo* table is queried using the *internal-emp-id* to obtain contact info fields. If the input record being processed has empty values for any of the fields, the old values (when present) are copied to those fields. Finally, new records are inserted or existing records updated in both *options* and *contactinfo* tables. Figure E.2 in Appendix E shows the procedure.

The rewritten program used an outer join in place of iterative selections, and performed batched updates and inserts. Figure 4.24 compares the performance of the rewritten program with the original program for varying number of input records.

In transforming the program, Rule-4 (control-dependencies to flow-dependencies), Rule-5 (reordering), Rule-2 (loop splitting) and Rule-1 (batching) were applied. After transformation there was a 17% increase in the program size.

### Experiment 3: Value Range Expansion

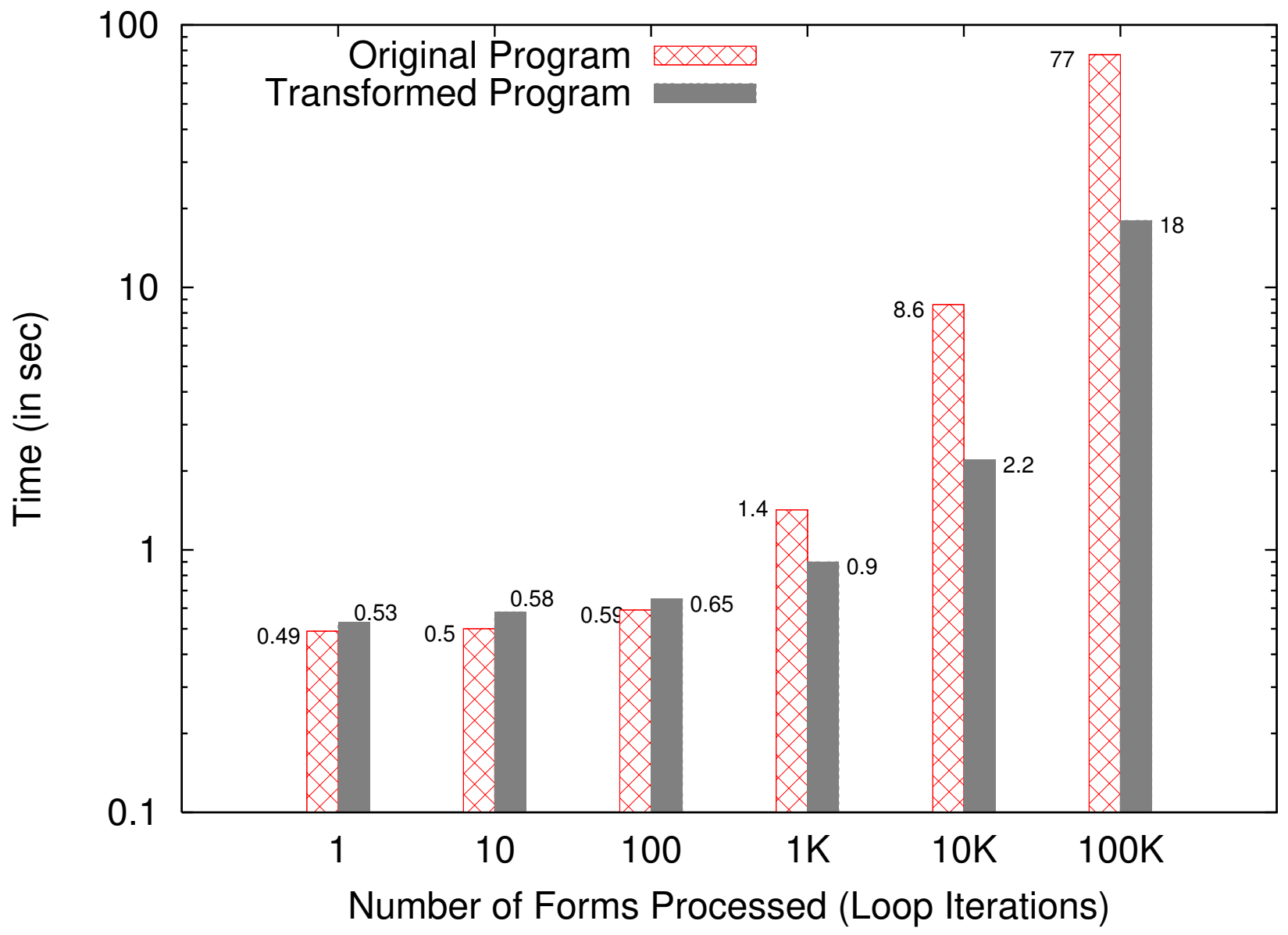

Figure 4.25: Performance Results for Experiment 3

In this application, data about *forms* issued to various agents would arrive in the format *(agent-id, start-form-number, end-form-number)*. The program (shown in Figure E.1 of Appendix E) would iterate over all the *form issue* records, expand the *issue* range and populate the *forms-master* table with entries corresponding to each individual form. The purpose was to be able to update and track the status of each individual form subsequent to its issue. The original program had an outer loop iterating over the *form issue* records and an inner loop iterating over the range *(start-form-number, end-form-number)*. An INSERT operation was performed inside the inner loop. The transformed program could pull the *insert* operation out of both the loops and perform a batched insert. The running times of the original and transformed program are shown in Figure 4.25. The batched version performs much better for large batch sizes. For small batch sizes (less than 1K) the computational overheads due to batch creation and nest/unnest operations cause the batched version to perform marginally slower than the original program.

In transforming the program, Rule-5 (reordering), Rule-2 (loop splitting), Rule-1 and Rule-6 (batching) were applied. The increase in program size was 16.5%.

### Time Taken for Program Transformation

Although the time taken for program transformation is usually not a concern (as it is a one-time activity), we note that, in our experiments the program transformation took very little time (less than a second).

## 4.7 Related Work

Queries such as those shown in Example 1.3 and Example 4.1 can be thought of as nested queries with complex inner blocks. The inner block in such cases contains subqueries embedded in procedural code. Known decorrelation techniques such as [31, 19, 49, 17, 9, 12] cannot be used to unnest such queries. The techniques proposed in this chapter make it possible to rewrite a procedure so as to enable set-oriented evaluation of the embedded sub-queries through batched bindings. This is an essential step in decorrelation [49]. Further optimizations, such as pipelining the output of the expression that produces the parameter batch into the expression that consumes it, are possible and should be considered for future work. Graefe [22] highlights the benefits of batched bindings for speeding up index nested loops joins. Batched bindings not only help in performing IO efficiently, but can also make it possible to employ a set-oriented strategy at the operator level. *Magic sets* decorrelation [49] employs parameter batches for decorrelation of nested queries. Our techniques can be thought of as extending this approach for complex procedures. Our work is a step towards combining query optimization with program analysis and transformation techniques; we believe this combination will give significant benefits for database applications.

Lieuwen and DeWitt [34] consider the problem of optimizing set iteration loops in database programming languages. Their techniques can convert nested set iteration loops into joins. However, their work does not address the issue of batching procedure calls. The program transformation rules in this chapter can work with complex control-flow including *if-then-else* and *while* loops. Rule-2 in this chapter is a more general version of Rule T4 in [34]. An earlier work similar to [34] is [50]. It describes a programming language called Theseus for manipulating relational databases, and proposes optimization transformations, some of which are special forms of those presented in [34]. To the best of our knowledge, reordering of statements in a loop to facilitate batched in-vocations of queries is not considered before. The works of Katz and Wong [27], and Demo and Kundu [10] address the problem of converting programs which make use of CODASYL FIND-statements into equivalent programs using relational queries. Similar to our work, their solution employs dataflow analysis. The aim of dataflow analysis in their work is to find dependencies between CODASYL FIND-statements so as to group the statements that access the same logically definable set of records. Fegaras [14] addresses query optimization in object oriented databases, in the presence of object identity and in-place updates. Poulovassilis and Small [43] consider algebraic optimization of declarative, functional database programming languages. Their work considers a computationally complete functional language and addresses issues such as termination, infinite data structures, but does not consider imperative language constructs and side-effects. Ceri and Widom [4] address the problem of deriving production rules for incremental maintenance of materialized views. The generated rules are set-oriented in the sense that they can process a set of changes performed on the base tables at once. In their work, the view definitions are assumed to be in SQL without any procedural constructs.

Some of the program transformation techniques we employ are derived from those proposed in the area of parallelizing compilers [29, 45, 2]. However, the problem of batching differs from the problem of parallelizing in the following ways: *(i)* presence of flow-dependencies (described in Section 4.2) does not allow parallelization. However, batching is possible even if the order of two operations cannot be changed due to flow-dependencies, and *(ii)* as the aim of batching is to improve the performance of expensive IO bound queries and other database operations, it may be acceptable for the transformations to introduce additional CPU operations or use extra memory to make batching possible. However, such approaches do not generally yield significant benefits in the context of parallelizing the instructions and are not considered to the best of our knowledge.

## 4.8 Summary

In this chapter we considered parameter batching as a means to improve performance of iteratively invoked database procedures and user-defined functions. We presented a technique, based on program analysis and transformation, to automate the generation of batched forms of procedures and to replace calls to stored procedures within imperative program loops with a call to the batched form. Our implementation and performance study show the practicality and usefulness of the the proposed techniques.

Procedural extensions to SQL offer new challenges and opportunities for query optimization. To deal with these challenges query optimization must be augmented with program analysis and transformation techniques. Our work is a step towards combining query optimization with program analysis and transformation techniques; we believe this combination will give significant benefits for database applications.

# Chapter 5

# Conclusions and Future Work

In this thesis we looked at ways to speed up iterative execution of queries, user-defined functions and stored procedures. Iterative execution can happen either because known decorrelation techniques do not apply, for instance when the nested block is a complex procedure, or because an iterative plan has the least expected cost amongst the alternatives available to the query optimizer. We took a two pronged approach to address the problem. First, we looked at exploiting ordered parameter bindings to speed up iterative plans and presented extensions to a cost-based optimizer for choosing efficient sort orders. Next, we showed how to achieve set-oriented execution of queries and updates within complex procedures through parameter batching enabled by program rewrite.

In Chapter 2 we showed how state retention of operators allows us to exploit sorted parameter bindings to improve the efficiency of iterative query plans. We then showed how a Volcano style cost-based query optimizer can be extended to take into account state retention and effects of sorted parameter bindings. An important problem, which arises in optimizing both nested queries as well as queries containing joins, grouping and other set operations is that of deciding the optimal sort order of parameters and inputs. We addressed this problem in Chapter 3. We showed that even a simplified version of the problem of choosing optimal sort orders is *NP-Hard* and gave principled heuristics, which take into account partially available sort orders and attributes common to multiple join predicates. We presented experimental results, carried out by forcing plans generated by our optimizer extensions on widely used database systems. The results showed significant improvements in actual query execution time due to the proposed techniques when compared to the default plans chosen by the respective systems. Although it is dif-ficult to quantify the percentage of real-world queries that can benefit from the proposed techniques, our experiments carried on queries taken from the TPC-H benchmark and a financial application, and our experience with ETL tasks in data warehousing indicate that a significant number of queries can benefit from the proposed techniques.

In Chapter 4 we considered iterative execution involving complex user-defined functions and stored procedures, which use a mix of procedural constructs and SQL. We proposed a program analysis and transformation based approach to enable set-oriented execution of queries and updates inside such procedures. The approach consists of a set of program transformation rules, which are used to *(i)* automatically generate the batched form of a given procedure and *(ii)* replace iterative calls to the procedure with code to construct a parameter batch and invoke the batched form. Our program transformation rules can deal with a rich set of language constructs such as looping, conditional control transfer and assignments. With the help of our implementation of the rewrite rules for a subset of Java, we carried out an experimental study on cases chosen from three real-world applications. The results are very promising, and show up to 75% improvement in the actual execution time.

With increasing use of procedural extensions to SQL and emergence of language integrated querying paradigms (*e.g.,* Microsoft LINQ [35]), combined optimization of application code and database queries/updates becomes more and more important for improving the application performance. Our work is a step towards this goal.

The proposed program rewrite techniques are useful in two broad scenarios: (a) to transform programmer written loops with database access into potentially more efficient code and (b) to automatically generate batch processing routines from routines that were developed for one at a time request processing. There is a large body of code in languages like PL/SQL and Java, in the former category. For our experiments we considered two such examples (Experiment 1 and 3 in Chapter 4) from real-world applications[1]. Although it is hard to quantify the number of cases which have the second requirement, we have come across two cases in which many stored procedures were required to be rewritten for batch processing. One such example is considered in Experiment 2 of Chapter 4.

Whether batching a specific query invocation yields significant benefits depends on the number of iterations, data characteristics and the physical design of the database.

[1] Due to confidentiality reasons we cannot name the applications and organizations.

Applying the program transformation rules presented in this thesis in a cost-based manner is a future work.

## Future Work

There are several interesting challenges ahead to fully realize the goal of optimizing iterative invocation of complex procedural blocks and iterative invocation of queries/updates from application code. We briefly discuss some directions for future work.

### Sort Orders from Secondary Indices

In Chapter 3, while computing the set of approximate favorable orders on base relations, we consider only the clustering index and secondary indices that cover the given query. Although it is possible to use non query covering secondary indices to get ordered tuples (by traversing the index leaf pages), it is usually very inefficient due to random IO. However, if the actual tuple fetch can be deferred until a point where the extra attributes are needed in the query plan, it is possible that the approach can perform better. If a highly selective filter discards many rows before the extra attributes are needed, only a few tuple fetches happen. Evaluating such alternatives in a cost-based manner is an interesting future work.

### Cost-Based Choice of Queries to Batch

Our current implementation requires manual input on which operations to consider for batching. An interesting and important problem is to make a cost-based decision on which queries/updates to batch. The important parameters on which this decision depends include *(i)* cost model for the operation as a function of batch size, *(ii)* expected number of iterations of the program loop *(iii)* branch probabilities for the branching statements *(if-then-else)* in the program and *(iv)* overheads of the transformed code.

### Pipelined Execution of Queries Inside a UDF

The $L$-tables, discussed in Section 4.3, serve to hold the parameter batch with which the batched form of a procedure is invoked. Though small batches can be held in memory, in general we may need to materialize the batches and the cost of materialization must be taken into account while deciding to batch an operation. However, for procedures that run entirely inside the database engine (*e.g.,* UDFs) it may be possible to avoid materialization of batches by constructing a single dataflow containing both relational and procedural nodes. Our loop splitting transformation is designed to facilitate such an approach. Appendix C contains a few additional rules that can be used for *(i)* avoiding creation of intermediate batches by passing a relational expression, instead of a table, to the batched forms and *(ii)* mapping program statements that perform simple and inexpensive operations (*e.g.,* expression evaluation and variable assignment) to operators that work on sets. The later of these helps in eliminating loops such as the one left over in Example 4.4. When a single dataflow is thus built, code that cannot be mapped to relational operators is executed by procedural nodes in the dataflow. Variable bindings inside such nodes are obtained from input tuples.

**Set-Oriented Evaluation in the Presence of Cyclic Flow-Dependencies**

The program transformation rules presented in Chapter 4 and the rule application algorithm guarantee that a query execution statement $s$ can be batched *w.r.t.* a loop if $s$ does not lie on a true-dependence cycle in the DDG (Theorem 4.3). Intuitively, the only statements we fail to batch are the statements whose execution in an iteration of the loop depends on the result of their own execution in a previous iteration. In such cases, the set of parameters for a query execution statement cannot be computed up front (without invoking the query itself).

However, in certain cases, it may be possible to identify a superset of the parameter batch even if the query execution statement lies on a true-dependence cycle. For example, consider the iterative execution of a scalar aggregate query with a parameterized equality predicate in its *where* clause. An example of such a query would be: SELECT sum(balance) FROM account WHERE branch=:branch. If a true-dependence cycle makes it impossible to compute the exact set of parameters (in this example, *branches*) for the query, one can compute the query results for all the possible parameters by using a vector aggregate query, which for our example would be: SELECT sum(balance) FROM account GROUP BY branch. Note that such an approach can be used only for pure functions and not for operations with side-effects (even if the operation is batch-safe).

# Appendix A

# Optimality with Minimal Favorable Orders

The notion of *minimal favorable orders*, introduced in Section 3.3, served as the basis for our heuristics for selecting sort orders. Since it is hard to compute the exact set of minimal favorable orders, we used a heuristic approach to compute them approximately. However, it is interesting to study the properties of minimal favorable orders. In this section we give a proof of Theorem 3.6 stated earlier in Section 3.3.2. The theorem essentially states the following: to identify an optimal sort order, it is sufficient to consider only the minimal favorable orders and not the full set of favorable orders. Below, we repeat the formal statement of the theorem and present a proof. The proof makes use of notation introduced in Sections 3.1.1 and 3.3.1.

**Theorem 3** *The set $\mathcal{I}(e, o)$ computed with exact ford-min contains an optimal sort order $o_p$ for the optimization goal $e = (e_l \bowtie e_r)$ with $(o)$ as the required output sort order.*

We prove Theorem 3.6 under the following assumption: If $o_1$, $o_2$ are two sort orders on the same set of attributes (*i.e.,* $\text{attrs}(o_1) = \text{attrs}(o_2)$), then the CPU cost of sorting the result of an expression $e$ to obtain $o_1$ will be same as that for $o_2$, *i.e.,* $\text{cpu-cost}(e, o_1) = \text{cpu-cost}(e, o_2)$.

**Proof**: Consider the optimization goal for a join expression ($e = e_l \bowtie e_r$, with $(o)$ as the sort order required on the result of $e$. Let $S$ be the set of join attributes and $o'$ be any sort order on $S$. The cost of the best merge-join plan for $e$, when $o'$ is chosen as the sort order for $e_l$, $e_r$, is given by:

$$\begin{aligned} PC(e, o, o') &= cbp(e_l, o') + cbp(e_r, o') + coe(e, o', o) + CM(e_l, e_r), \\ &\quad \text{where } CM(e_l, e_r) \text{ is the cost of merging.} \end{aligned} \tag{A.1}$$

In Equation A.1, we note that $CM(e_l, e_r)$ is independent of the sort order $o'$.

Let $o_b$ be an *optimal sort order* for $el \bowtie e_r$. Assume $o_b \notin \mathcal{I}(e)$. We show that $\exists o_p \in \mathcal{I}(e)$ such that $PC(o_p)$ =$PC(o_b)$.

**Case 1:** Suppose $o_b$ is such that $o_b \notin ford(e_l) \cup ford(e_r)$.

$$\begin{aligned} PC(e, o, o_b) &= cbp(e_l, o_b) + cbp(e_r, o_b) + coe(e, o_b, o) + \\ &\quad CM(e_l, e_r) \qquad \text{(A.2)} \\ &\quad \text{Since, } o_b \notin ford(e_l) \cup ford(e_r) \text{ we can write} \\ &= cbp(e_l, \epsilon) + coe(e_l, \epsilon, o_b) + cbp(e_r, \epsilon) + \\ &\quad coe(e_r, \epsilon, o_b) + coe(e, o_b, o) + CM(e_l, e_r) \qquad \text{(A.3)} \end{aligned}$$

Let $o_p$ be a sort order in $\mathcal{I}(e)$ such that $o \wedge S \leq o_p$, where $o$ is the required output sort order in the optimization goal. The existence of such a sort order in $\mathcal{I}(e)$ directly follows from the construction of $\mathcal{I}(e)$, specifically, steps 1 and 2 in Section 3.3.2.

Since both $o_b$ and $o_p$ are sort orders on the same attribute set $S$, we have

$$coe(e_l, \epsilon, o_b) = coe(e_l, \epsilon, o_p) \text{ and } coe(e_r, \epsilon, o_b) = coe(e_r, \epsilon, o_p) \tag{A.4}$$

Substituting Equation A.4 in Equation A.3 we get:

$$\begin{aligned} PC(e, o, o_b) &= cbp(e_l, \epsilon) + coe(e_l, \epsilon, o_p) + cbp(e_r, \epsilon) + \\ &\quad coe(e_r, \epsilon, o_p) + coe(e, o_b, o) + CM(e_l, e_r) \qquad \text{(A.5)} \\ &\geq cbp(e_l, o_p) + cbp(e_r, o_p) + coe(e, o_b, o) + \\ &\quad CM(e_l, e_r) \qquad \text{(A.6)} \end{aligned}$$

As $(o \wedge S) \leq o_p$, we have $(o_b \wedge o) \leq (o_p \wedge o)$ (because $o_b$ is a permutation of $S$). Therefore, $coe(e, o_b, o) \geq coe(e, o_p, o)$. From this, we can rewrite Equation A.6 as:

$$\begin{aligned} PC(e, o, o_b) &\geq cbp(e_l, o_p) + cbp(e_r, o_p) + coe(e, o_p, o) + CM(e_l, e_r) \\ &\geq PC(e, o, o_p). \end{aligned}$$

By assumption $o_b$ is an optimal sort order. So we conclude $PC(e, o, o_b) = PC(e, o, o_p)$. In other words, $\mathcal{I}(e)$ contains a sort order $o_p$ having the same plan cost as the optimal sort order $o_b$.

**Case 2:** Suppose $o_b$ is such that $o_b \in ford(e_l)$ or $ford(e_r)$ but not both.
Without loss of generality we assume $o_b \in ford(e_l)$. This implies one of the following:

(i) $\exists o' \in \textit{ford-min}(e_l)$ such that $o_b \leq o'$ and $cbp(e_l, o_b) = cbp(e_l, o')$ or

(ii) $\exists o' \in \textit{ford-min}(e_l)$ such that $o' \leq o_b$ and $cbp(e_l, o') + coe(e_l, o', o_b) = cbp(e_l, o_b)$.

We now consider, each of these cases separately.
**Case 2-A**: Suppose condition (i), repeated below as Equation A.7, holds.

$$\exists o' \in \textit{ford-min}(e_l) \text{ such that } o_b \leq o' \text{ and } cbp(e_l, o_b) = cbp(e_l, o') \tag{A.7}$$

$o' \in \textit{ford-min}(e_l)$ implies $(o' \wedge S) \in \textit{ford-min}(e_l, S)$. Therefore, from the construction of set $\mathcal{I}(e)$, we know:

$$\exists o_p \in \mathcal{I}(e) \text{ such that } (o' \wedge S) \leq o_p \tag{A.8}$$

$$\text{Since } o_b \leq o', \text{ we know } (o_b \wedge S) \leq (o' \wedge S) \tag{A.9}$$

Substituting Equation A.9 in Equation A.8, we get $(o_b \wedge S) \leq o_p$. Since both $o_b$ and $o_p$ are permutations of the same attribute set $S$, we must have $o_b = o_p$. *i.e.,* the optimal sort order $o_b$ must be in $\mathcal{I}(e)$.
**Case 2-B**: Suppose condition (ii), repeated below as Equation A.10, holds.

$$\exists o' \in \textit{ford-min}(e_l) \text{ such that } o' \leq o_b \text{ and } cbp(e_l, o') + coe(e_l, o', o_b) = cbp(e_l, o_b) \tag{A.10}$$

The plan cost for $e$, with $o_b$ as as the chosen sort order, is given by:

$$\begin{aligned} PC(e, o_b) &= cbp(e_l, o_b) + cbp(e_r, o_b) + coe(e, o_b, o) + CM(e_l, e_r) \\ &\quad \text{Substituting for } cbp(e_l, o_b) \text{ from Equation A.10, we get} \\ &= cbp(e_l, o') + coe(e_l, o', o_b) + cbp(e_r, o_b) + coe(e, o_b, o) + CM(e_l, e_r) \quad \text{(A.11)} \end{aligned}$$

$o' \in \textit{ford-min}(e_l)$ implies $\exists o_p \in \mathcal{I}(e)$ such that $(o' \wedge S) \leq o_p$. Since $o' \leq o_b$, we know $\textit{attrs}(o') \subseteq S$. Therefore, we have $o' \wedge S = o'$. And hence, $o' \leq o_p$. Also, since both $o_p$ and $o_b$ are permutations of $S$, we have $|o_b| = |o_p|$.

Since, $o_b \notin ford(e_r)$, we have $cbp(e_r, o_b) = cbp(e_r, o_p)$. Substituting this in Equation A.11, we get:

$$PC(e, o_b) = cbp(e_l, o') + coe(e_l, o', o_b) + cbp(e_r, o_p) + coe(e, o_b, o) + CM(e_l, e_r) \quad \text{(A.12)}$$

Since $o' \leq o_b$ and $o' \leq o_p$ and $|o_b| = |o_p|$ we can write Equation A.12 as:

$$\begin{aligned} PC(e, o_b) &= cbp(e_l, o') + coe(e_l, o', o_p) + cbp(e_r, o_p) + coe(e, o_b, o) + CM(e_l, e_r) \\ &\geq cbp(e_l, o_p) + cbp(e_r, o_p) + coe(e, o_b, o) + CM(e_l, e_r) \quad \text{(A.13)} \end{aligned}$$

Now, we show that $coe(e, o_b, o) \geq coe(e, o_p, o)$ to complete the proof.

**Case (a)**: Suppose, $o' \leq o$.

Since $\mathcal{I}(e)$ contains a sort order which subsumes, $o \wedge S$, it is possible to choose $o_p$ from $\mathcal{I}(e)$ such that $(o \wedge S) \leq o_p$. This implies, $|o_b \wedge o| \leq |o_p \wedge o|$. Hence, $coe(e, o_b, o) \geq coe(e, o_p, o)$. Substituting this in Equation A.13, we get:

$$\begin{aligned} PC(e, o_b) &\geq cbp(e_l, o_p) + cbp(e_r, o_p) + coe(e, o_p, o) + CM(e_l, e_r) \\ &\geq PC(e, o_p) \end{aligned}$$

**Case (b)**: Suppose, $o' \not\leq o$.

Now, $o' \wedge o = o_b \wedge o = o_p \wedge o$ (because $o' \leq o_b$ and $o' \leq o_p$). Therefore, $coe(e, o_b, o) = coe(e, o_p, o)$. Substituting this in Equation A.13, we get:

$$\begin{aligned} PC(e, o_b) &\geq cbp(e_l, o_p) + cbp(e_r, o_p) + coe(e, o_p, o) + CM(e_l, e_r) \\ &\geq PC(e, o_p) \end{aligned}$$

**Case 3:** Suppose $o_b$ is present in both $ford(e_l)$ and $ford(e_r)$

This implies one of the following:

(i) $\exists o' \in$ *ford-min*$(e_l) \cup$ *ford-min*$(e_r)$ such that $o_b \leq o'$. In this case the proof can proceed as in Case 2-A.

(ii) $\exists o_1 \in$ *ford-min*$(e_l)$ and $\exists o_2 \in$ *ford-min*$(e_r)$ such that *(a)* $o_1 \leq o_b$ and $o_2 \leq o_b$ and *(b)* $cbp(e_l, o_1) + coe(e_l, o_1, o_b) = cbp(e_l, o_b)$ and *(c)* $cbp(e_r, o_2) + coe(e_r, o_2, o_b) = cbp(e_r, o_b)$.

Since $o_1 \leq o_b$ and $o_2 \leq o_b$, either $o_1 \leq o_2$ or $o_2 \leq o_1$. Hence, $\exists o_p \in \mathcal{I}(e)$ such that $o_1 \leq o_p$ and $o_2 \leq o_p$. Choosing such an $o_p$, and proceeding as in Case 2-B we can prove $PC(e, o_b) \geq PC(e, o_p)$

This completes the proof of Theorem 3.6. □

# Appendix B

# Correctness of Transformation Rules

In this section, we give a formal proof of correctness of all the program transformation rules presented in Chapter 4.

The *program state* $G$ comprises of values for all variables accessible at a program position $p$ and the *system state* $S$ comprises of the state of all external resources like database and file system. Let $P_L$ be a program fragment that matches the LHS of a rule and $P_R$ be the program fragment instantiated by the corresponding RHS. Let $p$ be the position in the program at which $P_L$ begins. Let $(G, S)$ be the pair of *any* valid program and system states at $p$. To prove the correctness of a transformation rule, we must show the following. If the execution of $P_L$ on $(G, S)$ results in the state $(G', S')$ then the execution of $P_R$ on $(G, S)$ will also result in the state $(G', S')$.

- **Rule 1A(i):** In Rule 1A(i), both $P_L$ and $P_R$ do not modify the program state (as we use a call by value semantics and there are no global variables).

  Consider the multiset $S = \Pi^d_{c_1, c_2, \ldots, c_m}$ with which $qb$ is invoked. Let $S'$ be the multiset of tuples constructed from parameters passed to each invocation of $q$ inside the loop. We can see that $S$ is multiset equivalent to $S'$. Now the equivalence of the two program fragments follows directly from the definition of *batch-safe* operation (when an operation is batch-safe the final system state depends only on the set of parameters and not the order of invocations).

- **Rule 1A(ii):** Proof is similar to that of Rule 1A(i).

- **Rule 1B:** The only program state $P_L$ and $P_R$ modify is the table $r$. Let the initial state (state at the point where $P_L/P_R$ begins) of the table be $r_{init}$. Let $r'$ be the state the table reaches if $P_L$ is executed and $r''$ be the state the table reaches if $P_R$ is executed. We show $r'$ and $r''$ to be multiset equivalent.

Since $q$ is a scalar query, from the definition of *batched forms* it follows that, during merge, each tuple in $r_{init}$ matches with exactly one tuple in the result of the batched invocation, $qb(\Pi_{c_{r1},\ldots,c_{rm}}(r))$. As a result, *(i)* cardinalities of $r', r''$ and $r_{init}$ are equal, i.e., $|r'| = |r''| = |r_{init}|$ and *(ii)* For each tuple $t \in r_{init}$, there exists a distinct tuple $t' \in r'$ and a distinct tuple $t'' \in r''$ such that $t$, $t'$ and $t''$ have the same values for all attributes except (possibly) the updated attributes *viz.,* $c_{w1}, c_{w2}, \ldots, c_{wn}$.

Let $t_r$ be the tuple in the result of the batched invocation ($qb$) that matches (during merge) with tuple $t$ of $r_{init}$. Let $(v_1, v_2, \ldots, v_m)$ be the values of attributes $c_{r1}, c_{r2}, \ldots, c_{rm}$ of $t$. Therefore, attributes $c_{r1}, c_{r2}, \ldots, c_{rm}$ of $t_r$ must also have the values $(v_1, v_2, \ldots, v_m)$. Let $(w_1, w_2, \ldots, w_n)$ be the values of the remaining attributes (named $c_{w1'}, c_{w2'}, \ldots, c_{wn'}$) of $t_r$. From the definition of *batched forms*, we have $(w_1, w_2, \ldots, w_n) = q(v_1, v_2, \ldots, v_m)$. *i.e.,* the tuple resulting from the merge ($t''$ in $r''$) has values assigned from $q(v_1, v_2, \ldots, v_m)$ for its attributes $c_{w1}, c_{w2}, \ldots, c_{wn}$. From the LHS of the rule, it is clear that the corresponding tuple $t' \in r'$ also has the values of $q(v_1, v_2, \ldots, v_m)$ assigned for its attributes $c_{w1}, c_{w2}, \ldots,$ $c_{wn}$. This makes $t' = t''$ and hence $r' = r''$.

Since $q$ is a *pure* function the system state remains unaffected by both $P_L$ and $P_R$.

- **Rule 1C:** Proof is similar to the proof for 1B.

- **Rule 1D:** The equivalence directly follows from the definition of batch-safe operation.

- **Rule 2:** Let $P_L$ be the program fragment matching the LHS of the rule and $P_R$ be the program fragment instantiated by the RHS. Let us call the *while* loop in $P_L$ as $L$, the first (*while*) loop of $P_R$, which contains $ss'_1$, as $L_1$, the cursor loop after that, which contains $s'$, as $L_2$, and the last cursor loop of $P_R$, which contains $ss_2$, as $L_3$.

  First, we note that there exists a one-to-one correspondence between statements in $ss_1$ of $L$ and statements in $ss'_1$ of $L_1$. The correspondence follows from the construction of $ss'_1$. For every statement $s_x$ in $ss_1$, the corresponding statement $s'_x$ in $ss'_1$ performs exactly the same set of operations.

Let $v$ be the value of a variable $a$ at statement $s_x$ in $ss_1$ in the $i^{th}$ iteration of $L$. As there are no loop-carried flow dependence edges crossing the split boundaries, it is evident that the statement that assigns value $v$ to variable $a$ must also appear in $L_1$ or must precede $L_1$. Therefore, we can see that $v$ will be the value of $a$ at $s'_x$ (the statement corresponding to $s_x$) in the $i^{th}$ iteration of $L_1$.

Similarly, we can see that values read by statement $s'$ in the $i^{th}$ iteration of $L_2$ are same as the value read by statement $s$ in the corresponding iteration of $L$.

Now, consider a statement $s_y$ in $ss_2$ of loop $L$. Let $v$ be the value of a variable $a$ read by $s_y$ in the $i^{th}$ iteration of $L$. We now prove by induction on $i$ that $v$ will be the value read by the corresponding statement $s'_y$ in $L_3$.

Let $i = 1$ (the first iteration). In this case, the assignment of $v$ to $a$ must be performed by a statement that precedes $s_y$ in $L$. Therefore, we can observe that in the $i^{th}$ iteration of $L_3$, either a statement in $ss_r$ or a statement that precedes $s'_y$ in $ss_2$ will assign $v$ to $a$. And hence, $s'_y$ will read the value $v$ from $a$.

Now consider the $k^{th}$ iteration ($i = k$). In this case, $s_y$ will read the value of $a$ which is either assigned by a preceding statement in the $k^{th}$ iteration of $L$, or a the value of $a$, which was present at the end of previous ($(k-1)^{th}$) iteration. In the former case, we will again have a statement that precedes $s'_y$, which assigns value $v$ to $a$. In the later case, we know by the induction hypothesis that the value of $a$ at the end of $(k-1)^{th}$ iteration for both $L$ and $L_3$ will be the same. Since $a$ was not assigned in the $k^{th}$ iteration by any statement preceding $s_y$, no statement in $ss_r$ will overwrite $a$. And hence, we see that $v$ will be the value read by $s'_y$ in the $k^{th}$ iteration.

Since there are no inter-statement dependencies involving external system state, the output and change in system state produced by any statement $s_x$ and the corresponding statement $s'_x$ will be the same. Further, for every non local variable $gv$, if the last assignment in $P_L$ was made by a statement $s_x$ in the $i^{th}$ iteration of $L$, then in $P_R$ the last assignment will be made by the corresponding statement $s'_x$ in the $i^{th}$ iteration of $L_1$, $L_2$ or $L_2$. The only additional change introduced by $P_R$ to the program state is the new variable *loopkey*, which is not used after the point where $P_R$ ends and hence does not affect the program. Hence, the execution of both the $P_L$ and $P_R$ result in equivalent program and system states.

- **Rules 3, 4 and 5:** The equivalence of these rules is straight forward to infer.
- **Rule 6A:** As in the proof for *Rule 1A*, we can observe that the multiset of parameters passed to $qb$ in $P_R$ is equivalent to the multiset of parameters passed over all the iterations of $P_L$. Hence, from the definition of the batch-safe operation, the equivalence holds.
- **Rule 6B:** The proof is similar to that of *Rule 1B*.

# Appendix C

# Additional Transformation Rules

In Chapter 5, we mentioned that our program transformation techniques can be extended to build a single dataflow for queries with UDF invocations, and thereby avoid materialization of parameter batches. In this section, we present a few additional program transformation rules, which are useful for future work in this direction.

Rule 7 in Figure C.1 and Rule 8 in Figure C.2 are useful in replacing loops containing simple expressions and assignment with relational operators and avoiding materialization of intermediate results.

Batched forms of procedures are relation valued, *i.e.,* they return sets of tuples. For example, the batched form of Figure 4.1 constructs the table to be returned iteratively. Rule 9 in Figure C.3 is useful to convert such code into a set valued expression. The example in Appendix D illustrates the use of Rule 9.

**Rule 7**

for each t by ref in r [order by key] loop

    t.b = $\textit{arith-expr}(t.a_1, t.a_2, \ldots, t.a_n)$;

end loop;

$$\Updownarrow$$

$r = \Pi_{A, \textit{arith-expr}(t.a_1, t.a_2, \ldots, t.a_n) \textit{ as } b}(r)$,

where $A$=*schema(r)*$-\{b\}$

Figure C.1: Loops with Arithmetic Expressions and Assignment

**Rule 8**

Let expr1 be a side-effect free expression (*e.g.,* a query).

table r1 = expr1();

table r2 = expr2(r1);

*dead*(r1) // r1 unused hereafter

$\Updownarrow$

table r2 = expr2(expr1);

Figure C.2: Rule for Avoiding Materialization

**Rule 9**

table result;

for each t [by ref] in r [order by key] loop

result.addRecord(($c_1, c_2, \ldots, c_n$));

// where each $c_i$ is a function of attributes of $t$.

end loop;

return result;

$\Updownarrow$

return select $c_1, c_2, \ldots, c_n$ from r;

Figure C.3: Rule for *return* Statement

We omit a formal proof of correctness for these transformation rules, as their correctness is straight-forward to infer.

# Appendix D

# Transformation Examples

In this section, we illustrate the transformation of the UDFs in Example 1.3 and Example 4.1 as the rules get applied following the batching procedure in Figure 4.8. We call these two UDFs as UDF-1 and UDF-2 respectively. Here, we assume that every query needs to be batched (when possible) and with respect to all the loops enclosing it.

## Rewriting UDF-1

- **Generate the Trivial Batched Form:** First, we generate the *trivial batched form* of the procedure as explained in Section 4.1.3. Figure D.1 shows the resulting procedure.

- **Isolate the Query Execution:** We isolate the query (expression) to be batched using Rule-3 and convert the control-dependencies to flow-dependencies using Rule-4. The resulting procedure after applying these two rules is shown in Figure D.2.

- **Split the Loop:** Split the loop (by applying Rule-2) before and after the query execution statements. In this example, the pre-conditions for Rule-2 are directly satisfied and we do not need to reorder any statements. However, in some cases we may need to reorder the statements using Rule-5 to satisfy the pre-conditions for Rule-2. Figure D.3 shows the resulting program.

- **Replace Loops with Batched Calls:** Apply Rule-1D to remove the order-by clauses around batch-safe operations and then replace the iterations with batched calls using Rules 1B and 1C. We further apply Rule 9 (Figure C.3) for the RETURN

```
TABLE count-offers-batched(TABLE r1)
DECLARE
    TABLE result;
BEGIN
    FOR EACH t IN r1 LOOP
        FLOAT amount-usd;
        INT count-offers; // The return value named after the function
        IF (t.curcode == "USD")
            amount-usd := t.amount;
        ELSE
            amount-usd := t.amount * (SELECT exchrate FROM curexch
                                      WHERE ccode = t.curcode);
        END IF
        count-offers := SELECT count(*) FROM buyoffers
                        WHERE itemid = t.itemcode AND price >= amount-usd;
        result.addRecord((t.itemcode, t.amount, t.curcode, count-offers));
    END LOOP;
    RETURN result;
END;
```

Figure D.1: UDF-1: Trivial Batched Form

```
TABLE count-offers-batched(TABLE r1)
DECLARE
    TABLE result;
BEGIN
    FOR EACH t IN r1 LOOP
        FLOAT amount-usd; INT count-offers;
        BOOLEAN cond1; FLOAT exchrate;
        cond1 := (t.curcode == "USD");
        cond1 == true? amount-usd := t.amount;
        cond1 == false? exchrate := SELECT exchrate FROM curexch
                                    WHERE ccode = t.curcode;
        cond1 == false? amount-usd := t.amount * exchrate;
        count-offers := SELECT count(*) FROM buyoffers
                        WHERE itemid = t.itemcode AND price >= amount-usd;
        result.addRecord((t.itemcode, t.amount, t.curcode, count-offers));
    END LOOP;
    RETURN result;
END;
```

Figure D.2: UDF-1: After Applying Rules 3 and 4

statement. The resulting program is given in Figure D.4. Earlier, in Example 4.4 we had shown this batched form with minor simplifications for readability.

```
TABLE count-offers-batched(TABLE r1)
DECLARE
    TABLE result; INT loopkey = 0;
    TABLE (key, itemcode, amount, curcode, cond1, exchrate, amount-usd, count-offers) r2;
BEGIN
    FOR EACH t IN r1 LOOP
        FLOAT amount-usd; BOOLEAN cond1;
        RECORD rec;

        cond1 := (t.curcode == "USD");
        cond1 == true? amount-usd := t.amount;
        rec.key = loopkey++;
        rec.itemcode = t.itemcode;
        rec.amount = t.amount;
        rec.curcode = t.curcode;
        rec.cond1 = cond1;
        rec.amount-usd = amount-usd;
        r2.addRecord(rec);
    END LOOP;

    FOR EACH t BY REF IN r2 ORDER BY key LOOP
        t.cond1 == false? t.exchrate :=
                    SELECT exchrate FROM curexch WHERE ccode = t.curcode;
    END LOOP;

    FOR EACH t BY REF IN r2 ORDER BY key LOOP
        t.cond1 == false? t.amount-usd := t.amount * t.exchrate;
    END LOOP;

    FOR EACH t BY REF IN r2 ORDER BY key LOOP
        t.count-offers := SELECT count(*) FROM buyoffers
                        WHERE itemid = t.itemcode AND price >= t.amount-usd;
    END LOOP;

    FOR EACH t BY REF IN r2 ORDER BY key LOOP
        result.addRecord((t.itemcode, t.amount, t.curcode, t.count-offers));
    END LOOP;

    RETURN result;
END;
```

Figure D.3: UDF-1: After Loop Split

## Rewriting UDF-2

UDF-2 (Example 4.1) contains two queries, one in statement s8 and the other in statement s10. As mentioned in Section 4.5, we cannot batch the query in statement s10 due to cyclic flow dependence. However, we can batch the query in statement s8 with respect to the WHILE loop (of s5) as well as the outermost cursor loop, which iterates over all the

```
TABLE count-offers-batched(TABLE r1)
DECLARE
    TABLE (key, itemcode, amount, curcode, cond1, exchrate, amount-usd, count-offers) r2;
    INT loopkey = 0;
BEGIN
    FOR EACH t IN r1 LOOP
        FLOAT amount-usd; BOOLEAN cond1;
        RECORD rec;
        cond1 := (t.curcode == "USD");
        cond1 == true? amount-usd := t.amount;
        rec.key = loopkey++;
        rec.itemcode = t.itemcode;
        rec.amount = t.amount;
        rec.curcode = t.curcode;
        rec.cond1 = cond1;
        rec.amount-usd = amount-usd;
        r2.addRecord(rec);
    END LOOP;

    MERGE INTO r2 USING q1b(b1) AS q1b ON (r2.curcode = q1b.curcode)
    WHEN MATCHED THEN UPDATE SET exchrate = q1b.exchrate;

//  where the parameter batch b1 is constructed as:
//      SELECT distinct curcode FROM r2 WHERE cond1=false;
//  and the batched form q1b(b1) is defined as:
//      SELECT b1.curcode, c.exchrate FROM b1 JOIN curexch c ON b1.curcode=c.ccode;

    FOR EACH t BY REF IN r2 ORDER BY key LOOP
        t.cond1 == false? t.amount-usd := t.amount * t.exchrate;
    END LOOP;

    MERGE INTO r2 USING q2b(b2) AS q2b
    ON (r2.itemcode=q2b.itemcode AND r2.amount-usd=q2b.amount-usd)
    WHEN MATCHED THEN UPDATE SET count-offers = q2b.count-offers;

    where b2 = SELECT distinct itemcode, amount-usd FROM r2;
    and q2b(b2) = SELECT b2.itemcode, b2.amount-usd, count(o.itemcode) AS count-offers
                  FROM b2 LEFT OUTER JOIN buyoffers o ON o.itemid = b2.itemcode AND
                                                         o.price >= b2.amount-usd
                  GROUP BY b2.itemcode, b2.amount-usd;

    RETURN SELECT itemcode, curcode, amount, count-offers FROM r2;
END;
```

Figure D.4: UDF-1: The Final Batched Form

parameters (in the *trivial batched form*).

In Example 4.1, observe that splitting the WHILE loop (of s5) before and after the query execution statement (of s8) is not directly possible due to the loop-carried dependencies from s11 and s12 to s5, s6 and s7, which violate pre-condition c1 of Rule-2. We therefore, reorder of statements by moving statements s8 and s9 past s12 (using Rule-5). We then split the WHILE loop and batch the query execution. The batched query execution is further pulled out of the outermost cursor loop in the *trivial batched form* using Rule-6.

Figure D.5 shows the final batched form of UDF-2. The functions NEST and UNNEST implement the *nest* and *unnest* operations discussed in Section 4.3.6 and take the corresponding arguments. NEST takes as its arguments the table, columns to be nested and the name for the resulting table-valued column. Similarly, the UNNEST method takes the table and the name of the table-valued column that needs to be unnested.

```
TABLE count-items-batched(TABLE pb)
DECLARE
    TABLE (key, catid, loop-table2, totalcount) loop-table1;
    INT loopkey1 = 0;
BEGIN
    FOR EACH t IN pb LOOP
        INT totalcount := 0; INT top := 0; INT stack[100]; RECORD rec1;
        stack[top] := t.catid;
        top := top + 1;

        TABLE (key, curcat, catitems) loop-table2;
        int loopkey2 = 0;
        WHILE top > 0 LOOP
            RECORD rec2;
            top := top - 1;
            curcat := stack[top];
            // Now push all the subcategory ids onto the stack
            FOR catrec IN SELECT category-id FROM category
                          WHERE parent-category=curcat LOOP
                stack[top] := catrec.category-id;
                top := top + 1;
            END LOOP;
            rec2.key = loopkey2++; rec2.curcat = curcat; loop-table2.addRecord(rec2);
        END LOOP;

        rec1.key = loopkey1++; rec1.catid = t.catid; rec1.loop-table2 = loop-table2;
        loop-table1.addRecord(rec1);
    END LOOP;

    temp = UNNEST(loop-table1, "loop-table2");
    MERGE INTO temp USING qb(b) AS qbr(curcat, res) ON temp.curcat = qbr.curcat
    WHEN MATCHED THEN UPDATE SET catitems = res;

//  where the parameter batch b is constructed as:
//      SELECT distinct curcat FROM temp;
//  and the batched form qb(b) is defined as:
//      SELECT b.curcat, count(itemid) AS catitems
//      FROM b LEFT OUTER JOIN item ON category-id=curcat GROUP BY curcat;
    loop-table1 = NEST(temp, schema(loop-table1.loop-table2), "loop-table2");

    FOR EACH rec1 BY REF IN loop-table1 ORDER BY key LOOP
        FOR EACH rec2 BY REF IN rec1.loop-table2 ORDER BY key LOOP
            rec1.totalcount := rec1.totalcount + rec2.catitems;
        END LOOP;
    END LOOP;

    RETURN SELECT catid, totalcount FROM loop-table1;
END;
```

Figure D.5: UDF-2: The Final Batched Form

# Appendix E

# Procedures Used in Experiments

This section gives pseudocode for the additional procedures used for performance evaluation in Section 4.6. Figure E.1 shows the procedure for Experiment-3 and the procedure for Experiment-2 is given in Figure E.2. The functionality implemented by these procedures was explained earlier, in Section 4.6.

```
PROCEDURE expand-issued-forms(DATE issuedate)
DECLARE
    INT num;
BEGIN
    FOR EACH irec IN SELECT agent_id, start_no, end_no, issue_date
                    FROM issued_forms WHERE issue_date = issuedate LOOP
        num := irec.start_no;
        WHILE (num <= irec.end_no) LOOP
            INSERT INTO forms-master VALUES (num, irec.agent_id, irec.issue_date, 'NEW');
            num := num + 1;
        END LOOP;
    END LOOP;
END;
```

Figure E.1: Procedure for Experiment 3

```
PROCEDURE emp-upload(VARCHAR filename)
DECLARE
    // Data types of local variables omitted for brevity.
    empid, clientid, iempid, optcode, optinfo, termcode, taxinfo,
    city, state, zip, operation, curtaxinfo, curcity, curstate, curzip
BEGIN
    fd := open(filename);
    linestr := readline(fd);
    WHILE (linestr ! = null) LOOP
        tokenize linestr and extract empid, clientid, optcode, ... zip
        // some validation and pre-processing code
        if(optcode == 0)
            optinfo = ... ;
        ...

        SELECT internl-empid into iempid, tax-info into curtaxinfo,
        FROM options WHERE client-id=clientid AND emp-id=empid;

        // If options has no record for the employee
        if(iempid == null) {
            operation := 1; // we must insert
            iempid := gen-new-id();
        }
        else {
            operation := 2; // we must update

            SELECT city into curcity, state into curstate, zip into curzip
            FROM contactinfo WHERE internal-empid=iempid;

            // Retain the current values if new ones are blank
            if(taxinfo == "")
                taxinfo := curtaxinfo;
            if(city == "")
                city := curcity;
            ...
            ...
        }
        if(operation == 1) {
            INSERT INTO options VALUES (iempid, clientid, ... optinfo, ... taxinfo);
            INSERT INTO contactinfo VALUES(iempid, city, state, zip);
        }
        else {
            UPDATE options set option-info=optinfo, ... tax-info=taxinfo
            WHERE internal-empid=iempid;
            UPDATE contactinfo SET city=city, state=state, zip=zip
            WHERE internal-empid=iempid;
        }
        linestr := readline(fd);
    END LOOP;
END;
```

Figure E.2: Procedure for Experiment 2

# Appendix F

# API and Code Patterns

As mentioned in Section 4.6, we implemented the transformation rules for Java because tools for Java program analysis are available in public domain. We make use of the *SOOT* optimization framework for obtaining data dependence information. To simplify the task of recognizing query execution statements and code patterns that match a rule, our current implementation requires that queries and updates be performed using our API layer built on top of JDBC. During rewrite, we recognize these calls and transform them for batched bindings when possible. In this section, we give the details of our API layer and the Java code patterns, which map to constructs described earlier, in this thesis. *SOOT* uses an intermediate code representation called *Jimple*. Our implementation works on *Jimple* and transforms it back to Java. Recognizing *Jimple* code patterns corresponding to each of our API calls and Java code patters is relatively straight forward and we omit the details.

- **Query/Update Execution Statements**: The class *DBI* in our implementation provides various methods for executing queries and updates.

  - *Record executeScalarQuery(int queryId, Record params)*
  - *Table executeQuery(int queryId, Record params)*
  - *int executeUpdate(int queryId, Record params)*

  The *queryId* specifies a parameterized SQL query or update statement in a query registry. Unlike JDBC, where the query string is directly specified in the program, we make use of a registry. The query registry, in addition to the query string, also contains the manually written batched form for the query. Unlike position-based parameters in JDBC, we use named parameters. The *Record* class is used to pass parameters as name-value pairs and also to obtain the result of a scalar query. The class *Table* implements a tuple set and is used to retrieve query results, and for constructing parameter batches.

| | | |
|---|---|---|
| Table res = DBI.executeQuery(...);<br>Iterator resIter = res.iterator();<br>while(resIter.hasNext()) {<br>    Record r = (Record) resIter.next();<br>    ...<br>} | $\Longrightarrow$ | for each r in *query*<br>    ...<br>end loop; |

Figure F.1: Pattern for Cursor Loops

- **Looping Statements**: *while* loops have a direct mapping to Java. Unlike some of the procedural languages offered by database systems (e.g., PL/SQL), Java does not have cursor loops. Each cursor loop maps to a sequence of statements in Java. Figure F.1 shows the code pattern.

- **Batched Execution**: The class *Table* in our implementation can be used for constructing parameter batches row by row using the method *addRecord(Record r)*. The DBI class has methods for executing batched forms of queries by passing a batch of parameters and also for merging back the results. These methods are used by the transformed program (generated code).

  - *Table executeBatchedQuery(int queryId, Table paramBatch, FilterPred pred)*
  - *void executeBatchedUpdate(int queryId, Table paramBatch, FilterPred pred)*

- **Other Constructs**: Control flow, assignment and other constructs in the simple procedural language used in this thesis have a direct mapping to Java. Conditional statements, generated when control-dependencies are converted to flow-dependencies, are Java *if* statements, where the *if-block* contains a single statement and the predicate is a boolean variable.

# Publications based on this work

1. Ravindra Guravannavar, Ramanujam H.S. and S. Sudarshan, *Optimizing Nested Queries with Parameter Sort Orders*, Intl. Conf. on Very Large Databases (VLDB) 2005.

2. Ravindra Guravannavar and S. Sudarshan, *Reducing Order Enforcement Cost in Complex Query Plans*, IEEE 23rd Intl. Conf. on Data Engineering (ICDE) 2007.

3. Ravindra Guravannavar, S. Sudarshan, Ajit A. Diwan and Ch. Sobhan Babu, Technical Report: *Reducing Order Enforcement Cost in Complex Query Plans*, arXiv:cs.DB/0611094 v1 20 Nov 2006. (Extended version of the ICDE 07 paper).

4. Ravindra Guravannavar and S. Sudarshan, *Rewriting Procedures for Batched Bindings*, Proc. of VLDB Endowment, Vol 1, 2008.

# Acknowledgments

I have been very fortunate to have Prof. Sudarshan as my advisor. I am indebted to Sudarshan for motivating me to take up research in database systems, for guiding me at every stage and for his constant encouragement. It is only due to his patience and understanding that I could cross the hurdles of returning to academic life after a long gap. His guidance and appreciation were essential to drive things till the finishing line.

I thank Prof. Sunita Sarawagi, Prof. N.L. Sarda and Prof. Krithi Ramamritham for their feedback at various stages, and Prof. Ajit Diwan and Dr. Sobhan Babu for extending their help and collaborating on parts of this thesis. Thanks to Prasan Roy for the extensible Volcano implementation, it was a joy working with his code; to Ramanujam H. S. and C. Santosh Kumar for assisting me in parts of the implementation. I thank Yogesh Murarka for the several discussions, which educated me on the techniques and terminology used in compilers and gave me clarity on several issues, and Sobhan, Shetal and Aru for painstakingly reviewing parts of this thesis and my technical papers.

I thank all my friends at IIT Bombay for making my days here so joyful. Special thanks to Shetal, Sundar, Sita madam, my music teacher Shakuntala madam and Puru for being so nice to me, and for feeding me often with great food!, and to Sobhan for introducing me to the mountains and marathons. Infolab has always been a great place to work and I thank all the labmates, past and present, for having created such a wonderful environment. The members of CSE office staff, particularly Vijay Ambre, have provided timely help in all administrative tasks.

I thank Trivikram Nayak and my earlier colleagues at Aztecsoft, particularly Govi, for all the help during the initial years. I am highly grateful to Bell Laboratories for supporting me through a Ph.D. fellowship.

Finally, I thank all my family members, especially my little nieces, for their understanding through all these years.

Ravindra Guravannavar